\documentclass[longauth]{aa}  

\usepackage{graphicx}
\usepackage{txfonts}
\usepackage{natbib}
\usepackage{float}
\usepackage{graphicx}
\usepackage{pstricks}
\usepackage{enumitem}
\usepackage{amsmath}
\usepackage{amssymb}

\bibpunct{(}{)}{;}{a}{}{,}

\newcommand{\sophi}{{\sc SO/PHI}}
\newcommand{\bm}[1]{\mbox{\boldmath$#1$\unboldmath}}

\newcommand{\varcsec}{^{\prime\!\hskip0.4pt\prime}}
\renewcommand{\arcsec}{.\hspace{-0.9mm}'\!\hskip0.4pt'\hspace{-0.2mm}}
\newcommand{\kms}{${\rm km}\,{\rm s}^{-1}$}

\begin{document} 

   \title{The Polarimetric and Helioseismic Imager on Solar Orbiter}

   \author{S.K.~Solanki\inst{1,15}\thanks{\hbox{Corresponding author: Sami~K.~Solanki} \hbox{\email{solanki@mps.mpg.de}}}
   \and 
   J.C.~del Toro Iniesta\inst{2}
     \and J.~Woch\inst{1} \and A.~Gandorfer\inst{1} \and
     J.~Hirzberger\inst{1} \and A.~Alvarez-Herrero\inst{3} \and
     T.~Appourchaux\inst{4} \and V.~Mart\'\i nez Pillet\inst{5} \and
     I.~P\' erez-Grande\inst{6}
% ...
}

   \institute{
         Max-Planck-Institut f\"ur Sonnensystemforschung, Justus-von-Liebig-Weg 3,
         37077 G\"ottingen, Germany \\ \email{solanki@mps.mpg.de}
         \and
         Instituto de Astrofísica de Andalucía (IAA-CSIC), Apartado de Correos 3004,
         E-18080 Granada, Spain \\ \email{jti@iaa.es}
         \and
         Instituto Nacional de T\' ecnica Aeroespacial, Carretera de
         Ajalvir, km 4, E-28850 Torrej\' on de Ardoz, Spain
         \and
         Univ. Paris-Sud, Institut d’Astrophysique Spatiale, UMR 8617,
         CNRS, B\^ atiment 121, 91405 Orsay Cedex, France
         \and
         National Solar Observatory, 3665 Discovery Drive, Boulder, CO 80303, USA
         \and
         Universidad Polit\'ecnica de Madrid, IDR/UPM, Plaza Cardenal Cisneros 3, E-28040 Madrid, Spain
         \and
         Universitat de Valencia, Dpt. Ingenieria Electronica, Avd. de
         la Universitat s/n Burjassot, 46100 Spain
         \and
         Grupo de Astronom\'\i a y Ciencias del Espacio, Universidad de
         Valencia, E-46980 Paterna, Valencia, Spain
         \and
         Kiepenheuer-Institut für Sonnenphysik, Sch\" oneckstr. 6, D-79104 Freiburg, Germany
         \and
         University of Barcelona, Department of Electronics, Carrer de Mart\'\i\ i Franqu\`es, 1 - 11, 08028 Barcelona, Spain
         \and
         Institut f\"ur Datentechnik und Kommunikationsnetze der TU
         Braunschweig, Hans-Sommer-Str. 66, 38106 Braunschweig,
         Germany
         \and
         Instituto de Astrof\'isica de Canarias, Avda. V\'\i a L\'actea s/n, La Laguna, Spain
         \and
         European Space Agency, European Space Research and Technology Centre, Keplerlaan 1, 2201 AZ Noordwijk, The Netherlands 
         \and
        LESIA, CNRS, PSL Research University, Universit\'e Pierre et Marie Curie, Universit\'e Denis Diderot, Observatoire de Paris, 92195, Meudon Cedex, France
        \and
        School of Space Research, Kyung Hee University, Yongin, Gyeonggi-Do, 446-701, Korea
        \and
        Institut f\"ur Astrophysik, Georg-August-Universit\"at G\"ottingen, Friedrich-Hund-Platz 1, 37077 G\"ottingen, Germany}

\date{Received December 31, 2018; accepted January 1, 2019}

 
  \abstract
   {}
   {This paper describes the Polarimetric and Helioseismic Imager on the Solar Orbiter mission (\sophi ), the first magnetograph and helioseismology instrument to observe the Sun from outside the Sun-Earth line. It is the key instrument meant to address the top-level science question: How does the solar dynamo work and drive connections between the Sun and the heliosphere? \sophi\ will also play an important role in answering the other top-level science questions of Solar Orbiter, as well as hosting the potential of a rich return in further science. }
   {\sophi\ measures the Zeeman effect and the Doppler shift in the Fe~{\sc i} 617.3 nm spectral line. To this end, the instrument carries out narrow-band imaging spectro-polarimetry using a tunable LiNbO$_3$ Fabry-Perot etalon, while the polarisation modulation is done with liquid crystal variable retarders (LCVRs). The line and the nearby continuum are sampled at six wavelength points and the data are recorded by a 2k $\times$ 2k CMOS detector. To save valuable telemetry, the raw data are reduced on board, including being inverted under the assumption of a Milne-Eddington atmosphere, although simpler reduction methods are also available on board. \sophi\ is composed of two telescopes; one, the Full Disc Telescope (FDT), covers the full solar disc at all phases of the orbit, while the other, the High Resolution Telescope (HRT), can resolve  structures as small as 200\,km on the Sun at closest perihelion. The high heat load generated through proximity to the Sun is greatly reduced by the multilayer-coated entrance windows to the two telescopes that allow less than 4\%\ of the total sunlight to enter the instrument, most of it in a narrow wavelength band around the chosen spectral line. }
   {\sophi\ was designed and built by a consortium having partners in Germany, Spain, and France. The flight model was delivered to Airbus Defence and Space, Stevenage, and successfully integrated into the Solar Orbiter spacecraft. 
   A number of innovations were introduced compared with earlier space-based spectropolarimeters, thus allowing \sophi\ to fit into the tight mass, volume, power and telemetry budgets provided by the Solar Orbiter spacecraft and to meet the (e.g. thermal) challenges posed by the mission's highly elliptical orbit.}
   {}


\maketitle


\section{Introduction}

The Sun's magnetic field is to a large extent responsible for driving a host of active phenomena, ranging from sunspots at its surface to coronal mass ejections propagating through the heliosphere \citep{solanki06,wiegelmann14}. The magnetic field also couples the various layers of the solar atmosphere, connecting the solar surface to the chromosphere and corona and transporting  the energy needed to heat the upper atmosphere and to accelerate the solar wind. Hence it is imperative to measure the Sun's magnetic field if we are to follow, understand and model the active phenomena on the Sun and in the heliosphere.  Consequently, polarimeters aimed at measuring the magnetic field have become increasingly central to solar physics, although only few have flown in space so far, mainly because of their complexity and the technical challenges involved. 

Whereas a polarimeter can measure the field in the solar atmosphere, usually close to the solar surface, the magnetic field itself is produced in the solar interior, which is opaque to electromagnetic radiation  \citep[e.g.][]{charbonneau10}. To gain insight into the structures and forces acting there we must take recourse to helioseismology, meaning the study of the acoustic waves that are excited profusely in the convection zone of the Sun \citep[e.g.][]{gizon05,basu16}. 

In this paper we describe the \sophi\ instrument, the Polarimetric and Helioseismic Imager on board the Solar Orbiter Mission \citep[][]{Garcia2019}. This instrument aims at achieving both tasks outlined above, in other words, measuring the magnetic field at the solar surface and probing the solar interior by measuring oscillations seen in the line-of-sight velocity. It is one of the suite of remote sensing instruments on Solar Orbiter, the first medium class mission of the European Space Agency's Cosmic Vision programme \citep[][]{Mueller2019a}.

\sophi\ is a magnetograph, the fifth   instrument aimed at measuring the solar magnetic field in space, after SOHO/MDI \citep{scherrer95}, SDO/HMI \citep{schou12}, Hinode/SP \citep{lites13} and Hinode/NFI \citep{tsuneta08}. It is also an instrument designed to do helioseismology from space, after SOHO/MDI \citep{scherrer95}, SOHO/GOLF \citep{gabriel95}, SOHO/VIRGO \citep{froehlich95}, SDO/HMI \citep{schou12} and Picard \citep{corbard13}.

The capabilities of \sophi\ differ from these earlier instruments in a number of ways. Firstly, \sophi\ is the first magnetograph that will observe the Sun from outside the Sun-Earth line. Secondly, it is the first such instrument planned to leave the ecliptic and get a clear view of the solar poles. It has two channels, one to observe the full solar disc and another to observe the Sun at high resolution (which is qualitatively similar to SOHO/MDI, although \sophi\ will reach considerably higher spatial resolution around perihelion). 

The paper is structured as follows. In Section~2 the science objectives of the \sophi\ instrument are given, which significantly overlap with those of the Solar Orbiter mission as a whole. An overview of the instrument is provided in Section~\ref{instrument_overview}, with more details on various aspects of the instrument being provided in Section~\ref{o-unit} (optical unit), Section~\ref{e-unit} (electronics), Section~\ref{calibration} (instrument characterisation and calibration) and Section~\ref{sc_operations} (science operations). Finally, a summary is given in Section~\ref{summary}. 


\section{Science objectives}

\subsection{Top level science questions}

The overarching science goal of Solar Orbiter is to answer the question: {\it How does the Sun create and control the heliosphere? } This umbrella encompasses four top-level science questions: 
\begin{enumerate}
    \item How does the solar dynamo work and drive connections between the Sun and the heliosphere?    \item What drives the solar wind and where does the coronal magnetic field originate from?
    \item How do solar transients drive heliospheric variability?
    \item How do solar eruptions produce energetic particle radiation that fills the heliosphere?

 \end{enumerate}
A more detailed discussion of these questions is given in the Solar Orbiter Red Book \citep[see][]{marsden11a} and in \cite{Mueller:2013a}.

The magnetograms and helioseismic data recorded by \sophi\ will provide significant, often vital information to answer the above questions. We consider these questions individually in Sects.~2.2, 2.3, 2.4 and 2.5, respectively and consider how \sophi\ will help to address them. Finally, in Section~\ref{additional_science} we also present and discuss additional science questions. \sophi\ is unique in being the first ever magnetograph and helioseismology instrument to observe the Sun from outside the Sun-Earth line, enabling it to do science that goes beyond the specific aims identified in the Solar Orbiter Red Book. This will allow \sophi\ to greatly enhance the science output from Solar Orbiter.

\subsection{How does the solar dynamo work and drive connections between the Sun and heliosphere?}

The magnetic field of the Sun is, in one way or another, the main driver of solar activity. It structures the solar chromosphere and corona and is responsible for coronal heating, it leads to flares and CMEs besides playing an important role in driving the solar wind \citep{solanki06,priest14}. The magnetic field, its large-scale structure and its roughly 11-year cycle \citep[e.g.][]{hathaway10}, are clearly results of a dynamo mechanism \citep[e.g.][]{charbonneau10,cameron17}. Nonetheless, there are still many open questions surrounding the nature of this dynamo. For example, there is no consensus on the depth below the solar surface at which the dynamo responsible for sunspots and the solar cycle is located; there is not even agreement if it is mainly restricted to the overshoot layer below the convection zone, or if it is distributed over (a part of) the convection zone \citep{babcock61,leighton69,cameron15}. Proposals for the location of the dynamo cover the two main radial shear layers, one  at the bottom of the convection zone and one near the solar surface \citep{howe09,brandenburg05}, cf. \citet{charbonneau13}. There is also a debate on whether the dynamo responsible for the solar cycle is the only solar dynamo, or if a separate small-scale turbulent dynamo is also acting closer to the solar surface \citep[][]{voegler07}. The variety of approaches and models of the dynamo responsible for the solar cycle have been reviewed by \citet{charbonneau10,charbonneau14}, who also discusses some of the major open questions.

Critical unknowns entering solar dynamo models are the structures of the Sun's magnetic and flow fields at high latitudes. In particular, the poloidal field at solar activity minimum is the source of the toroidal magnetic flux that dominates during the high activity phase of the solar cycle \citep[][]{babcock61,cameron15}. Thus, the polar magnetic flux at activity minimum is the parameter that best predicts the strength of the next solar cycle \citep[e.g.][]{schatten78,petrovay10}, so that it clearly plays a critical role in seeding the solar dynamo. However, because all magnetographs built so far have observed from within the ecliptic plane, they have only limited sensitivity to the polar field. High-resolution images and magnetograms of the polar region taken with the SOT/SP on board Hinode \citep[][]{tsuneta08a,tsuneta08,shiota12} show a rich and evolving landscape of magnetic features around the poles. However, at the poles themselves the results are less clear-cut, largely due to the very strong foreshortening, but also because the nearly vertical magnetic features close to the poles are almost perpendicular to the line-of-sight (LOS), leading to a small signal in the magnetograms showing the line-of-sight magnetic field. Measurements and models of solar polar magnetic fields are reviewed by \citet{petrie15}. 

The overarching question in the title of this subsection leads to a series of more detailed questions: How is the surface magnetic field transported in latitude by the meridional flow? Are there multiple cells in latitude? Where is the return flow located and what role does it play in the evolution of the field? What is the rotation rate near the poles and how does that effect the evolution of the field? How is the field reprocessed at high latitudes? Is there a significant local solar dynamo? In the following subsections we will discuss these questions and demonstrate how \sophi\ will address them, especially by using the high latitude passes.

\subsubsection{What is the structure of the solar rotation?}

The transport of the magnetic flux near the poles by convection, differential rotation and meridional flows is important for the polarity reversal of the global magnetic field \citep[see][]{wang89,sheeley91,makarov03,jiang14}. To gain insight into surface magnetic flux transport, the driving flows must be studied and the motions of the magnetic flux elements followed. 

Thanks to global helioseismology \citep[e.g.][]{christensen02,basu16}, the solar differential rotation has been mapped as a function of latitude and radius throughout most of the Sun \citep{schou98}. However, uncertainties in the results become large at high latitudes, so that there is a gap in our knowledge for heliographic latitudes above $\approx 70^\circ$. 
 
 While current data above $\approx 70^{\circ}$ heliographic latitude are uncertain, the bands of faster and slower rotation (torsional oscillations) moving towards the poles above $45^{\circ}$ latitude show a very dynamic behaviour in the near-polar regions \citep{2018ApJ...862L...5H}. 
 
 An alternative to global helioseismology is local helioseismology \citep[e.g.][]{gizon05}, which aims to measure the 3D velocity vectors of the material flows in the solar interior, allowing studies of convective, rotational and meridional flows, as well as of the subsurface structures of sunspots and active regions. Local helioseismology, using \sophi\ observations will enable the study of all major subsurface flows at high heliographic latitudes. By repeating both, global and local helioseismology measurements over the
course of the mission, it will be possible to deduce solar-cycle variations in these flows.


\begin{figure}[h]
    \centering
    \includegraphics[width=\columnwidth]{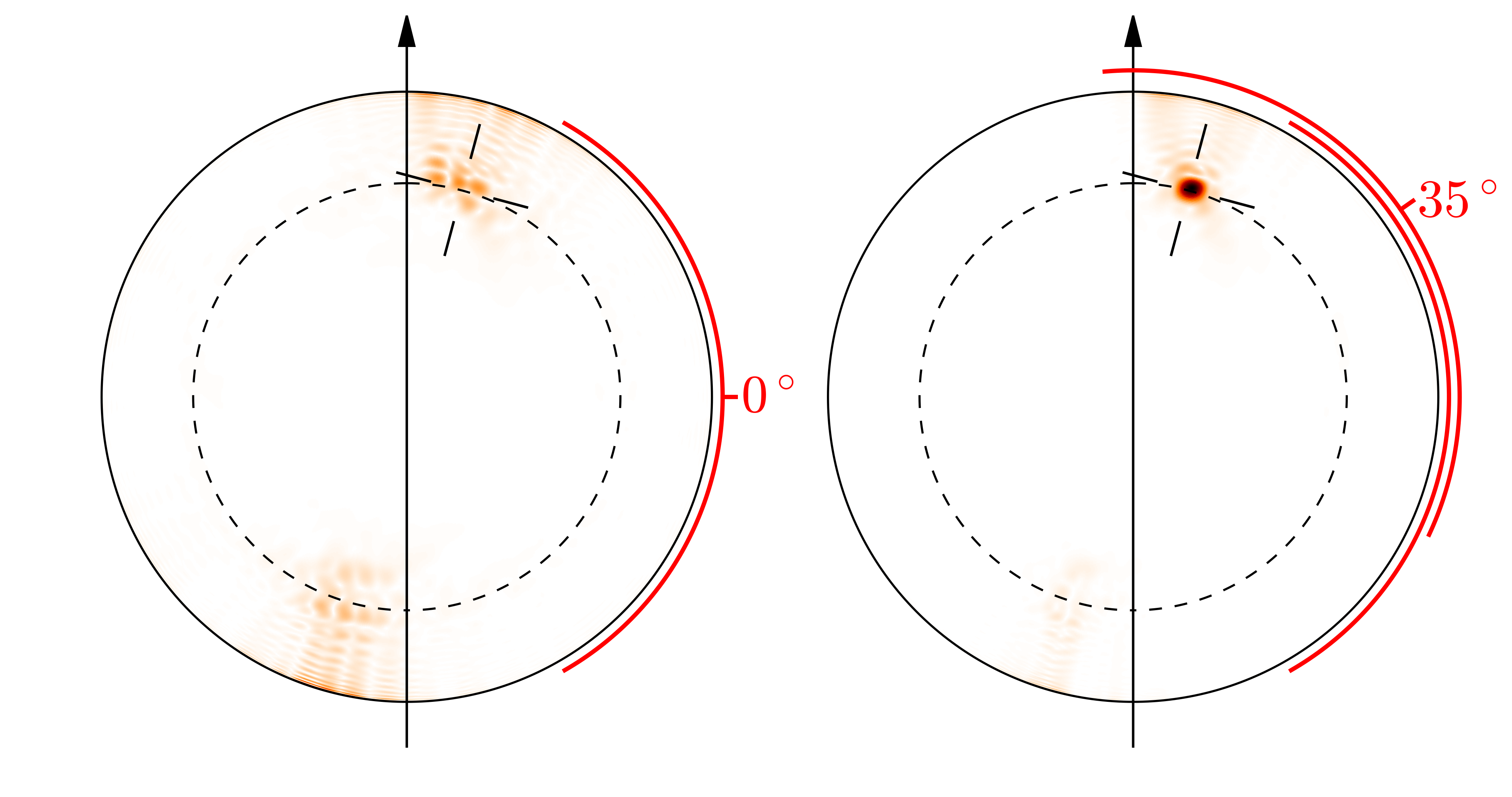}
    \caption{Stereoscopic helioseismology. Numerical simulations showing the spatial sensitivity of helioseismic holography to the meridional flow at latitude 75$^\circ$ and radius  $0.7 R_\odot$. The resolution is poor when using only observations of solar oscillations in the equatorial plane (left panel, data coverage indicated in red). The resolution approaches the diffraction limit of $\lambda/2 \approx 38$~Mm when combining the previous data with observations from a line of sight inclined by $35^\circ$. The noise in the measurements may be very high; it depends on the total duration of the observations. See \cite{2018arXiv181000402G} for a discussion of  signal and noise in helioseismic holography.}
   \label{F:stereoseismology}
\end{figure}


To this end \sophi\ will enable us to use stereoscopic helioseismology by combining \sophi\ data with Doppler measurements from Earth-based, or Earth-orbiting instruments, e.g. GONG \citep[][]{harvey96} or SDO/HMI (see Fig.~\ref{F:stereoseismology}). Local helioseismic inversions from techniques such as time-distance helioseismology, or helioseismic holography will be able to probe deep into the Sun using observations from widely separated vantage points, because skip distances (distances between the surface end points of the ray paths) of order half a circumference will at last become accessible \citep{loeptien15}. This will be important for probing the tachocline at the base of the convection zone, where the dynamo has been surmised to be situated. 

A discussion of the helioseismic investigations that can be done with \sophi\ has been published by \citet{loeptien15}.

\begin{figure}[]
  \centering
  \includegraphics[width=90mm]{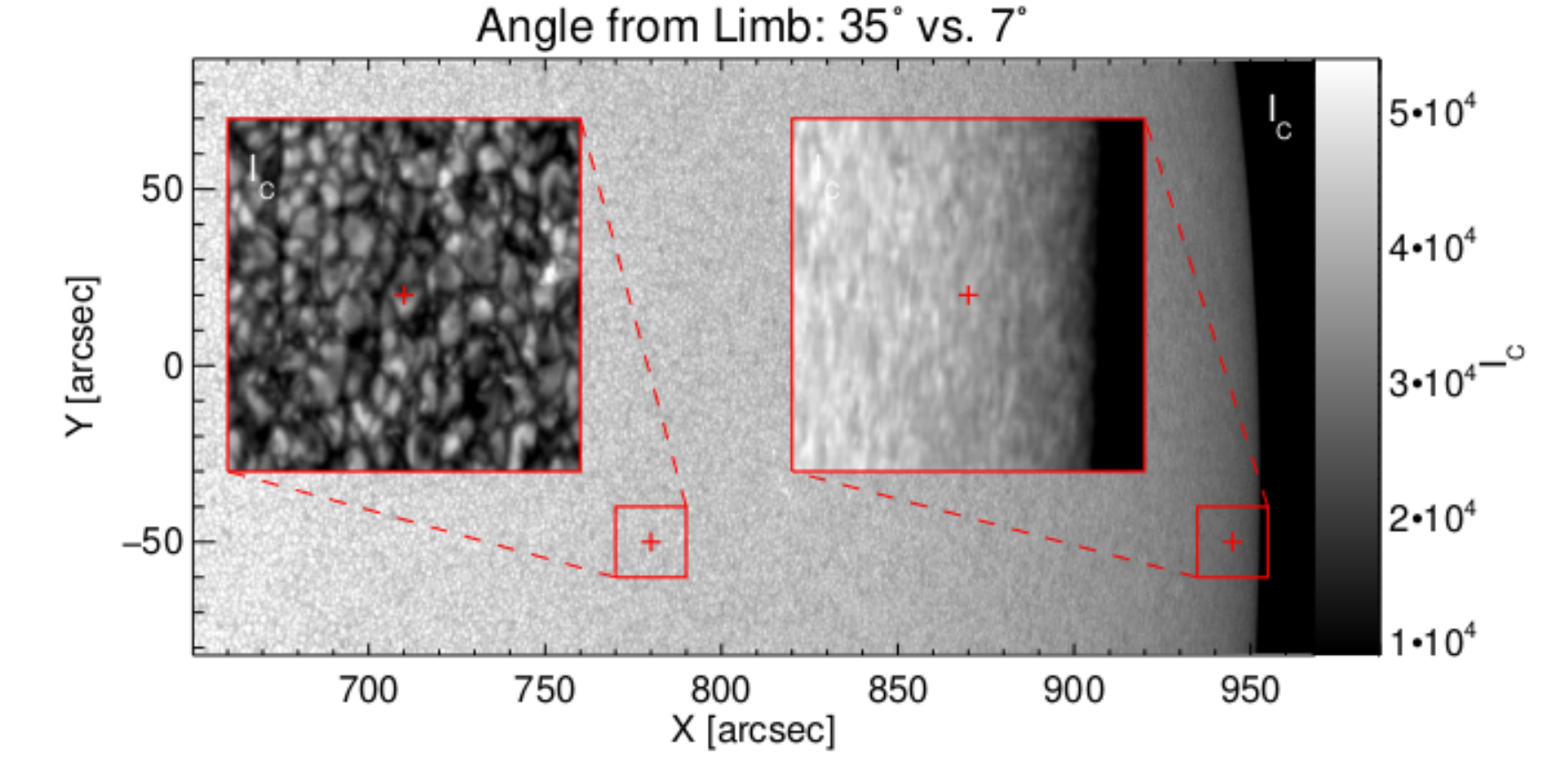}
  
  \includegraphics[width=90mm]{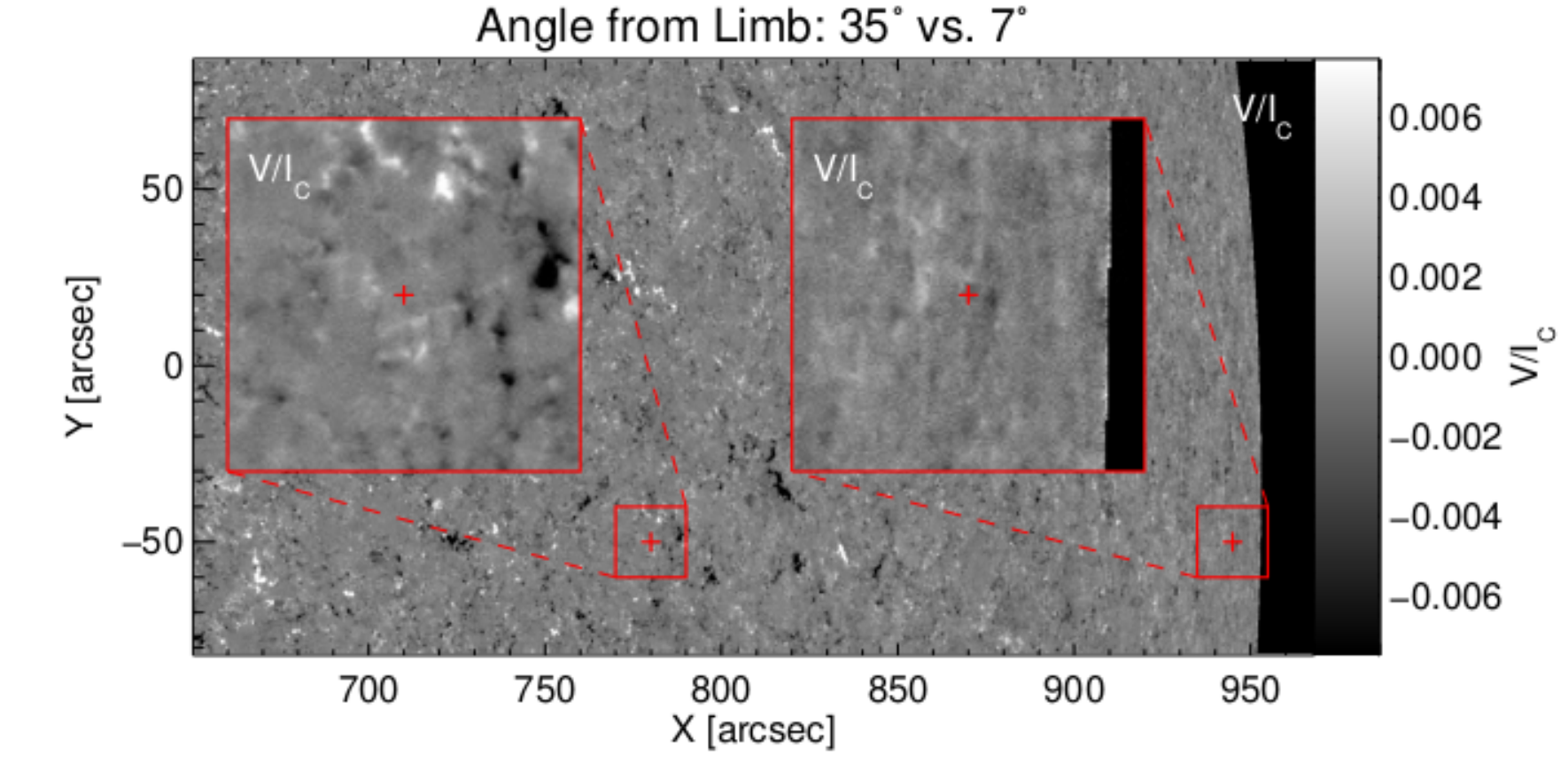}
  \caption[]{Continuum intensity map (upper panel) and Stokes~$V$ map (lower panel) of a quiet-Sun region near the limb observed by Hinode/SOT/SP.  The inserts at $7^\circ$ and $35^\circ$ from the limb (red crosses) are centred close to the maximum viewing angles of the solar poles from the ecliptic and from Solar Orbiter, respectively. The enhanced contrast signal at $35^\circ$ viewing angle allows for a superior determination of the atmospheric parameters. The grey scales of both inserts cover the same dynamic range, but are individually centred to their mean intensity values. \label{F:view37vs7} %
 }
\end{figure}

During the high-latitude phases of the mission's orbit, \sophi\ will also  determine surface flows at and around the poles with unprecedented accuracy by tracking small-scale features, such as granules or magnetic elements \citep[e.g. using local correlation tracking;][]{november88} complemented by Doppler-shift measurements. This will be possible thanks to the high spatial resolution achieved by the High Resolution Telescope (HRT, see Section~\ref{sec:HRT}) over most of the orbit. As Fig.~\ref{F:view37vs7} shows, although granules are hard to identify at $7^\circ$ from the solar limb (which is the most favourable angle at which a solar pole can be seen from  Earth), they become clearly visible at $35^\circ$, close to the highest heliographic latitude to be reached by Solar Orbiter. Also, magnetic features are far more clearly visible in Stokes~$V$ at $35^\circ$ than at $7^\circ$ from the limb. 

\subsubsection{What is the structure of the meridional flow?}

Local helioseismology is also able to measure the meridional flow and provides evidence for temporal variations \citep[][]{2018arXiv180808874L,2018SoPh..293..145K,2017ApJ...849..144C,2017ApJ...845....2B}. Unfortunately the various measurements continue to be inconsistent leading to significant uncertainty regarding the dynamics near the poles.  \sophi\ will make a fundamental contribution to our understanding of the solar dynamo by observing the meridional flow in the polar regions using helioseismic techniques.

\subsubsection{How is magnetic flux reprocessed at high solar latitudes?}

\sophi\ will explore, from vantage points at  different heliographic latitudes, the transport processes of magnetic flux from the activity belts towards the poles and the interaction of this flux with the already present polar magnetic field. This includes the often small-scale cancellation effects  whose combined effect causes the reversal of the dominant polarity at the poles leading to the next activity cycle. \sophi\ will use a multi-pronged approach, employing Doppler, proper motion and helioseismic measurements to determine convective flows, the poleward meridional flow at the surface and surface differential rotation at high latitudes. It will follow the short-term evolution of individual magnetic features at high latitudes, but also the evolution of the distribution of the field at the poles over the lifetime of the mission. 

\sophi\ will obtain a much clearer view of the solar polar magnetic fields during the high latitude passes than currently possible. This will be true particularly in the later phases of the mission when the highest heliographic latitudes will be reached.

\subsubsection{Is a small-scale turbulent dynamo process acting on the Sun?}

In addition to the cancellation effects just mentioned, MHD simulations indicate that a small-scale turbulent dynamo is acting in the solar interior and at its surface \citep[][]{brun04,voegler07,rempel14}. In spite of tantalising observational evidence for a surface dynamo  \citep[e.g.][]{danilovic10,danilovic16,buehler13,lites14} {\it and} for a single source of all magnetic flux on the Sun \citep[e.g.][]{parnell09}, the case for either local dynamo action or a single source for all solar magnetic flux is still not settled. 

Here \sophi\ will be able to make a unique contribution to determining the origin of small-scale fields by observing the Sun from different latitudinal vantage points. By reaching heliographic latitudes higher than $25^\circ$ \sophi\ will measure magnetic fields equally reliably at all latitudes on the Sun \citep{martinezpillet07}. By measuring the properties of freshly emerged small-scale magnetic features over all  heliographic latitudes, \sophi\ will be able to distinguish between their formation by a small-scale turbulent dynamo, which is independent of solar rotation, or by the differential rotation-driven global solar dynamo. In the former case the emergence rate and properties of the small scale magnetic features should be largely independent of heliographic latitude, while in the latter case there should be a clear latitudinal dependence. If, e.g. smaller magnetic features are mainly formed by a small-scale dynamo, while larger features carrying more magnetic flux per feature are largely a  product of the global dynamo, \sophi\ will provide an estimate for where the magnetic flux or size boundary between such features of different origin lies.

\subsection{What drives the solar wind and where does the coronal magnetic field originate?}

\subsubsection{Pinpoint the origins of the solar wind streams and the heliospheric magnetic field}

The origin and acceleration of the solar wind is intimately linked to the magnetic field \citep[][]{marsch06}.
Of the two types of solar wind, the rather homogeneous and steady fast wind (with speeds in excess of 600\,\kms) originates in the open magnetic configuration of coronal holes. Thus, \citet{tu05} have identified coronal funnels anchored in the magnetic network as the source regions of the fast wind. 

The source of the highly variable --- in speed, composition and charge state ---  slow wind component is more complex and less certain, although its origin tends to lie in the dominantly closed-field regions.  It
has been proposed to originate from boundary layers of small coronal holes, from the tops of streamers  \citep[][]{sheeley97}, or from opening loops \citep[][]{fisk03}. Embedded in the solar wind are magnetic field lines that are dragged out with it. One of the aims of Solar Orbiter is to determine the origins of both, the solar wind plasma and the embedded magnetic field.

The Solar Orbiter mission will establish the possibility of detecting slow solar wind streams by its in-situ instruments in co-ordination with measurements of the photospheric field below the wind stream detection site. 

\sophi\ will provide the distribution and evolution of the vector magnetic and velocity fields in the photosphere at a spatial and temporal resolution commensurate with the other remote sensing instruments on board Solar Orbiter \citep[][]{Auchere2019a}. From the data products delivered by \sophi, the magnetic field geometry in the upper solar atmosphere  responsible for accelerating the solar wind can be derived \citep{schrijver03, wiegelmann12, wiegelmann14}.

Thus, \sophi\ will supply the magnetic and dynamic boundary conditions for the plasma processes observed in the higher layers of the solar atmosphere by the EUI, SPICE, Metis and STIX instruments \citep[][]{Rochus2019a,SpiceConsortium2019,Antonucci2019a,Krucker2019a} on board Solar Orbiter and in the inner heliosphere by SoloHI \citep[][]{Howard2019a}. In addition, such plasma processes will be measured in situ by EPD, SWA, MAG and RPW \citep[][]{Rodriguez2019a,Owen2019a,Horbury2019a,Maksimovic2019a} In addition, studies of the dynamical connections between the solar interior and the atmosphere will benefit from the subsurface flows derived from local helioseismology with \sophi\ \citep{duvall93, gizon05}.

\paragraph{How is the polar high-speed wind generated and how does this relate to the polar plume phenomenon?}

There are still considerable gaps in our knowledge of how the fast solar wind is accelerated. Thus, although it was shown that the wind emanates mainly from coronal funnels, network regions where the field lines are open and reach out into the heliosphere \citep[][]{tu05}, their  structure and properties are still only vaguely known. Observations from high latitudes will uncover the detailed magnetic structure responsible for these features in the polar coronal holes.  Due to the $\cos\theta$ dependence (with $\theta$ being the heliocentric angle) of the longitudinal magnetograph signals of a vertical magnetic field, we expect that the signal seen by \sophi\ will be 4 times stronger than that obtained by, e.g. the narrow-band filter imager of Hinode SOT.  

More generally, models of fast solar wind acceleration can be divided into two classes \citep{Cranmer09}. In one group of models, waves propagating along flux tubes into the corona and the associated turbulence are the drivers. These waves are in turn excited by convection at and below the solar surface, which jostles the flux tubes. Differences in wind speed are due to the amount by which the flux tubes expand with height over several solar radii \citep[][and references therein]{cranmer07,ofman10}. 

In the second class of models, the  interchange reconnection models, magnetic reconnection between previously closed magnetic field lines and flux tubes with open fields (i.e. fields connected with the solar wind) provides the energy to accelerate the fast wind. The reconnection is fed by the emergence, decay or evolution of the loop-like closed flux \citep[e.g.][]{fisk99, fisk03}. 

\sophi\ will provide the time series of measurements of the photospheric magnetic field, which can then be extrapolated into the corona and heliosphere, needed to distinguish between these two families of models. Answering this question will require combining data from \sophi\ with imaging and spectroscopic measurements of the overlying coronal gas, as well as recordings of the solar wind properties close to the Sun by the in-situ instruments on Solar Orbiter \citep[][]{Horbury2019b}. 

Polar plumes are enigmatic bright structures in coronal holes reaching far into the corona, which harbour gas moving slowly, compared to the fast solar wind in the interplume regions \citep[][and references therein]{poletto15}. Plumes have been proposed to form by reconnection between freshly emerged closed and pre-existing open field regions \citep{wang95}. However, current magnetograms miss most of the important details. Observations of the magnetic field at the plume's footpoints obtained by \sophi\ in its high latitude phase will be crucial for an understanding of their origin by allowing high quality extrapolations of the field.
Similarly, in combination with the EUI and SPICE instruments, \sophi\ will also provide fresh  insights into the mechanisms leading to the formation of the polar coronal holes and into the nature of their boundaries.

\paragraph{What are the solar sources of the heliospheric magnetic field?}

The heliospheric magnetic field is anchored at the solar surface and is fed by field lines transported from the solar corona into the heliosphere \citep[see e.g.][]{gilbert07}. 
In particular, the roots of those field lines  that are embedded in the slow solar wind are enigmatic. To probe the complex structure of this field, \sophi\ will record vector magnetograms at the solar surface. These provide the lower boundary for non-linear force-free extrapolations of the  magnetic field into the corona. For comparisons with measurements by the MAG instrument, needed to identify the source of the detected heliospheric field, models of the magnetic field in the heliosphere, such as EUHFORIA \citep[see][]{pomoell18}, will also have to be used.  

There is also a mismatch between the heliospheric magnetic flux   as deduced from spacecraft-based in-situ recordings and the Sun's open magnetic flux computed from magnetograms. Typically, the heliospheric magnetic flux is found to be larger than the open magnetic flux measured at the solar surface. One reason for this could be that the polar fields (which provide the dominant contribution to the open magnetic flux over most of the solar cycle) are not well measured by magnetographs located in the ecliptic \citep[see, e.g.][]{linker17}. By measuring  the polar magnetic field more reliably, \sophi\ will be able to test this and other possible explanations. 

\subsection{How do solar transients drive heliospheric variability?}

Solar transients, such as flares, coronal mass ejections (CMEs), eruptive prominences, coronal jets etc., are of particular importance as they can influence the Earth's space environment and upper atmosphere, causing effects that are subsumed under the heading of space weather. 
Solar transients are often driven by instabilities in the magnetic field sometimes triggered by magnetic reconnection \citep[e.g.][]{klimchuk01,priest02,shibata11,chen11}. Basically, magnetic energy is thought to be partially converted into kinetic energy of the erupting/ejected plasma. For example, CMEs are associated with  erupting filaments, and in particular with the presence of filament channels, that are regions of highly sheared magnetic field. Therefore, identifying the sources and uncovering the drivers of solar eruptions requires a good knowledge of the vector magnetic field. This will be provided by \sophi\ in the solar photosphere. 

Models of CMEs predict a flux rope structure in the CME, with a current sheet following it \citep[e.g.][]{lin00,lynch04,kilpua17}. Extrapolations from the measured magnetograms into the corona and the heliosphere will allow estimating the magnetic structure of the interplanetary coronal mass ejection (ICME) that can then be tested by the in situ instruments on Solar Orbiter.

Close to perihelion, that is when Solar Orbiter is partially co-rotating with the Sun, \sophi\ will follow the helicity content of individual active regions for longer than possible from the ground. The evolution of the helicity at the solar surface provides the connection with the helicity carried away from the Sun by CMEs.

\subsubsection{Unravel the evolution of coronal mass ejections in the inner heliosphere}


The evolution of Interplanetary Coronal Mass Ejections \citep[ICMEs, for reviews see][]{linker03,kilpua17} in the inner heliosphere depends on the structure of the magnetic field in both, the atmosphere near the source region and in the heliosphere. Both can be obtained by extrapolating from photospheric magnetograms.  However, an accurate extrapolation requires a precise, co-temporal measurement of the field. 
Magnetograms of the whole solar surface that allow extrapolation into the direction in which the ICME propagates are ideally suited for this. \sophi\ will provide the necessary magnetograms for the ICMEs that will be sampled by the in-situ instruments on board Solar Orbiter.

\subsection{How do solar eruptions produce the energetic particle radiation that fills the heliosphere?}

Energetic particles are typically accelerated during solar flares and coronal mass ejections, for example as a consequence of magnetic reconnection during a flare, or of shock waves excited during a coronal mass ejection \citep[][]{desai16,Benz17}. These particles either travel towards the denser lower solar atmosphere where they produce secondary phenomena such as chromospheric evaporation, or escape into the heliosphere, depending on their direction of propagation and the magnetic field geometry. Unlike presently available observations, Solar Orbiter will be in the unique position to investigate both, the source regions of these particles using remote-sensing instruments and the properties of the particles themselves while still in the inner heliosphere, using in situ instrumentation. 

\sophi\ will provide high resolution, high cadence vector magnetograms from which the magnetic field structure at the acceleration site and the surrounding corona can be determined via extrapolations. This will help identify the physical process underlying the acceleration and provide a complete picture of the particle acceleration and release. Those energetic particles propagating downward can generate heating and shocks in the chromosphere, with observable consequences, possibly including compact sunquakes in the underlying photosphere and solar interior \citep[][]{martinez07,kosovichev98, kosovichev14}.  The Doppler capabilities of \sophi\ will help locate and quantify the effects of downward streaming particles at the solar surface and in subsurface layers.

\paragraph{How are solar energetic particles released and distributed in space and time?}

Solar energetic particles follow magnetic field lines during their propagation through the heliosphere \citep{desai16}. \sophi\ will provide the large-scale photospheric field from which the heliospheric field can be computed and consequently the propagation of particles traced \citep[][]{Luhmann07}, establishing the connectivity between Solar Orbiter and the source of the particles on the Sun. In particular, by combining with resources along the Sun-Earth line, \sophi\ will be able to produce synoptic charts much faster than done conventionally (see Section~\ref{global_mag_structure}). \sophi\ will show changes in the field that precede changes in particle flux measured in situ by Solar Orbiter, for instance when small-scale magnetic flux emergence is followed by magnetic reconnection.

\subsection{Science going beyond the core science aims of Solar Orbiter}\label{additional_science}

\sophi\ will be the first magnetograph to observe the Sun from vantage points away from the Sun-Earth line. This will allow it to provide unique information that will address a series of fundamental solar physics questions that are not addressed in the Solar Orbiter Red Book, in other words science questions that go beyond the four top-level science goals of Solar Orbiter. Examples of such additional important science questions are described below.

\subsubsection{Solar irradiance and luminosity variations}

\paragraph{How do solar irradiance variations depend on the viewing latitude?}

The Sun is the main source of external energy entering the Earth's climate system and variations in solar irradiance are a potential driver of climate change \citep[][]{haigh07,solanki13}. In addition, they serve as a prototype of brightness variability of other cool stars, which can now be studied with high precision thanks to space missions such as Kepler, TESS and PLATO \citep[][]{borucki10, ricker16, rauer14}. Besides being of intrinsic interest, stellar variability or stellar "noise" hides planetary transits, hindering the detection of small, rocky planets \citep[e.g.][]{meunier15}. It also hides the signal of stellar oscillations \citep{rabello97}. Because solar and cool-star variability is caused by magnetic features and granulation at the stellar surface \citep{shapiro17}, which can be spatially resolved only on the Sun, it serves to validate and constrain any successful model of stellar variability. The main limitation so far of solar irradiance observations as a guide to other stars has been that they have all been restricted to the ecliptic, while stars are observed from all latitudes. 

\sophi\ will measure the Sun's magnetic field and continuum intensity from different heliographic latitudes. This will enable computing the Sun's irradiance as it would be visible from different  latitudes, for instance using the successful SATIRE model \citep[][]{fligge00,krivova03}. The most recent version (SATIRE-3D) of this model  accurately reproduces measured total solar irradiance (if given a magnetogram and a continuum image) without having to adjust the computed irradiance variability to the observations \citep[][]{yeo17}. 

Reconstructing the irradiance from different heliographic latitudes is important for testing model predictions
\citep[][]{vieira12,shapiro16}. In addition, it has considerable implications for stellar and exoplanet research by helping improve the detection of exoplanets via transit photometry. It is also key to establishing why the Sun displays a smaller variability than other, similarly active sun-like stars on both, solar rotation \citep[][]{Reinhold13, McQuillan14} and solar cycle time scales \citep[e.g.][]{Lockwood92, radick18}. One proposal to explain this difference is that, unlike the Sun, stars are typically not seen from their equatorial planes. In such a geometry, sunspots (starspots) compensate the brightening produced by faculae more poorly than in an equator-on view. By measuring the magnetic field and brightness from different latitudes, \sophi\ will distinguish between the geometry-based mechanism \citep[cf.][]{schatten93,knaack01} and other proposals \citep[e.g. by][]{witzke18} to explain the Sun's too low variability. If no latitude dependence of irradiance variations is found, then the stellar observations imply that the Sun may in future display a factor of 2-3 larger irradiance variations, with a correspondingly enhanced  influence on climate.

\paragraph{How strongly does the solar luminosity vary?}

Although the irradiance of the Sun (i.e.\ the Sun's radiative flux in the direction of the Earth) and its variations are well measured \citep[see][]{lockwood05,ermolli13}, the variation of its luminosity (i.e.\ the integral of intensity radiated in all directions) is largely unconstrained by observations. The importance of luminosity variations has been discussed by \citet{Foukal06} and \citet{vieira12}.

\sophi\ will provide the data with which irradiance from significantly different directions than the Sun-Earth line can be determined and will thus allow a first estimate of the Sun’s luminosity and its variations on a  time-scale of years.  

Since Solar Orbiter has no dedicated irradiance monitor on board, the irradiances modelled from \sophi\ data products will have to be calibrated during spacecraft passages across (or close by) the Sun-Earth line against measurements from an irradiance instrument in near-Earth orbit. 

\subsubsection{What is the nature of solar  magnetoconvection?}

Magnetoconvection, that is the interaction of magnetic field and convection (e.g.\ in the photosphere), drives many of the Sun’s active phenomena and is not only of fundamental importance for solar physics as a whole, but also is a physical process worthy to study in its own right \citep[interaction of turbulent convection with a magnetic field in the regime of plasma $\beta\sim 1$; ][]{stein12,borrero17}. 

To gain a better knowledge and understanding of magnetoconvection, it is important to know the full velocity and magnetic field vector with as few assumptions as possible. The LOS velocity is determined from Doppler shifts, while proper motions are generally obtained by tracking structures such as granules \citep{november88}. However,   the spatial and temporal resolutions of the proper motions obtained from tracking are considerably lower than of the LOS velocity. In addition, the proper motion of brightness structures need not correspond to actual motions of the plasma. For example, the bright grains in sunspot penumbrae are seen to move inward, whereas Doppler shift measurements show an outward flow of gas \citep[Evershed flow; ][]{solanki03}. 

Simultaneous spectropolarimetric imaging  of convective and magnetic features with \sophi\ and an instrument in approximate quadrature observing along the Sun-Earth line will allow velocities based on proper motions to be validated and calibrated. 

Observations of the same feature from two directions can resolve the $180^{\circ}$  ambiguity in the magnetic azimuth inherent to magnetic field measurements based on the Zeeman effect, without any prior assumption (unlike the techniques currently used). The measurements carried out from each direction (one by \sophi\, one by an instrument along the Sun-Earth line) each suffer from the ambiguity, but only the correct solution will in general be common to both. Besides their value in cleaning the measured magnetograms, such data can test and validate the various ambiguity-resolving techniques \citep{metcalf06}. More details are given by \citet{Rouillard2019a}.

\subsubsection{What is the 3D geometry of the solar surface?}

The SECCHI instrument suite on the two STEREO spacecraft \citep[][]{howard08}  obtained the first true stereoscopic view of the solar corona. \sophi\ will allow  carrying out the first stereoscopic imaging and polarimetry of the solar photosphere by co-ordinated observations with an instrument in near-Earth orbit (such as SDO/HMI) or on the ground, for instance by telescopes such as the Swedish 1-m Solar Telescope \citep[SST; ][]{scharmer03}, Gregor \citep{schmidt12}, the Goode Solar Telescope \citep[GST;][]{cao10}, or the Daniel K.\ Inouye Solar Telescope \citep[DKIST; ][]{warner18}.

High-resolution imaging observations of  regions away from solar disc centre indicate an undulating 3D structure of the visible solar surface \citep[cf.][]{lites04,schmidt04}. Granulation shows what appears to be bright granular hills surrounded by darker intergranular trenches. Similarly, the solar surface is depressed in magnetic structures such as faculae, pores and sunspots, with the difference in height to the surface in the quiet Sun being called the Wilson depression \citep[see, e.g. the review by][]{solanki03}. The undulating solar surface is predominantly an effect of a geometrical shift of the iso-optical-depth surfaces \citep[see numerical simulations of][]{carlsson04,keller04}. Therefore, measurements of the height of the solar surface in different structures is important for an understanding of magnetoconvection. Co-ordinated observations at the correct phase of its orbit by \sophi\ and by an instrument in the Sun-Earth line observing at the same wavelength (e.g.\ SDO/HMI) will give direct measurements of photospheric height differences. Such observations will also allow testing the more indirect techniques for determining the Wilson depression  used so far \citep[e.g.][]{martinezpillet93,solanki93,mathew04,loeptien18}.

A related problem involves the interaction of solar oscillations with the near-surface convection. Significant variations in the phase and amplitude of the oscillations with position in the convection cells \citep{Schou15} are expected, and a direct measurement of the heights, as well as the radial and horizontal velocities (ideally also obtained by combining Doppler shifts from two vantage points) will help us understand the centre to limb effects seen in helioseismology \citep{Zhao12, Baldner12}, which introduces systematic errors, for example, in helioseismic measurements of meridional flow in the solar interior.

Improvements in flow measurements made by local correlation tracking (LCT) of granules are also expected. LCT exhibits a centre-to-limb effect caused by the apparent asymmetry of granules close to the limb \citep{Lisle04,loeptien16}. Stereoscopic observations of granulation will provide the information needed to remove this systematic effect, thus improving LCT measurement of, for instance, meridional flows at the solar surface.

\subsubsection{How does the brightness of magnetic features change over the solar disc?}

The contrast of magnetic features relative to the quiet Sun varies across the solar disc \citep[][]{topka97, ortiz02, hirzberger05, yeo13}. Although this provides a very sensitive test of models of photospheric magnetic features, it suffers from the fact that most magnetic features evolve much faster than the time it takes for the Sun to rotate by a sufficiently large angle to see the same feature at a strongly different limb distance. Therefore, only much less sensitive statistical analyses can so far be conducted. Such analyses may suffer from significant biases, since different populations of features may be selected near the limb and at disc centre. 

Detecting the same feature from two directions by combining observations from \sophi\ with observations made from the Sun-Earth line will allow determining the brightness of the same feature simultaneously from different directions. This will greatly increase the sensitivity of tests of flux tube models. 

\subsubsection{How do active regions and sunspots evolve?}

Our knowledge of the evolution of active regions at the solar surface is still far from satisfactory. As an active region rotates across the solar disc, projection effects limit the length of time over which the evolution of the magnetic vector and line-of-sight velocity of an active region can be reliably followed. Geometrical effects (foreshortening, changing visibility of the corrugated solar surface) also contribute.  Disentangling real solar evolution from projection effects is not straightforward. 

Measurements by \sophi\ during Solar Orbiter's near co-rotation phases will greatly simplify determining the evolution of the magnetic flux, the brightness and velocity in the solar photosphere. Insight will be gained from tracking almost any active region or sunspot. However, small active regions have the advantage that they go through their full evolution (emergence to decay) within Solar Orbiter's typical near co-rotation time span of 5 to 10 days. We note that in the fortuitous circumstance that the angle between Solar Orbiter and Earth is not too large, the period over which a solar feature can be followed can be extended even further. 

\subsubsection{What is the global structure of the solar magnetic field?}\label{global_mag_structure}

The global structure of the coronal magnetic field is obtained by computing it from synoptic charts that are assembled typically from daily magnetograms. Currently it takes a full solar rotation to produce a synoptic chart, during which the field evolves quite significantly. Thus, the lifetime of most sunspots is shorter than a solar rotation. 

During a given phase in almost every science orbit, the remote-sensing instruments on Solar Orbiter will be able to observe partially or completely the side of the Sun facing away from Earth. Consequently, coordinated full-disc observations made at such times by \sophi\ and instruments on or around Earth allow “synoptic charts” to be constructed within, say, two weeks, or with a slight loss of accuracy over only eight days. This will greatly lower the influence of evolution and provide new insights into the global structure of the magnetic field. 

In addition, observations made by \sophi\ during the phase when Solar Orbiter is observing the far side of the Sun will provide an opportunity to calibrate holographic far-side imaging of solar active regions \citep[][]{2000Sci...287.1799L,2017SoPh..292..146L}.

\subsubsection{Where does magnetic reconnection of importance for coronal heating take place?}

Pinning down the main physical process leading to the high coronal temperatures remains one of the most fundamental open tasks in solar physics. Nanoflares associated with magnetic reconnection and Ohmic dissipation at current sheets are prime candidate drivers of coronal heating \citep{parker88, priest02, klimchuk06, reale14}, but it is not settled where the reconnection takes place.  

The traditional view is that the braiding of field lines by the horizontal motions of their photospheric footpoints leads to the formation of current sheets in the corona \citep{parker88, klimchuk06, priest14}. However, recently an alternative scenario has been proposed, involving cancellation in the lower atmosphere between the dominant polarity field at the footpoint of a magnetic loop and a small opposite polarity patch \citep{chitta17, chitta18, priest18}. Distinguishing between the two views requires magnetograms recorded at high spatial resolution as offered, for instance, by the IMaX magnetograph \citep{martinezpillet11} on board the Sunrise balloon-borne solar observatory \citep{solanki10, solanki17,barthol11,berkefeld11,gandorfer11}. Due to the brevity of the Sunrise science flight and the limited targets, the statistics are relatively poor, so that it is  not clear just how common this second mechanism is. 

Close to perihelion, the \sophi\ High Resolution Telescope will provide magnetograms having sufficient resolution to reveal how common magnetic cancellation is at the footpoints of brightening coronal loops. This will help determine the relative importance of the two mechanisms, in particular when combined with high-resolution data from the EUI instrument. 

\subsubsection{Improving space weather forecasting}

Solar transients such as coronal mass ejections and flares can influence the Earth's space environment and man-made resources in a variety of ways \citep{schwenn06, pulkkinen07}. Predictions of space weather events are therefore of considerable societal importance, but are limited by shortcomings in our current knowledge. One of these is our limited ability to sense activity and magnetism behind the solar limb, which will be further reduced if the remaining STEREO spacecraft \citep[][]{kaiser08} stops operating. 

Solar Orbiter will provide the necessary data, in its low latency mode. Of particular interest will be the first ever magnetograms of the far side of the Sun recorded by \sophi. From these data active regions can be detected before they become visible from Earth and an improved estimate of the structure of the interplanetary field obtained, which will help to make better predictions of the propagation of CMEs.


\section{Instrument overview}\label{instrument_overview}

\subsection{Physical effects underlying the \sophi \ functional principle} 

\begin{figure}[h]
    \centering
    \includegraphics[width=\columnwidth]{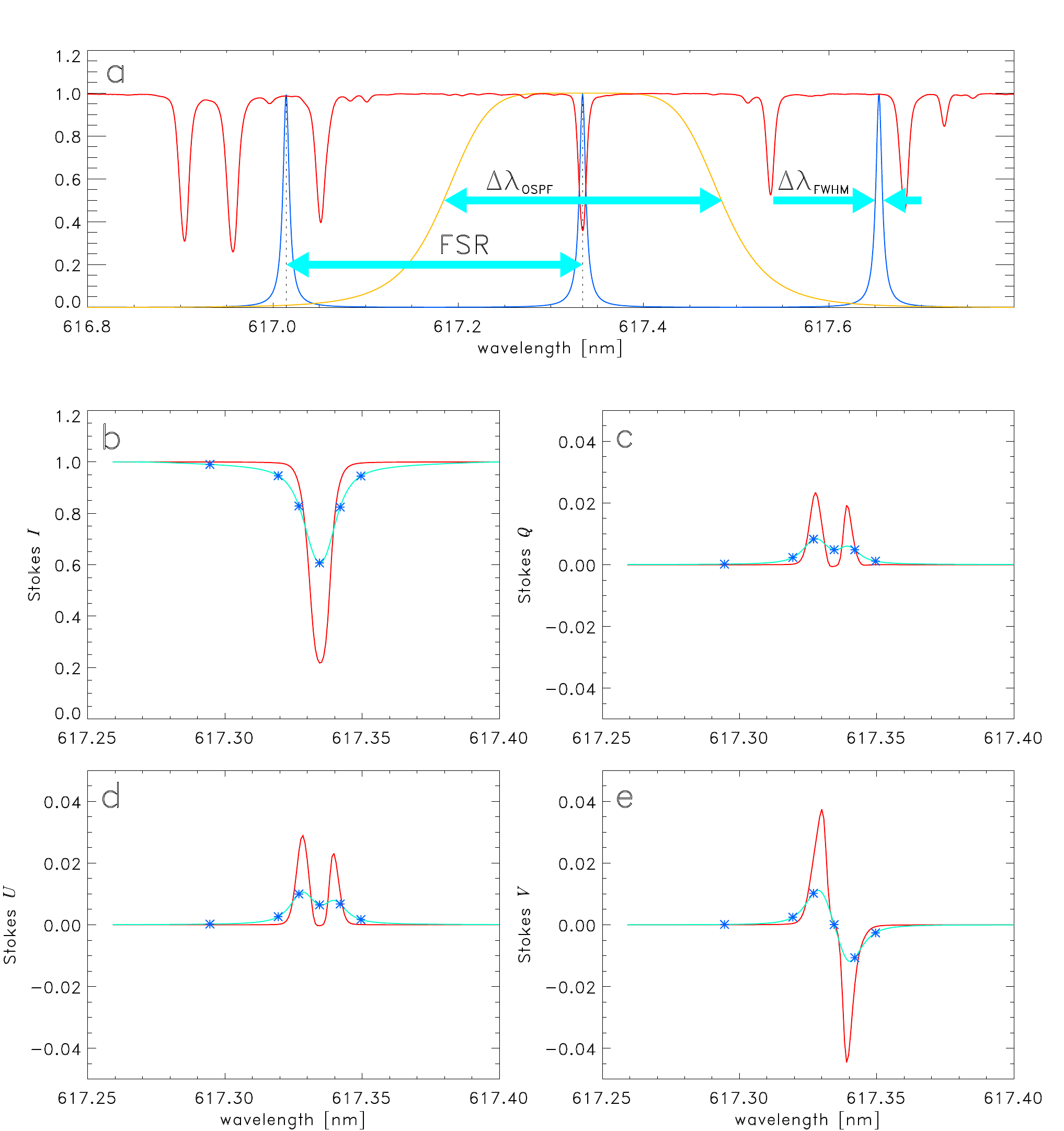}
    \caption{Measurement principle of \sophi . a: solar spectrum around 617\,nm \citep[red; FTS atlas; see][]{neckel84}, tunable filter profile (blue) and prefilter bandpass (yellow); {\sf FSR} and $\Delta\lambda_{\sf FWHM}$ denote the free spectral range and the full width at half maximum of the Filtergraph, $\Delta\lambda_{\sf OSPF}$ is the full width of the order-sorting prefilter; b-e: \ion{Fe}{i}\,6173\,\AA\ Stokes profiles obtained from one point of an MHD simulation (red) and ideally simulated \sophi\ primary observables (light blue). The blue asterisks denote the expected \sophi\ measurements when tuning the filter pass-band to the dedicated wavelength positions.}
    \label{Fig. meas_princ}
\end{figure}

\sophi\ will map the continuum intensity, $I_{\rm c}$, the LOS velocity of the photospheric plasma, $v_{\rm LOS}$, and the vector magnetic field, ${\bm B}=(B,\gamma,\phi)$, embedded in it. While the continuum intensity can be considered a good proxy of the photospheric temperature at optical depth $\tau = 1$, the other two physical quantities are derived from the imprints that physical mechanisms leave on the shapes of the four Stokes profiles of the \ion{Fe}{i}\,6173\,\AA\ line probed by \sophi\ (see Fig.~\ref{Fig. meas_princ}). We derive those quantities by inverting the radiative transfer equation (RTE) for polarised light under the assumption of Milne-Eddington atmospheric conditions. To achieve those maps, the instrument must combine four basic features: it must be an imager to make maps, a spectrometer to record the spectroscopic consequences of both the Doppler and Zeeman effects, a polarimeter to measure the polarisation induced in spectral lines by the Zeeman effect, and must incorporate sophisticated processing capabilities to carry out the inversion of the measured profiles on board. While the last feature is presented in Section~\ref{sec:RTE}, we discuss here the other three.

\begin{figure*}[hbt]
    \centering
    
    \includegraphics[width=1.9\columnwidth]{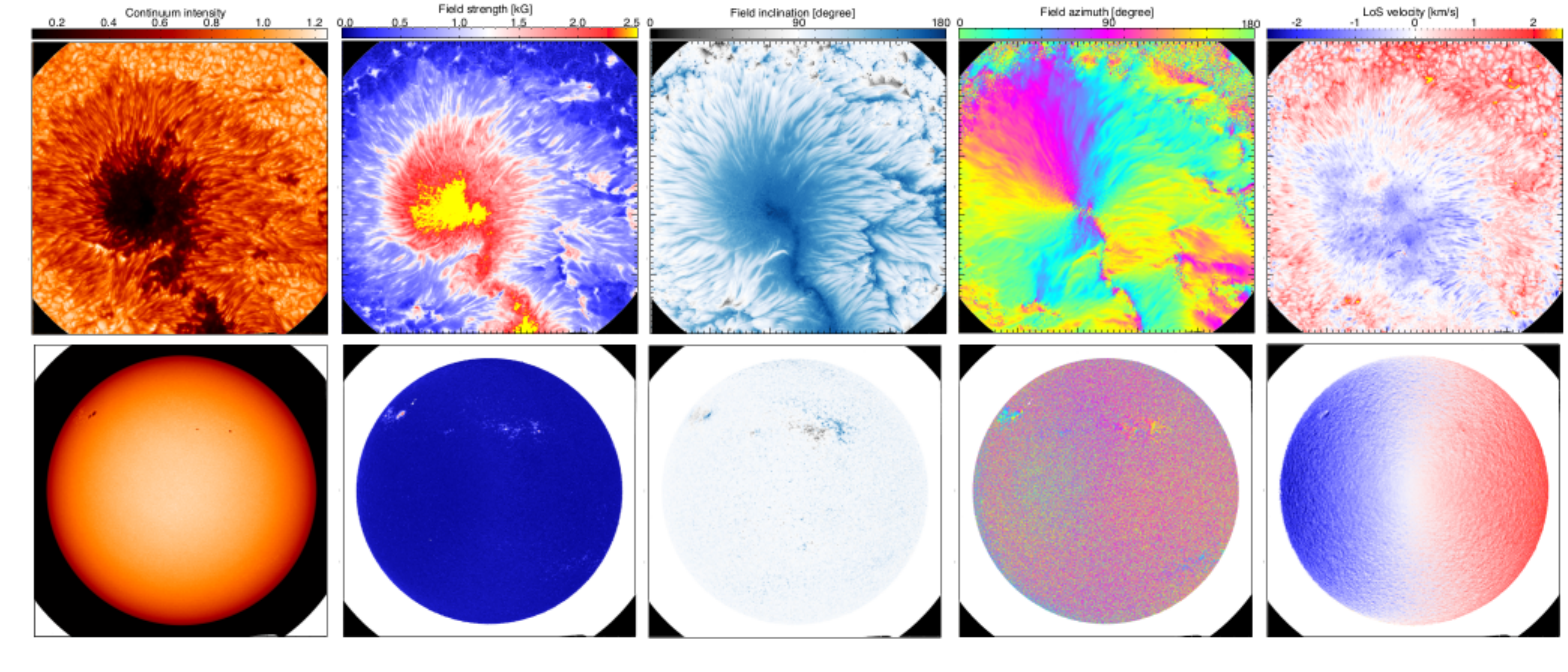}
    \caption{Final data products (from left to right: continuum intensity, $I_{\rm c}$, magnetic field strength, $B$, magnetic field inclination, $\gamma$, magnetic field azimuth, $\phi$ and LOS velocity, $v_{\rm LOS}$) after on-board data analysis. The inversion code used is a software version of the flight one. Upper row: simulated high resolution channel data results. The original data come from the Swedish 1-m Solar Telescope \citep[][]{scharmer03} in the Observatorio de La Palma (Canary Islands) in 2012. Lower row: Simulated full disc data results. The original images have been taken with the Helioseismic
Magnetic Imager \citep[HMI; see][]{scherrer12} on the Solar Dynamics Observatory (SDO) with a resolution of 1 arcsecond (December 2, 2011).}
    \label{Fig. data products}
\end{figure*}

\paragraph{{\sc SO/PHI} {\sf as an imager}}

\sophi\ is able to record images of the solar surface with two different combinations of   field-of-view on the plane of the sky and angular resolution: 
The Full Disc Telescope (FDT) covers the full Sun even at closest perihelion and thus has a field of view (FOV) of 2 degrees, with a sampling of $3\arcsec 75\,\mbox{pixel}^{-1}$. The High Resolution Telescope (HRT) maps a solar scene over a FOV of $0.28^\circ\times 0.28^\circ$, with an angular sampling of $0\arcsec 5\,\mbox{pixel}^{-1}$. Both channels can be used alternatively, not simultaneously. The optical paths aim for diffraction limited performance over the full range of observing conditions along the mission. Images are quasi-monochromatic linear combinations of all four Stokes parameters.

\paragraph{{\sc SO/PHI} {\sf as a spectrometer}} 

\sophi\ samples the \ion{Fe}{i}\,6173\,\AA\ spectral line profile in each pixel of the image by sequentially recording 
quasi-monochromatic images and tuning the transmission 
band-pass of the filter from one image to the next. Six images are recorded sequentially with 6 different wavelength settings for the transmission band-pass. See also Fig.~\ref{Fig. meas_princ}.   

\paragraph{{\sc SO/PHI} {\sf as a polarimeter}} 

In each of these wavelength samples, the full polarisation state of the quasi-monochromatic light is measured by differential photometry (see also Fig.~\ref{Fig. meas_princ}b-e). Four images are taken in different linear combinations of the  Stokes parameters, which are selected by an electro-optic polarisation analyser. Such a device is made up of two Liquid Crystal Variable Retarders (LCVRs) plus a linear polariser, all mounted in a single block, called the Polarisation Modulation Package (PMP).  The four polarised images are later {\em demodulated} to provide the four Stokes parameters. 

In summary, we can say that --- from a technical point-of-view --- \sophi\ is a diffraction-limited, quasi-monochromatic, wavelength-tunable, polarisation-sensitive, imager, with sophisticated processing capabilities.

\subsection{Data products }
\label{sec:products}

The \sophi\ data products (see Fig.~\ref{Fig. data products}) will be extracted from the primary observables by on-board
processing, that is by inverting the RTE for the measured
Stokes profiles (cf. Section~\ref{sec:RTE}). To optimise the processing speed, the atmospheric models are chosen to satisfy the Milne-Eddington approximation.

This inversion can be computed autonomously on board in order to optimise the science return within the limited telemetry of Solar Orbiter. Alternatively, raw data and partially processed data can be compressed on board and downlinked, although at a reduced rate to satisfy the telemetry bounds.

\subsection{Technical implementation and subsystems}

From the above we can identify the following necessary set of key functionalities:

\begin{itemize}
\item imaging (both high resolution and full disc), 
\item photon detection, 
\item monochromatic filtering and tuning, 
\item polarisation analysis, 
\item data acquisition and analysis.  
\end{itemize}

In addition to these basic functionalities the system requires further technical infrastructure: 

\begin{itemize}
\item  refocus mechanisms, 
\item  a feed select mechanism for switching between high resolution and full disc view, 
\item an image stabilisation system, 
\item false light protection and thermal architecture.  
\end{itemize}

The full set of functionalities must be provided under  extreme limitations in terms of mass, volume, and power, dictated by the Solar Orbiter mission.  Tables~\ref{tab:Unit_Ressources} and~\ref{tab:Inst_Ressources} list the main boundary conditions and instrument parameters.

The above listed functional devices are associated with technical subsystems, which are included in the four main instrument units:  

\begin{itemize}
\item  The heat rejection system based on two Heat Rejecting Entrance Windows (HREWs) 
\item The Optics Unit (O-Unit) 
\item The Electronics Unit (E-Unit) 
\item The harness connecting O-Unit and E-Unit. 
\end{itemize}

While the O-Unit, the E-Unit and the Harness are mounted within the spacecraft, the two HREWs are directly mounted into the corresponding feed-throughs within the spacecraft heat shield.

\begin{figure}
\centering
\includegraphics[width=\columnwidth]{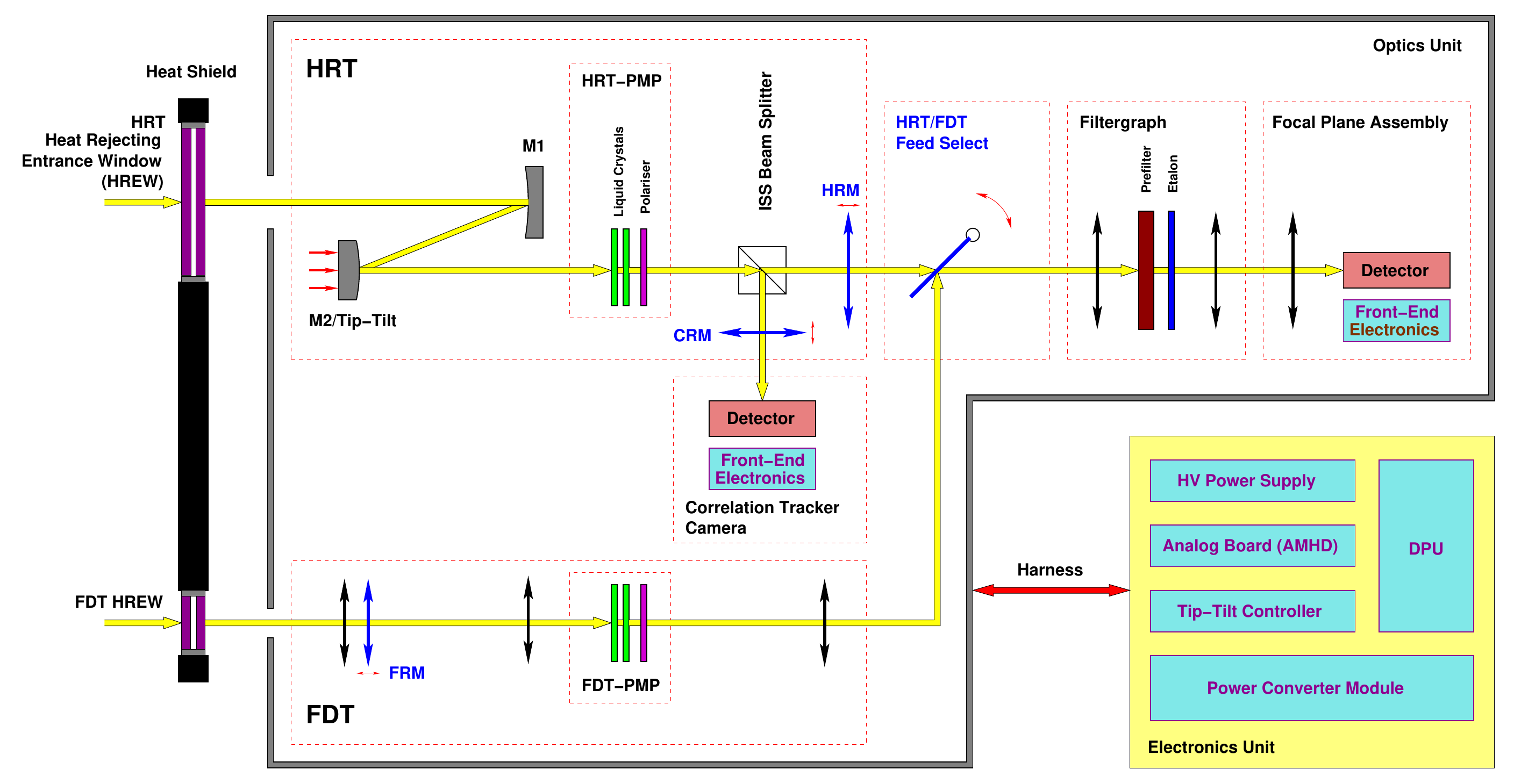}
\caption{Schematic overview of the \sophi\ units and functionalities. \label{hirzberger_fuctional}}
\end{figure}

The O-Unit and its subsystems are described in Section~\ref{o-unit}, together with the heat rejecting entrance windows. 
The Thermal subsystems and the thermal architecture of the Optics Unit are described in Section~\ref{thermomechanical}. 
The E-Unit with its electronic subsystems are described in Section~\ref{e-unit}. 
A schematic overview of the different subsystems is depicted in Fig.~\ref{hirzberger_fuctional}.

\begin{table*}
   \centering
   \caption{Summary of the \sophi\ as-built unit properties (mass, volume and temperature range). \label{tab:Unit_Ressources}}
    \begin{tabular}{lrrrr}
    \hline
    Unit & Mass & Volume & $T$ Non-Operational & $T$ Operational \\
    \hline
    O-Unit & 21.5\,kg & $820 \times 420 \times 350$\,mm$^3$ & $-30\,^\circ$C to $+60\,^\circ$C & $+10\,^\circ$C to $+50\,^\circ$C\\
    O-Unit (cold interface) &  &  & $-35\,^\circ$C to $+60\,^\circ$C & $-35\,^\circ$C to $-25\,^\circ$C \\
    E-Unit & 6.0\,kg & $232 \times 228 \times 175$\,mm$^3$ & $-30\,^\circ$C to $+60\,^\circ$C & $-20\,^\circ$C to $+50\,^\circ$C\\
    Harness & 1.2\,kg & & &\\
    HREW (high resolution) & 2.5\,kg & $\varnothing\, 262 \times 33$\,mm$^3$ & $-91\,^\circ$C to $+253\,^\circ$C & $-29\,^\circ$C to $+204\,^\circ$C \\
    HREW (full disc) & 0.3\,kg & $\varnothing\, 95 \times 23$\,mm$^3$ & $-95\,^\circ$C to $+243\,^\circ$C & $-55\,^\circ$C to $+243\,^\circ$C \\
    \hline
    \end{tabular}
\end{table*}

\begin{table}
    \centering
    \caption{\sophi\ instrument resources. \label{tab:Inst_Ressources}}
    \begin{tabular}{lr}
    \hline
    Power Idle & 8.7\,W \\
    Power Nominal & 33.0\,W \\
    Alloc. Telemetry Rate$^\ast$ & 20\,kbits\,s$^{-1}$ \\
    Alloc. Telemetry Volume per Orbit & 52\,Gbits \\
    On-board Storage & 4\,Tbits \\
    Highest Observation Cadence$^{\ast\ast}$ & 1\,min$^{-1}$ \\
    Detector frame rate & 11\,s$^{-1}$ \\
    \hline 
    \multicolumn{2}{l}{} \\
    \multicolumn{2}{l}{$^\ast$\,for $3\times 10$\,days per orbit} \\
    \multicolumn{2}{l}{$^{\ast\ast}$\,see also Table~\ref{T:cycle_times}} \\
    \end{tabular}
\end{table}


\section{Optics Unit}\label{o-unit}

\subsection{Optical design}   
\subsubsection{Choice of the monochromatic filter} 
The selection of a narrow-band transmission pass-band of around 100\,m\AA\ (10\,pm) width requires either a 
resonance absorption cell \citep[for example a magneto-optical filter; see e.g.][]{cacciani90}, or a device that makes use of interference. 
Since in the case of Solar Orbiter, the pass-band needs to be tuned over several band-pass widths 
(in order to compensate for the strong line-of-sight shifts along the elliptical orbit), 
resonance absorption cells are not an option for \sophi. 

Possible implementations of interferometric devices are Michelson interferometers (as used successfully in both, 
the MDI and the HMI instruments on SoHO and SDO, respectively), or polarisation interferometers 
(also called Lyot filters or birefringent filters). 
Both options had to be abandoned for the \sophi\ instrument because of the severe 
limitations on the instrument mass and volume.  

Fabry-Perot interferometers are common in ground-based narrow-band imagers. 
Classical Fabry-Perot systems make use 
of high-order  interference (involving several thousand reflections) of the transmitted light 
within a cavity between two plane parallel mirror surfaces. Tuning is achieved by 
changing the separation between these mirrors. Those devices are therefore not only heavy 
(the substrates must be significantly 
oversized relative to the optically used area and must be 
thick in order to guarantee the extremely smooth surface figure of the mirror surfaces), 
but also extremely vulnerable to external forces and vibrations during launch, and therefore pose a high risk 
for a space mission. 

Solid state Fabry-Perot etalons consist of a single plane parallel piece of material and are therefore not only 
much lighter, but also inherently resilient to misalignment. 
In this configuration, tuning can be achieved only if the optical thickness of the plate can be changed, either by changing the refractive index, the mechanical thickness, or both. A tilting of the etalon also shifts the pass-band, but the tilting additionally leads to an asymmetric broadening of the transmission curve and is therefore not preferable. 

The situation becomes much more interesting when the substrate is made from an electro-optic material. LiNbO$_3$ was  identified as a 
suitable material for electrically tunable etalons already by  \citet{rust86} \citep[see also][]{rust88}. The effect of applying an electrical voltage is two-fold: Firstly, the refractive index of the material depends on an external electric 
field applied to the anisotropic crystal. Secondly, the piezo-electric effect leads to a mechanical deformation 
of the crystal structure and thus to a change in mechanical thickness. The optical repercussions of both effects are on 
the same order of magnitude and cannot be disentangled from each other. 

Etalons made from LiNbO$_3$ have been successfully used as tunable narrow-band filters for solar
magnetometry in stratospheric balloon missions.  The Flare Genesis Experiment \citep[][]{bernasconi00} employed a commercial etalon in a pressurised vessel.  The development of the LiNbO$_3$ etalon for the IMaX magnetograph \citep[][]{martinezpillet11} aboard Sunrise \citep[][]{barthol11} was fuelled by the wish to use LiNbO$_3$ etalons in \sophi.

\subsubsection{Choice of the optical arrangement of the filtergraph} 

Etalons can be used as quasi-monochromatic filters in different optical configurations, 
each having its specific spectral characteristics. 
In the so-called collimated (or spectroscopic) setup, the etalon is placed in the pupil 
of the imaging path. 
In this configuration, each image point sees the high-order-interference pattern of the full etalon. 
Therefore the optical thickness of the etalon 
must be extremely homogeneous over the full optically used surface, since both, the spectral purity, 
and the imaging performance would suffer from 
deviations of the optical thickness in the pupil plane. 
In the collimated setup the etalon maps the field angle into optical thickness, and therefore the 
centre wavelength of the band-pass depends on the position within the FOV. 
The larger the angle between the line-of-sight and the etalon normal, the shorter is the wavelength, 
which is passed through the etalon. 
This so-called "etalon blueshift" limits the size of the field-of-view, which can be reasonably 
sampled simultaneously. Collimated etalon setups have in principle a spectral purity which is only 
limited by the etalon finesse (sort of optical thickness homogeneity), but the centre wavelength of the 
pass-band varies over the FOV.   
 
An alternative optical arrangement places the etalon in the image plane of the instrument. Each image position sees only a small area of the etalon. 
The spectral purity is limited by the fact that the homofocal bundle of light rays, that are interfering in the etalon, has an angular variation 
due to the non vanishing optical aperture of the system. The larger the optical aperture, the larger is the smearing of the spectral band-pass. 
If the etalon is placed in an image plane with infinite pupil distance (telecentric image), then each point in the entire field of view will 
see the same angle geometry. The spectral characteristics of the etalon is then homogeneous for the entire field, but the spectral purity will be 
limited by the F-ratio of the imaging system. 

The choice between the collimated or the telecentric arrangement has been discussed extensively 
over the last decades. For most ground-based applications in solar physics, the final choice is a question of taste, since both 
configurations have quite a balanced set of advantages and drawbacks. 

For \sophi, however, a telecentric configuration is clearly favourable because the wide observed FOV produces an unacceptably large etalon blueshift in a collimated arrangement.

\subsubsection{Implications for the telescopes}\label{telescope_implications}

From a purely optical point of view, it is at least conceivable to place the etalon in front of the detector, provided that the science focal plane is telecentric. This is, however, not optimum for two reasons. Firstly, the F-ratio in the science focal plane is dictated by the wish for diffraction limited sampling. For typical detectors having pixel pitches of order $10\,\mu$m this corresponds to F/30 in the visible. At this fast illumination, the spectral smearing due to the angle variation inside the etalon is not acceptable any more. 
For the \sophi\ etalon a total bandwidth of 100\,m\AA\ was envisaged, which requires an F-ratio of about F/55. This number is only possible thanks to the high refractive index of LiNbO$_3$ (around 2.3). As a consequence of this, the  acceptance angle of the etalon is significantly higher than of a classical air-spaced Fabry-Perot interferometer, which would require an F-ratio of about 150. Only thanks to this  a compact and light-weight design can be achieved within the mass and volume limits. 

The second reason is that both, the detector and the etalon, need to be temperature controlled, but at very different temperature levels. 

For these two reasons the \sophi\ etalon is located in an intermediate real focal plane, which is then subsequently re-imaged to the science focal 
plane for optimum angular sampling on the detector. The intermediate image plane is called "etalon focus". It is equipped with a field stop, 
which physically limits the FOV of \sophi. This will be described in detail in Section~\ref{sec:OPT_Baffling}. 

Etalon size: 
The severe restrictions within \sophi\ in terms of mass and volume, in combination with the high demands on temperature 
stability required by the etalon (0.05\,K) set a technical limit on the diameter of the etalon. 
In the \sophi\ case, a useful area of 40\,mm times 40\,mm was chosen, with cut corners, corresponding to a circle of 50\,mm diameter. 
This was considered an optimum compromise between FOV loss and technical resources. 

The dimensions of the science focal plane are determined by the detector: 2048\,pixels with 10 microns pitch gives a square with $20.5\,\mbox{mm}\times 20.5$\,mm. 
Thus the required demagnification from the etalon focus to the detector is 1.95. 

This common path is identical for both telescopes; the role of the telescopes is 
to provide an image of the solar scene on the etalon focus, with an F-ratio of F/55. 
The focal length of the telescopes must be chosen in accordance with the desired field of view on sky.


\begin{figure}[h]
    \centering
    \includegraphics[width=\columnwidth]{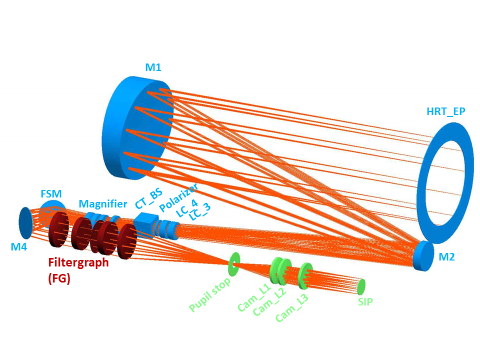}
    \caption{Optical scheme of the HRT path and the FG (common path).}
    \label{Fig. folding }
\end{figure}


\subsection{Opto-mechanical layout}

\subsubsection{High Resolution Telescope (HRT)} 
\label{sec:HRT}

The  High Resolution Telescope (HRT) has a square FOV of $0.28^\circ$ $\times$ $0.28^\circ$  
on the sky with an angular sampling of $0\arcsec 5$ per pixel (at closest perihelion; this is equivalent to $0\arcsec15 \, {\rm pixel}^{-1}$ for a ground-based instrument). 
This value corresponds to optimal sampling at the diffraction limit of a 140\,mm aperture telescope in the red. 
For this sampling, the effective focal length of the HRT path must be 4125\,mm in the science focal plane, and 7920\,mm in the etalon focus, respectively. 
Since this is more than 10 times the  physical length of the \sophi\ instrument, the optical system needs two internal magnifications. 
The HRT therefore consists of a two-mirror telescope, which is combined with a negative magnifying lens (Barlow lens). 
The stand-alone two-mirror system has a focal length of 2475\,mm, the Barlow type magnifier brings the value to the required 7920\,mm. The magnifier consists of 4 lenses in one group, which can be shifted in order to act as a defocus compensator during flight. An overview of the HRT design parameters is  given in Table~\ref{tab:opt_parameters}.

\begin{figure*}[hbt]
    \centering
    
    \includegraphics[width=2.0\columnwidth]{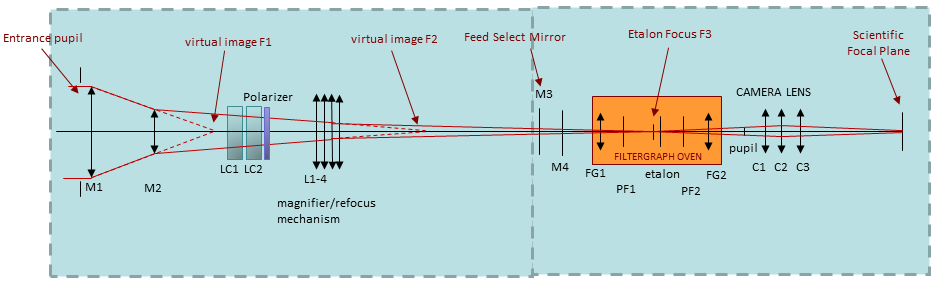}
    
    \caption{Optics Block Diagram of the High Resolution Telescope path. M1 and M2 are the mirrors of the Ritchey-Chrétien telescope; LC 1 and LC2 are the liquid crystal cells; L1, L2, L3 and L4 are the  lenses of the magnifying Barlow lens system, which can be shifted to act as the refocus compensator.  M3 (feed select mirror) and M4 are the folding mirrors; the Filtergraph oven includes the two lenses FG1 and FG2, the two components of the prefilter (PF1 and PF2) and the etalon; finally, the camera lenses (C1, C2 and C3) provide the image at the scientific focal plane. The beam splitter providing light to the Image Stabilisation System (ISS) is not shown.}
    \label{Fig. HRT_block }
\end{figure*}

The two-mirror system is a decentred Ritchey-Chr\' etien configuration with a hyperbolic primary and a hyperbolic secondary. 
The Ritchey-Chr\' etien configuration was chosen for two reasons. Firstly, the performance of the Gregory mirror system suffers from the large science FOV of $0.4^\circ$. Secondly, the placing of a prime focus field stop, which is the main advantage of a Gregory telescope, is inhibited, since -- due to the large apparent size of the Sun at perihelion -- the size of the field stop becomes too large and represents a major obstruction of the optical system. 

The system has a decentred aperture (off-axis configuration), which avoids any physical obstruction of the aperture. 
This not only provides a better contrast transmission at intermediate spatial frequencies, but also avoids that the mounting of the secondary mirror is exposed to direct sunlight, which would aggravate the thermal situation in the telescope. 
Another advantage of the decentred design is the very efficient suppression of optical ghost images in the optical path after the telescope. 
A decentred optical system must be regarded as an asymmetric part of a larger, symmetric "parent" system. 
The decentration of the 140\,mm aperture of the HRT is 170\,mm. 
This means that the parent telescope would have an aperture of 480\,mm. 

All optical components up to the science focal plane are symmetrically placed with respect to the optical axis of the parent telescope (with the  etalon itself being the only exception). 
Due to the decentred aperture, the light rays are all oblique to the components, and thus the ghost reflections do not coincide with their original beams. 
The only optical component which should be used with normal incidence is the etalon. For this reason it is inclined relative to the optical axis of the system.
Since the image plane is normal to the optical axis, this implies that there are no ghosts beating between the etalon and the detector.  Further information about ghost/false light suppression will be given in Section~\ref{sec:OPT_Baffling}.


\begin{table*}[]
   \centering
   \caption{\sophi\ nominal optical parameters. \label{tab:opt_parameters}}
    \begin{tabular}{lcrr}
    \hline
                                    &    & HRT                              & FDT \\
    \hline
    \hline
    Working wavelength   & $\lambda_0$   & $617.3\,\mbox{nm}\pm 0.1$\,nm    & $617.3\,\mbox{nm}\pm 0.1$\,nm \\
    Effective focal length  & $f_{\rm eff}$ & 4125.3\,mm                      & 579\,mm \\
    Field of view           & $\alpha_{\rm FOV}$ & $0.28^\circ\times 0.28^\circ$ & $\varnothing\,2^\circ$ \\
    Entrance pupil diameter & $D$        & 140\,mm                          & 17.5\,mm \\
    Effective F-ratio      & F\#        & 29.5                             & 33.1 \\
    PMP telecentricity       &            & $<0.3^\circ$                     & $<0.3^\circ$ \\
    Effective focal length at etalon focus & $f_{\rm etalon}$ & 7920\,mm    & 1111.7\,mm \\
    F\# at etalon focus     & F\#$_{\rm etalon}$ & 56.6                     & 63.5 \\
    Etalon telecentricity   &            & $<0.23^\circ$                    & $<0.23^\circ$ \\
    Used detector size      &            & $2048\times 2048$\,pixels        & $2048\times 2048$\,pixels \\
    Pixel size              & $d_{\rm pixel}$   & $10\,\mu$m                & $10\,\mu$m \\
    Plate scale   & $\alpha_{\rm pixel}$ & $0\arcsec 50$                    & $3\arcsec 75$ \\
    Image quality & $\sigma_{\rm WFE}$   & $\leq \lambda_0/14$ (diffraction limit) & $\leq \lambda_0/14$ (diffraction limit) \\
    \hline
    \end{tabular}
\end{table*}


\paragraph{Athermalisation of the HRT system:} 

The HRT telescope path was designed for an initially required operational temperature range from $-20\,^\circ$C to $+60\,^\circ$C, thus $\pm 40\,^\circ$C  colder and hotter than the temperature during the alignment. 

In order to keep the mechanical co-alignment between the primary and the secondary mirrors, special care was taken in the design of the mirror cells and the material choice of the main telescope structure (see below). Since the primary mirror is attached to the back part of the main structure, while the front piece of the structure holds the secondary mirror, the thermal expansion of the main structure should be minimised; also internal temperature gradients within the main structure are to be avoided as much as possible. 

The mirrors themselves are made from ZERODUR\textsuperscript{\textregistered}  
and can be considered as sufficiently athermal; however, any  mismatch in the coefficient of thermal expansion (CTE) to the metallic mirror cell would lead to a negative impact on the optical performance. Therefore the mirror cells use a tripod mount with 
tripods from hardened Invar. The  primary mirror uses a cell that is machined from hardened Invar to a large extent, which is mounted to the structure with Invar mounting brackets. 

The secondary mirror also uses tripods and a special core structure from hardened Invar, which connects to a titanium adapter on the mirror holder. 

The mirrors are coated with enhanced silver, optimised for very high reflectance (about 99.5\,\%) at 617\,nm. This implies that there is very little linear polarisation induced by the non-normal reflections. The phase retardance of this coating is small and the residual effect is calibrated during the overall polarimetric end-to-end calibrations (see Section~\ref{calibration}).

The design and fabrication of both mirror cells was managed by Carl Zeiss Optronics (later named Cassidian Optronics, then Airbus Defence and Space Optronics), the lightweighting, grinding, and polishing was done by Carl Zeiss Jena. Some details about the mirror specification and testing can be found in \citet{bischoff14}. \citet{gandorfer18} report on the alignment and the metrology of the mirror system in the HRT channel.  

\paragraph{Magnifying lens system and refocus mechanism (HRM):} 

The magnifying lens group introduced in Section~\ref{sec:HRT} is based on a near athermal opto-mechanical design. As part of the HRT Re-Focus Mechanism (HRM), it is used as de-focus compensator within the HRT. It compensates for manufacturing, alignment and in-flight focus shifts, which are mainly due to thermal lensing effects in the entrance window.

The HRM provides a total range of $\pm 14.7$\,mm along the optical path. The magnifier lens group is part of a translation stage, which is guided through three linear bearings and driven through a stepper motor via a  miniature ball screw. The mechanism has an average mechanical resolution of $0.375\,\mu\mbox{m}/\mbox{full step}$. 
In order not to rely only on step counting during mechanism operation, an absolute position sensor has been implemented to obtain complementary position information. The position information is obtained through combining the highly resolved positional information from a
rotary potentiometer which provides $\sim 0.37\,\mu$m per Digital Unit (DU), flanged directly to the ball screw
axis, with the reading of a coarse linear potentiometer 
($\sim 20\,\mu\mbox{m}/\mbox{DU}$).

 \subsubsection{Architecture of baffling system and false light control} 
 \label{sec:OPT_Baffling}

A decentred system like the Ritchey-Chr\'etien, which we use in the HRT path, must be carefully baffled in 
order to prevent sneak paths towards the detector. 
Since we used a negative magnifier (Barlow system), neither the primary  nor the secondary focus are real foci, which would allow placing a field stop (see Fig.~\ref{Fig. HRT_block }). 
In order to block unwanted portions of the solar disc and in order to reduce the thermal load in the instrument, a series of unsharp stops is used. At each stop, a fraction of the solar beam is absorbed, without vignetting the science beam. 
The closer the stops are located towards the secondary focus, the sharper they become, 
but the less energy they get. Before the magnifying lens group, the primary role of the stops is to trap the unwanted optical energy and thus reduce the heat load to the optical components downstream. During the elliptical orbit, the fraction and the total amount of the solar load, which is absorbed by the vanes, changes, while the fraction and the total value of the energy being passed by the vanes is constant. This is of high importance, since the polarimeter and the filtergraph need constant working temperatures. 

It is not the primary role of these vanes to act as stray-light vanes; this part is taken over by the baffling system in the common optical path. 
The first accessible image plane is the etalon focus, which is equipped with a physical field stop. The transfer path between the etalon and the detector is designed to contain a real image of the pupil of the system. In this pupil, a Lyot stop is placed, which is the most efficient and important part of the stray-light suppression system of \sophi. 
 
For more information on the HRT we refer to \citet{gandorfer18}.

\subsubsection{Structural design}\label{sec:struct_design}

The main drivers of the \sophi\ O-Unit structural design are the tight positional tolerances between the primary and secondary mirror of the HRT telescope, which have to be ensured over the full operational temperature range. 

\begin{figure}
    \centering
    \includegraphics[width=\columnwidth]{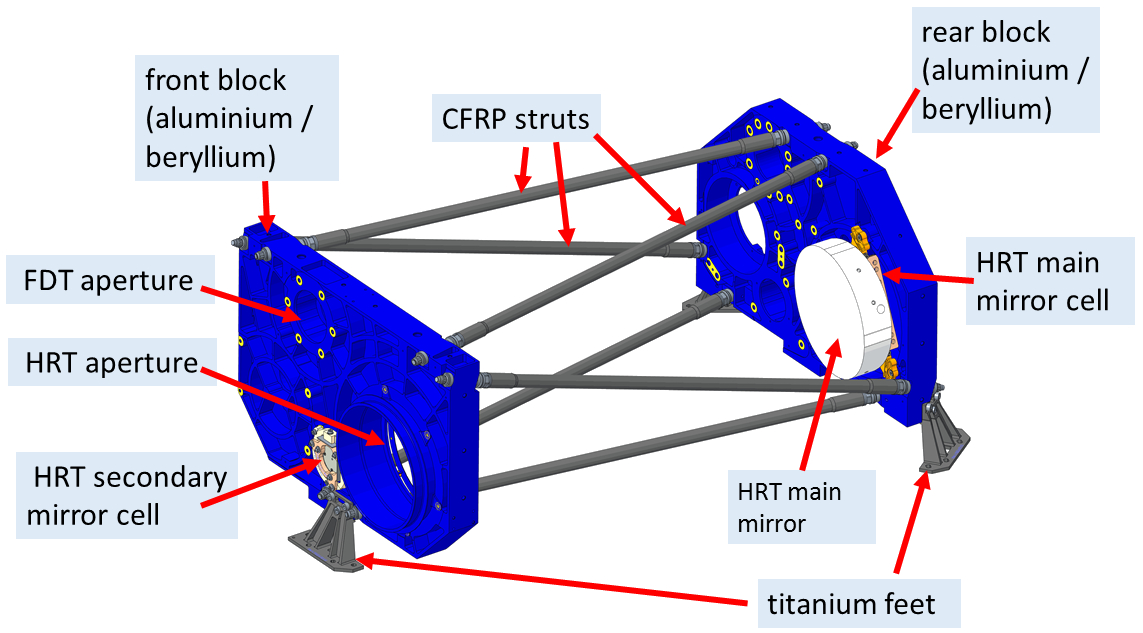}
    \caption{HRT main structure, based on two end blocks and six CFRP struts.}
    \label{fig:HRT_structure}
\end{figure}

\begin{figure*}
    \centering
 
    \vspace*{2mm}

    \includegraphics[width=2.0\columnwidth]{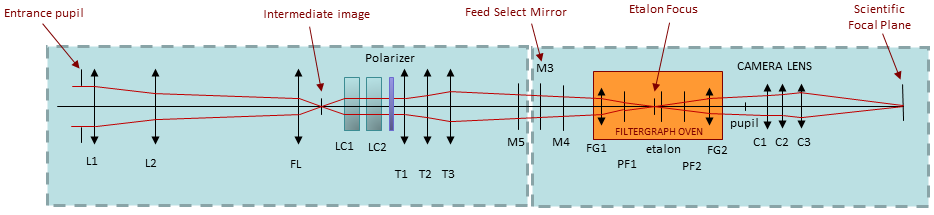}
    
    \vspace*{0mm}

    \caption{Optics Block Diagram of the Full Disc Telescope path. L1 and L2 are lens1 and lens 2; FL is the field lens; LC1 and LC2 are the liquid crystal cells; T1, T2 and T3 are the triplet lenses 1, 2 and 3 respectively; M5, M3 and M4 are the folding mirrors; the Filtergraph oven includes the two lenses FG1 and FG2, the two components of the prefilter (PF1 and PF2) and the etalon; finally, the camera lenses (C1, C2 and C3) provide the image at the scientific focal plane.}
    \label{Fig. FDT_block}
\end{figure*}

Ensuring this and providing the accommodation for 13 subsystems, while keeping the global stiffness of the unit above 140\,Hz, led to the selection of a more classical telescope structure with six struts and two end blocks, rather than an optical bench concept (see Fig.~\ref{fig:HRT_structure}). The six struts drive the induced focus error between the primary and secondary mirror of the HRT. In addition, the struts are the sizing elements for the global bending mode of the unit along its $x$-axis (along the line of sight). Consequently, high stiffness carbon fibre reinforced plastic (CFRP) struts have been selected. 

The very stringent requirement on the decentre between the primary and secondary HRT mirrors is tackled through a trifold approach: 1) mirror mounting concept, 2) material selection of mirror cells, 3) material selection of primary structure end block. The basic idea behind the mounting concept of the two mirrors is to define a common fixation point for both mirrors with 0\,mm offset in $y$-direction (complimentary axis with the z-axis pointing normal to the unit mounting plane) \citep[for details see][]{gandorfer18}. With this approach the problem can be split into two remaining issues: the intrinsic expansion of the mirror assemblies with temperature and a potential differential expansion between the two end blocks due to temperature gradients within the unit. The contribution of the mirror assembly itself has been mitigated by selecting ultra low expansion materials such as Zerodur and Invar  (Section~\ref{sec:HRT}). Selection of similar materials to cope with the differential expansion between the blocks is not realistic due to mass restrictions (Invar) or the high accommodation needs (CFRP). In the end, AlBeMet\textsuperscript{\textregistered}~AM162 has been found to be the optimal compromise between low thermal expansion coefficient, mass and machinability.

A CFRP sandwich plate has been added as secondary structural element in order to accommodate the O-Unit main baffle (Section~\ref{sec:OPT_Baffling}) and the HRT PMP (Section~\ref{pmp_design}). As a side effect, the addition of the sandwich plate  contributes to the torsional stiffness of the unit. The sandwich plate is mounted to the end blocks through flexures in order to avoid stress in the structure due to differential expansion of sandwich plate and CFRP struts.

\subsubsection{Full Disc Telescope (FDT)}  

Full disc observations require a dedicated telescope with short focal length and large unobstructed field of view (see Fig.~\ref{Fig. FDT_block}). The FDT has an effective focal length of 579\,mm with a round FOV of 2 degrees. An external entrance pupil of 17.5\,mm is placed at 10\,mm in front of the first lens of the instrument. Therefore, the FDT F-ratio is 33.1 with an angular sampling on the \sophi\ detector of $3\arcsec75$ (corresponding to $761$\,km at 0.28\,AU).


\begin{figure}[h]
    \centering
    \includegraphics[width=\columnwidth]{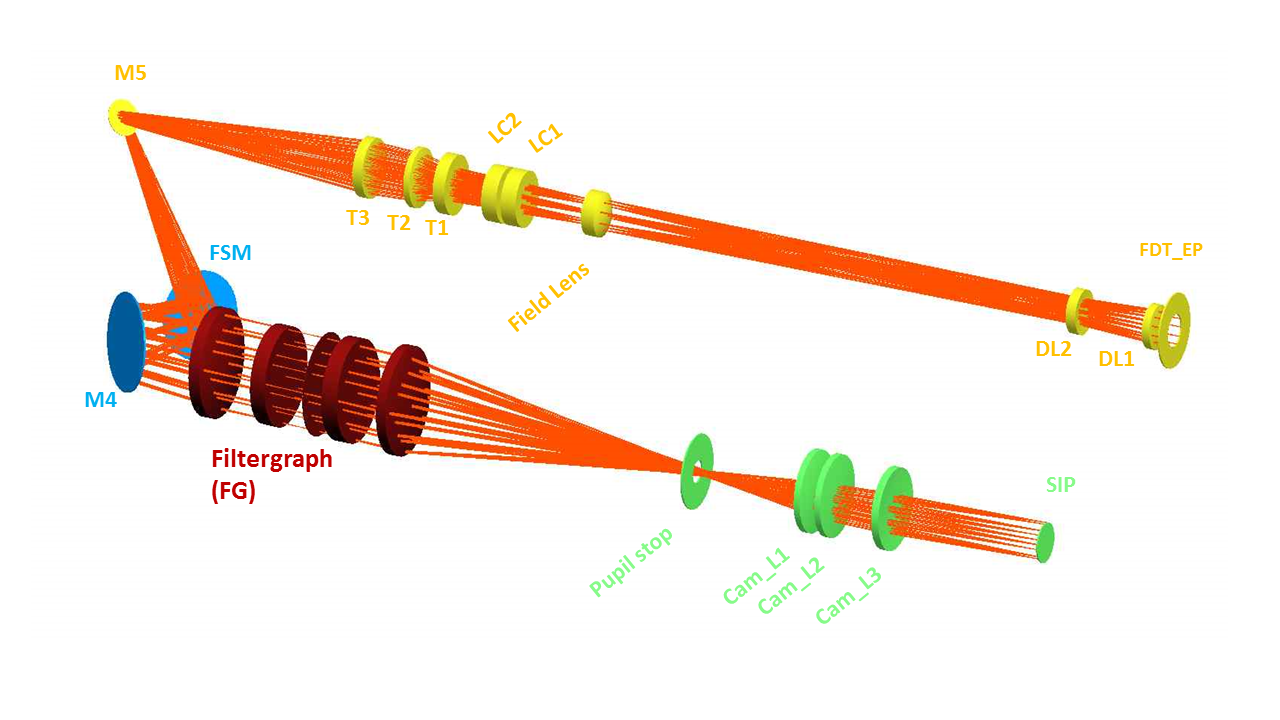}
    \caption{Optical scheme of the FDT path.}
    \label{Fig. fdt_optical_path }
\end{figure}



\begin{figure}[h]

    \centering
    \includegraphics[width=\columnwidth]{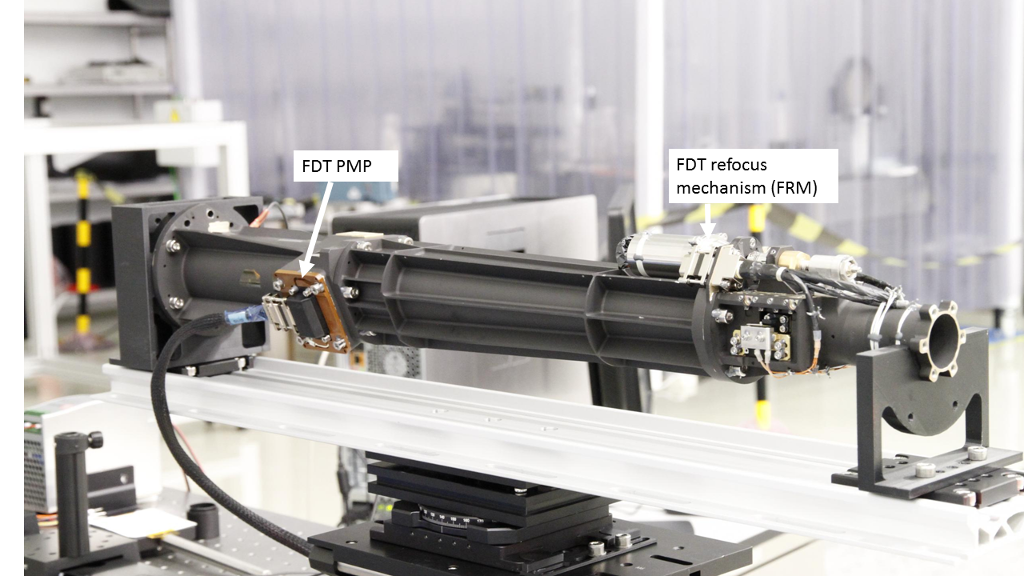}
    \caption{FDT tube with PMP included.}
    \label{Fig. FDT_phot }
\end{figure}


The FDT assembly is an on-axis refractive system. The first optical elements of the FDT assembly are a split doublet (L1 and L2) and a field lens (FL) forming an  uncorrected intermediate image, which is subsequently re-imaged by a triplet (T1, T2 and T3). FL acts as field lens which forms the exit pupil of the assembly in order to match the entrance pupil of the filtergraph (FG) system. In this way, both telescopes, the HRT and the FDT can have a common beam geometry and fulfill the requirement of telecentricity at the etalon focus. 

The FDT has its own refocus mechanism (FDT Refocusing Mechanism, FRM) for in-flight focus control. 
The FRM is based on axial movement of the second lens of the doublet (L2). It is used to compensate the manufacturing tolerances during the Assembly-Integration-Verification (AIV) phase as well as the thermal/vacuum environment defocusing produced during the mission mainly from the HREW. The FDT assembly incorporates a field stop at the intermediate image in order to block unwanted retro-reflected rays coming from the FDT HREW.
A summary of the main FDT features is given in Table~\ref{tab:opt_parameters}.


\begin{figure}
    \centering
    \includegraphics[width=\columnwidth]{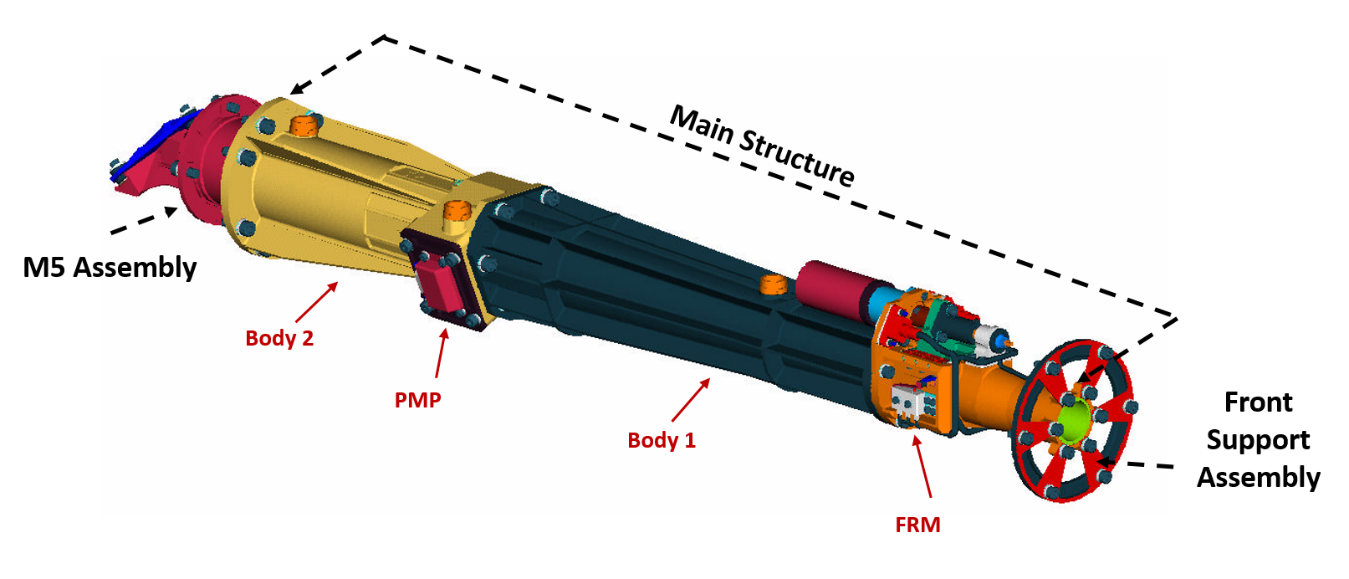}
    \caption{Mechanical design of the FDT assembly. The FDT assembly is mechanically composed of three elements: the Front Support Assembly, the Main Structure and the M5 assembly. The Main Structure includes the FRM, the PMP, body~1 and body~2.}
    \label{Fig. FDT_CAD}
\end{figure}


Mechanically, the FDT assembly is composed of three parts: the front support assembly (FSA), the main structure (MS) and the M5 assembly as shown in Fig.~\ref{Fig. FDT_CAD}.
The FDT MS is the main part of the FDT assembly. It includes two important subsystems with their own relevance, the FDT refocusing mechanism (FRM) and the PMP. The MS is placed in between the two structural blocks of the O-Unit structure. The FSA is placed at the front block, while MS and the M5 assembly are placed at the back block of the O-Unit structure, respectively. The role of the FSA is to provide support for the front section of the MS in such a way that no or negligible forces or moments will be induced into the \sophi\ O-Unit front block.

In the MS, the FRM \citep[][]{silvalopez15} assembly contains the entrance diaphragm barrel (with lens L1) and the L2 main mount (with the L2 lens barrel) that is mounted in a motorised platform. Body~1 is an empty tube. The PMP assembly contains the two LCVRs and the polariser (see Section~\ref{pmp_design}). Body 2 contains the field lens diaphragm sub-assembly (FL lens and the field diaphragm), the triplet sub-assembly (T1, T2 and T3 lens and triplet mount) and the housing for inserting the PMP. The M5 assembly contains the M5 folding mirror in its mount and the M5 housing. The material used for the MS was aluminum and the lens mounts are made of titanium 
in order to have similar thermal expansion coefficients for the optics and the mounts.

The FDT Re-focusing Mechanism (FRM) is an opto-mechanical assembly designed to hold 2 lenses, plus the entrance diaphragm of the FDT (see Fig.~\ref{Fig. FDT_phot }). The L2 lens is mounted on a motorised sliding platform and acts as focus compensator for the FDT path. L2 maximum excursion is approximately $\pm\, 2$\,mm from its nominal reference position with a precision of this movement of $\pm\, 25 \, \mu$m. 

\subsubsection{Folding scheme} 

\begin{figure}
    \centering
    \includegraphics[width=\columnwidth]{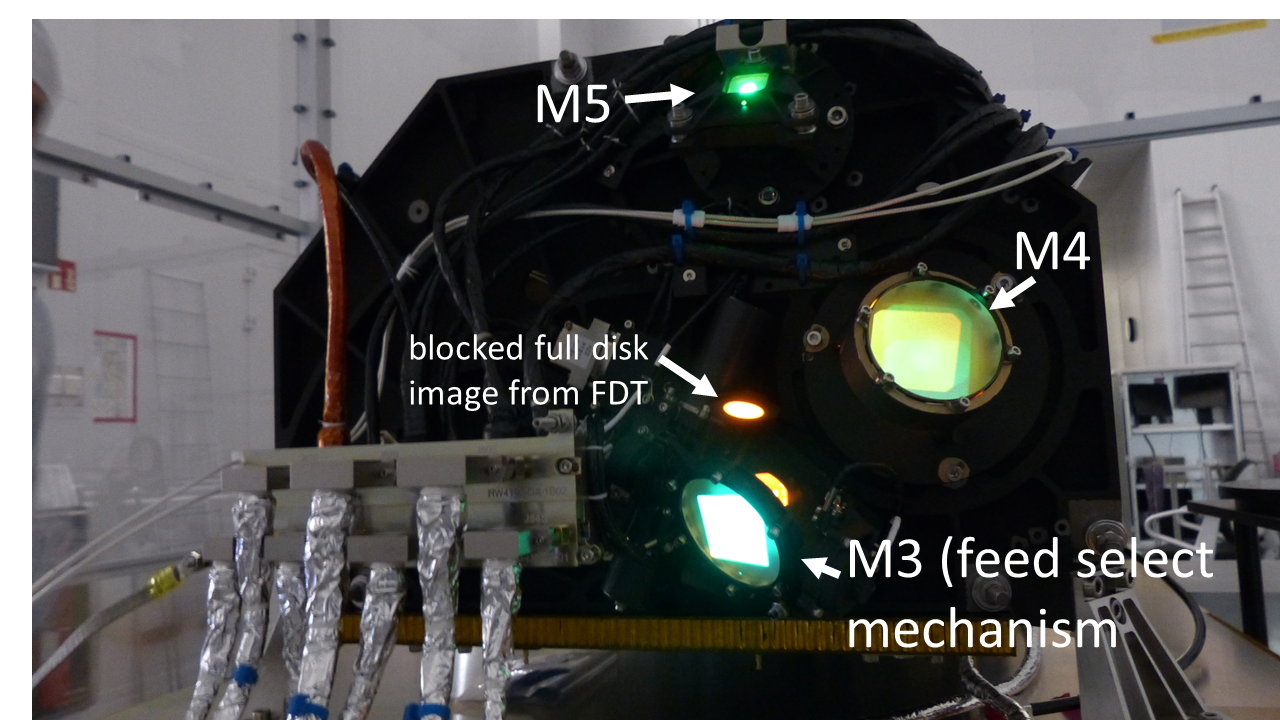}
    \caption{Rear side of the \sophi\ O-Unit with the folding mirrors M3 (feed select mechanism), M4 (common) and M5 (FDT).}
    \label{Fig. folding_photo}
\end{figure}

Both telescopes  are individually aligned with respect to the O-Unit Structure such that their intrinsic lines of sight are co-linear (see Figs.~\ref{Fig. folding } and~\ref{Fig. fdt_optical_path }). The selection of one of the telescopes is done by a movable folding mirror M3 (see Figs.~\ref{Fig. HRT_block } and~\ref{Fig. FDT_block}), which is actuated by the Feed-Select Mechanism (FSM). This fold mirror picks up the light from the HRT directly and sends it to the FG system via a fixed fold mirror M4. 
The light exiting the FDT assembly via the fold mirror M5, in picked up by the FSM, and the common fold mirror M4. 
M3, M4, and M5 (see Fig.~\ref{Fig. folding_photo}) are used as the compensators during the alignment of the instrument, in order to guarantee the co-alignment of both channels, as well as the telecentricity of both channels at  the etalon focus. It also serves to adapt the on-axis geometry of the FDT assembly to the off-axis geometry of the common path, which follows the geometry of the decentred HRT. Due to the folding scheme the images of the HRT and the FDT channel are rotated against each other by an angle of $45^\circ$. Also the FDT image plane is tilted with respect to the HRT image plane by an angle of $0.41^\circ$. This angle is small enough to ensure that the axial image shift is within the focus depth.  

\begin{figure}
    \centering
    \includegraphics[width=\columnwidth]{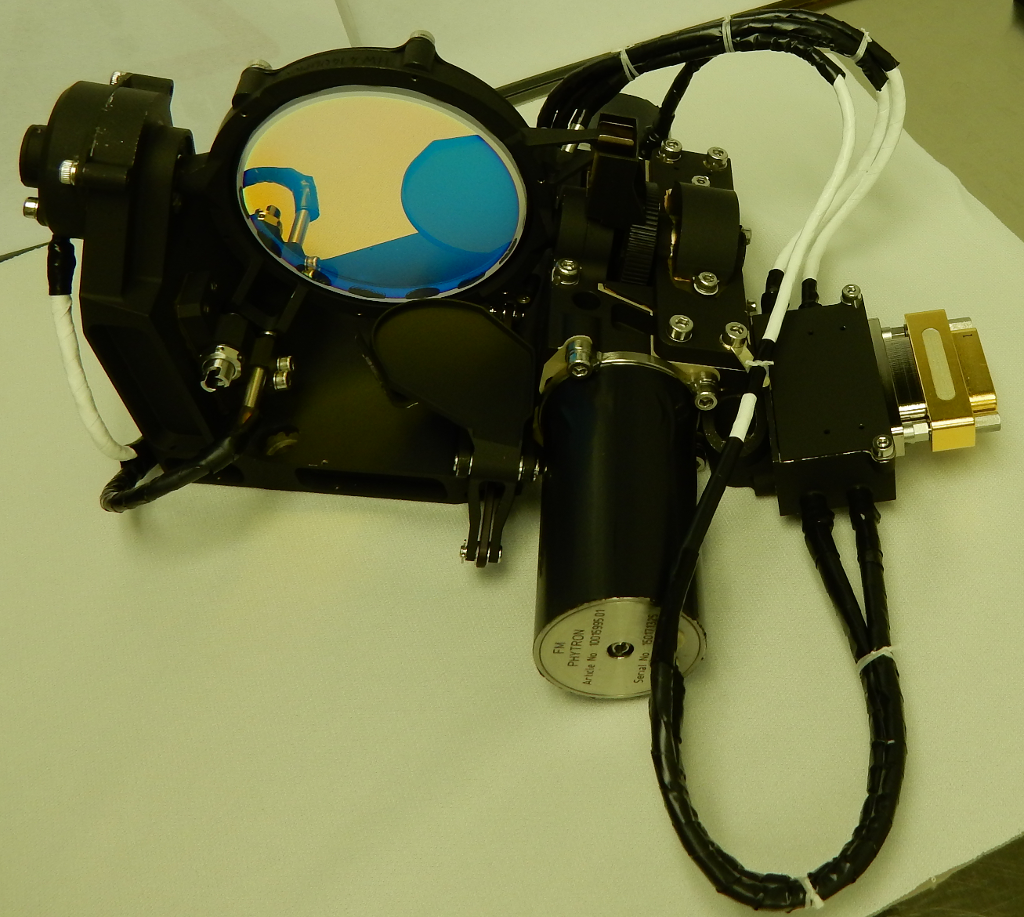}
    \caption{\sophi\ Feed-Select Mechanism flight model.}
    \label{Fig. fsm}
\end{figure}

\paragraph{Feed-Select Mechanism (FSM)}
As the junction  between the two optical feeds and the common optical path, the FSM (see Fig.~\ref{Fig. fsm}) serves a critical role inside the \sophi\ instrument. In nominal operation mode, the mechanism selects one of the optical feeds (HRT or FDT) with a mechanical positioning repeatability of at least $0.004^\circ$. Together with the mirror movement, the mechanism positions shutters such that the direct light path of the unused channel towards the focal plane assembly is blocked. To mitigate a potential source for single point failure, the mechanism is equipped with a fail-safe functionality. Through an auxiliary actuator, the fail-safe mechanism sets the M3 mirror into the FDT position. For more information on the FSM we refer to \citet{staub19}.


\subsubsection{Protection from the intense solar load}\label{hrew_design} 

\begin{figure}
    \centering
    \includegraphics[width=\columnwidth]{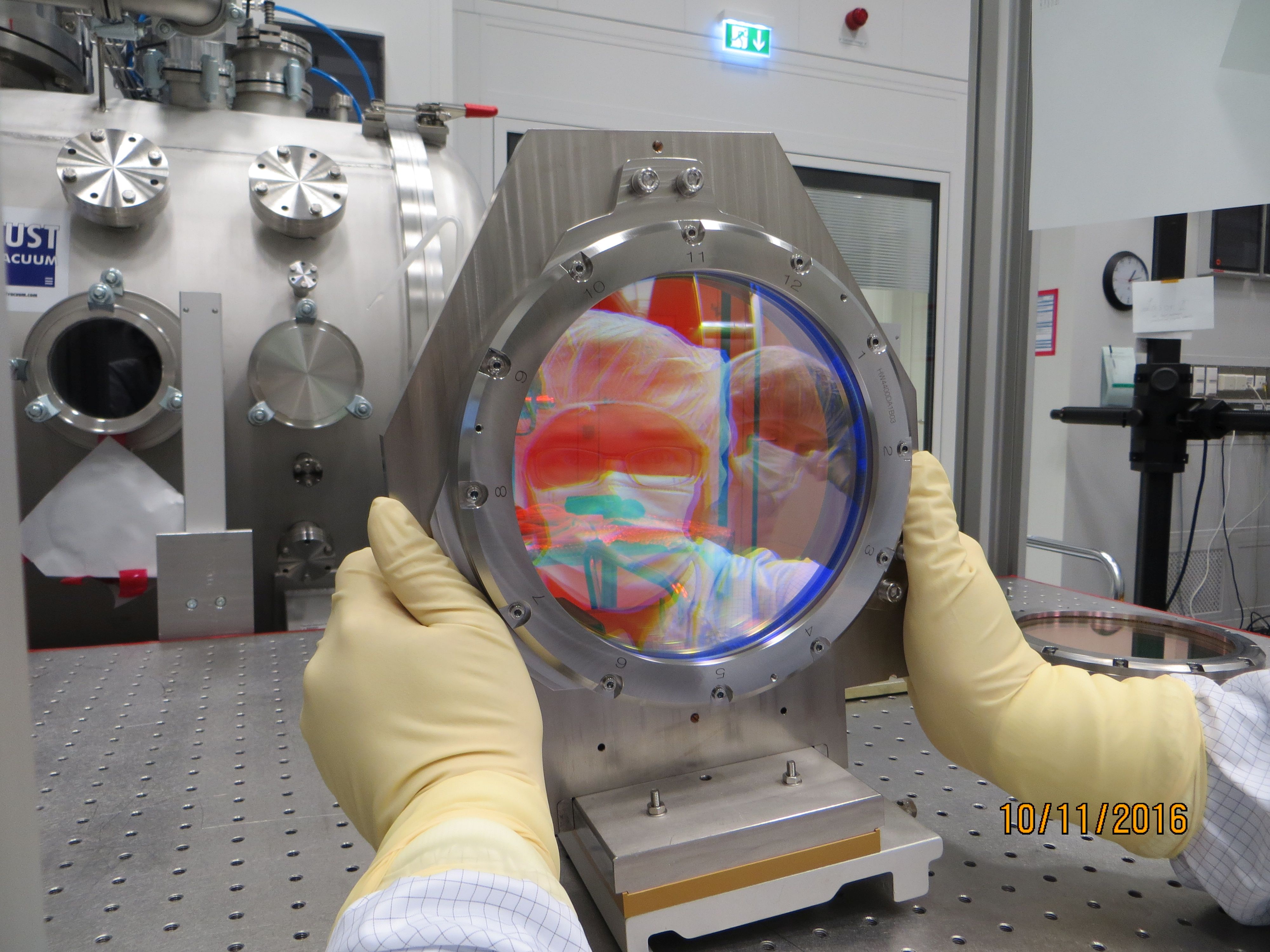}
    \caption{Heat rejecting entrance window of the High Resolution channel after successful thermal cycling test.}
    \label{Fig. HEW_test}
\end{figure}

Both telescopes are protected from the intense solar radiative flux by special Heat-Rejecting Entrance Windows (HREWs), 
which are part of the heat-shield assembly of the spacecraft. These multi-layer filters (see Fig.~\ref{Fig. HEW_test}) have more than 80\,\%\ transmittance in a narrow notch around the science wavelength, while effectively blocking 
the remaining parts of the spectrum from 200\,nm to the far infrared (IR) by reflection. Only a small fraction 
of the total energy is absorbed in the window, which acts as a passive thermal element by emitting part of the thermal radiation to cold space; emission of IR radiation into the instrument cavity is minimised by a low emissivity coating on the backside of the window (acting at the same time as an anti-reflection coating for the science wavelength). 
Thus the heat load into the instruments can be substantially reduced, while preserving the high photometric and polarimetric accuracy of \sophi.
Each window consists of two glass plates, carrying four different multi-layer coatings. The coatings have been developed and qualified for the use on Solar Orbiter by the company Leonardo in Carsoli, Italy, within the framework of an ESA technology development activity. 
The first coating is a ultraviolet (UV) shield. The second coating acts as a high pass filter, the third one as a low pass filter.
Together, these two coatings define a 30\,nm wide transmission pass-band around the science wavelength. The last coating on the inner side of the assembly is an IR blocker, which blocks the near IR portions of the incoming solar spectrum, and at the same time acts as a thermal IR mirror, thus minimising the thermal emission of the hot window into the instrument cavity (the window reaches temperatures $>200\,^\circ$C  in perihelion conditions).  
The design of the coatings and the material choice of the substrates (Suprasil\textsuperscript{\textregistered} 300) are identical for the HRT Entrance Window (HEW) and the FDT Entrance Window (FEW), while the mechanical designs of the windows differ slightly.  
 
\paragraph{HEW mechanical implementation} 

In the HEW, both Suprasil\textsuperscript{\textregistered} substrates are plane parallel and are 9.5\,mm thick. They are mounted in a titanium 
flange and held by two steel spiral springs against a spacer made of Tecasint (pure polyimide). 
The design is almost athermal, with 
the tendency to release the mounting force in hot operational conditions. 
The rms wave front error of the full assembly in ambient conditions has been measured to be 29\,nm, while the effective birefringence was measured to be 0.9\,nm over the full optical aperture. 


\paragraph{FEW mechanical implementation} 
The two SUPRASIL\textsuperscript{\textregistered} plates of the FEW
have a central thickness of 9\,mm and a wedge of 30\,arcsec each. The two glasses are mounted into a titanium holder with their wedges facing opposite directions. 
The mounting design aims for minimum mechanical loads and thermo-mechanical deformations in the substrates \citep{barandiaran17}.

\subsubsection{Polarisation analysis system}\label{pmp_design}

\begin{figure}
    \centering
    \includegraphics[width=\columnwidth]{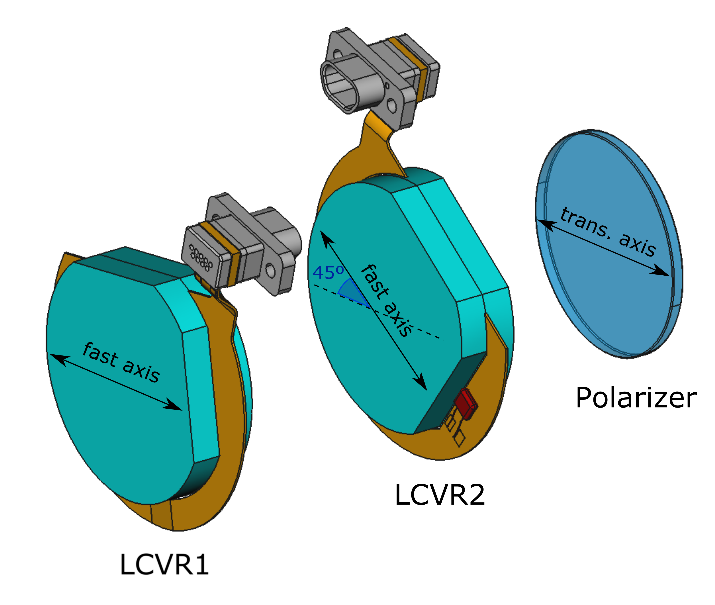}
    \caption{Scheme of the Polarisation State Analyser.}
    \label{Fig. PMP}
\end{figure}

The FDT and the HRT each include a Polarisation Modulation Package (PMP) to generate four known modulations of the polarisation state in order to extract the Stokes parameters of the incoming sunlight. This polarisation modulation is done using LCVRs. The complete device is a Polarisation State Analyser (PSA) since it includes a polariser as analyser.
The Polarisation Modulation Package scheme used for the \sophi\ instrument is employed in a number of ground-based polarimeters based on LCVRs \citep[cf.][]{deltoroiniesta03}; however, this is the first time that liquid crystals are used on board a space mission for polarimetric measurements \citep{alvarez15}. The LCVR technology was validated for the Solar Orbiter mission before the technology was included in the \sophi\ baseline. \cite{alvarez11} and \cite{uribe11} describe the main results of that work. One more instrument of the Solar Orbiter payload, Metis \citep[see][]{Antonucci2019a}, also employs these PMPs based on LCVRs. Each PSA consists of two anti-parallel Nematic LCVRs oriented with their fast axes at $45^\circ$ with respect to each other followed by a linear polariser (the polarisation analyser) at $0^\circ$ with respect to the fast axis of the first LCVR (see Figure~\ref{Fig. PMP}). 

\begin{figure}
    \centering
    \includegraphics[width=\columnwidth]{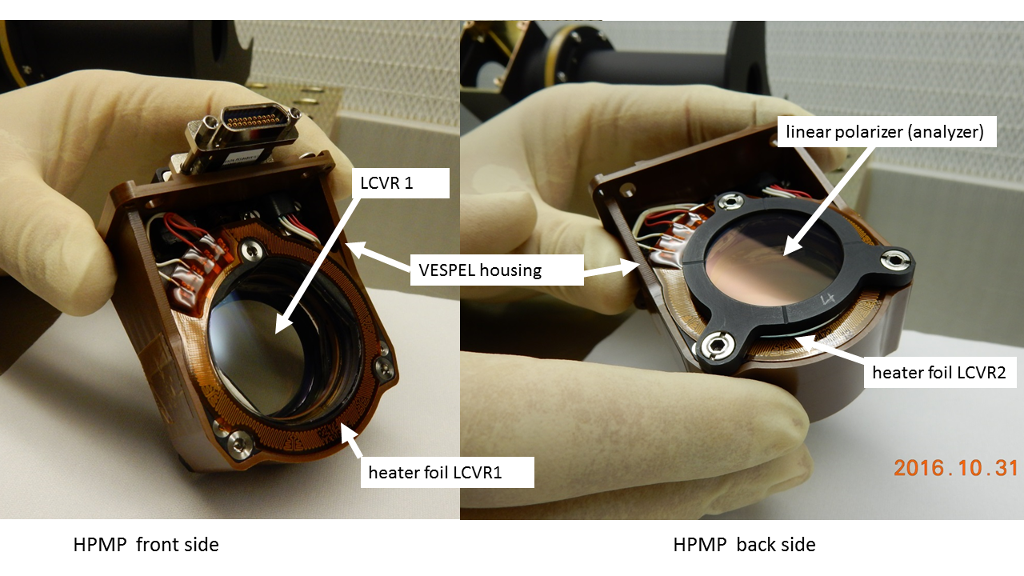}
    \caption{Polarisation State Analyser of the HRT path.}
    \label{Fig. HPMP}
\end{figure}

The PMP main structure (see Fig.~\ref{Fig. HPMP}) is manufactured of Vespel\textsuperscript{\textregistered} SP-1 in order to have enough mechanical stiffness and good thermal insulation. The voltage is applied to the LCVRs through a Kapton cable attached to the cells using an electrical conductor glue. This flexible cable also carries the signal of the temperature sensor (PT100) glued on it and located close to the clear aperture. This sensor provides the temperature of the cell for use by the integrated active thermal control. The cells are mounted into two aluminum rings. A heater with a maximum power of 4\,W is attached to them to provide the heating power. The temperature sensor and the heater have redundancy. The active thermal control has been designed to thermally stabilise the PMP to within $\pm0.5\,^\circ$C during data acquisition. For that, a proportional-integral-derivative electronic driver has been implemented in the \sophi\ E-Unit. The opto-mechanical mount guarantees that no stressing loads are applied to the LCVRs. This avoids breaking them, increasing the wavefront error, or inducing extra (asymmetric) retardance on either the substrates or the liquid-crystal material itself.

\subsubsection{Filtergraph}\label{fg_design}

The filtergraph subsystem contains the etalon, the order-sorting prefilter, and a blocking filter. Together with the field lenses FGL1 and FGL2 it forms an oven, which regulates the temperature of the sensitive etalon and prefilters. 


\begin{figure}[h]
    \centering
    \includegraphics[width=\columnwidth]{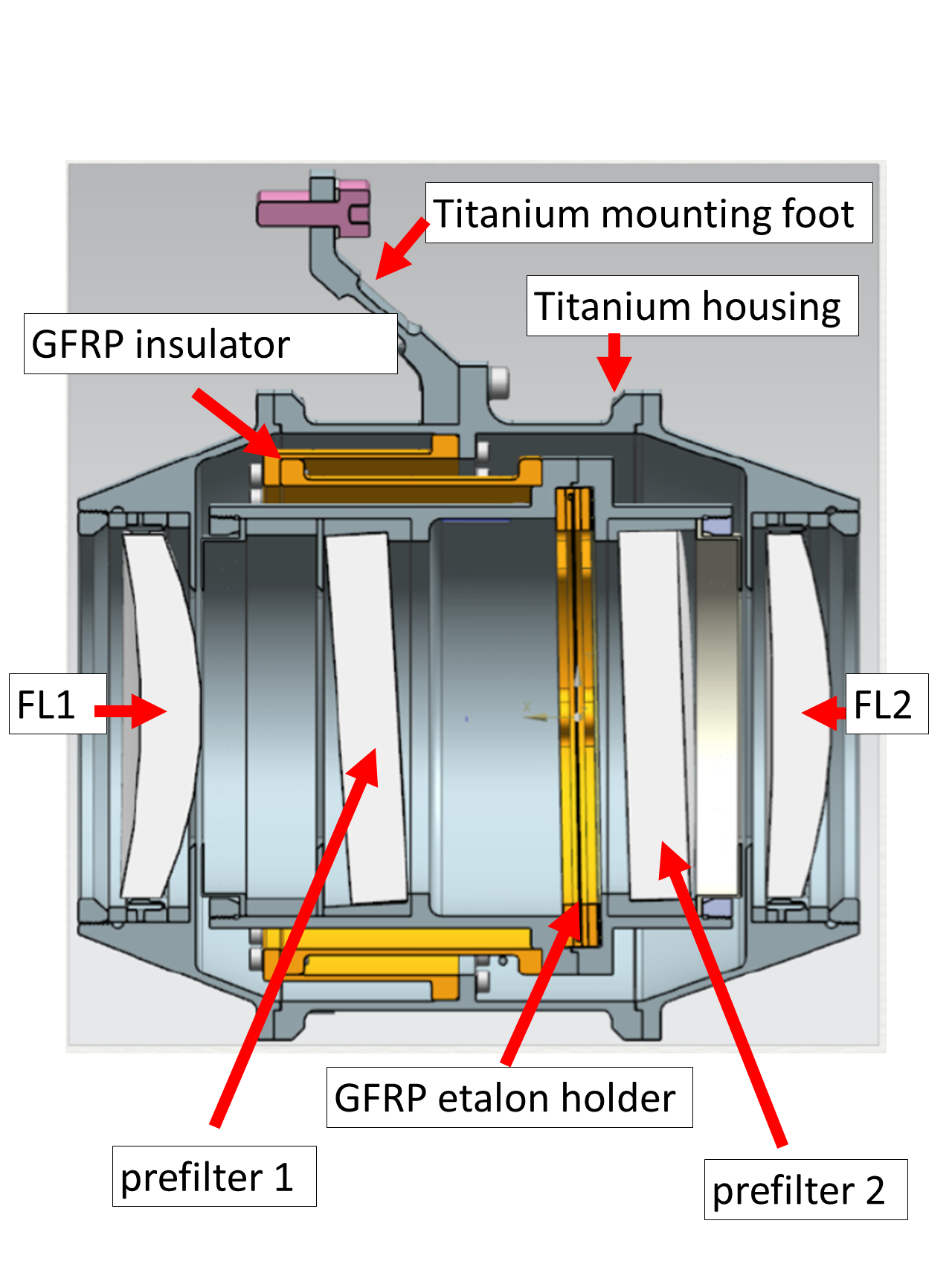}
    \caption{Optomechanical arrangement of the Filtergraph.}
    \label{Fig. FG_mech }
\end{figure}



\begin{figure}[h]
    \centering
    \includegraphics[width=\columnwidth]{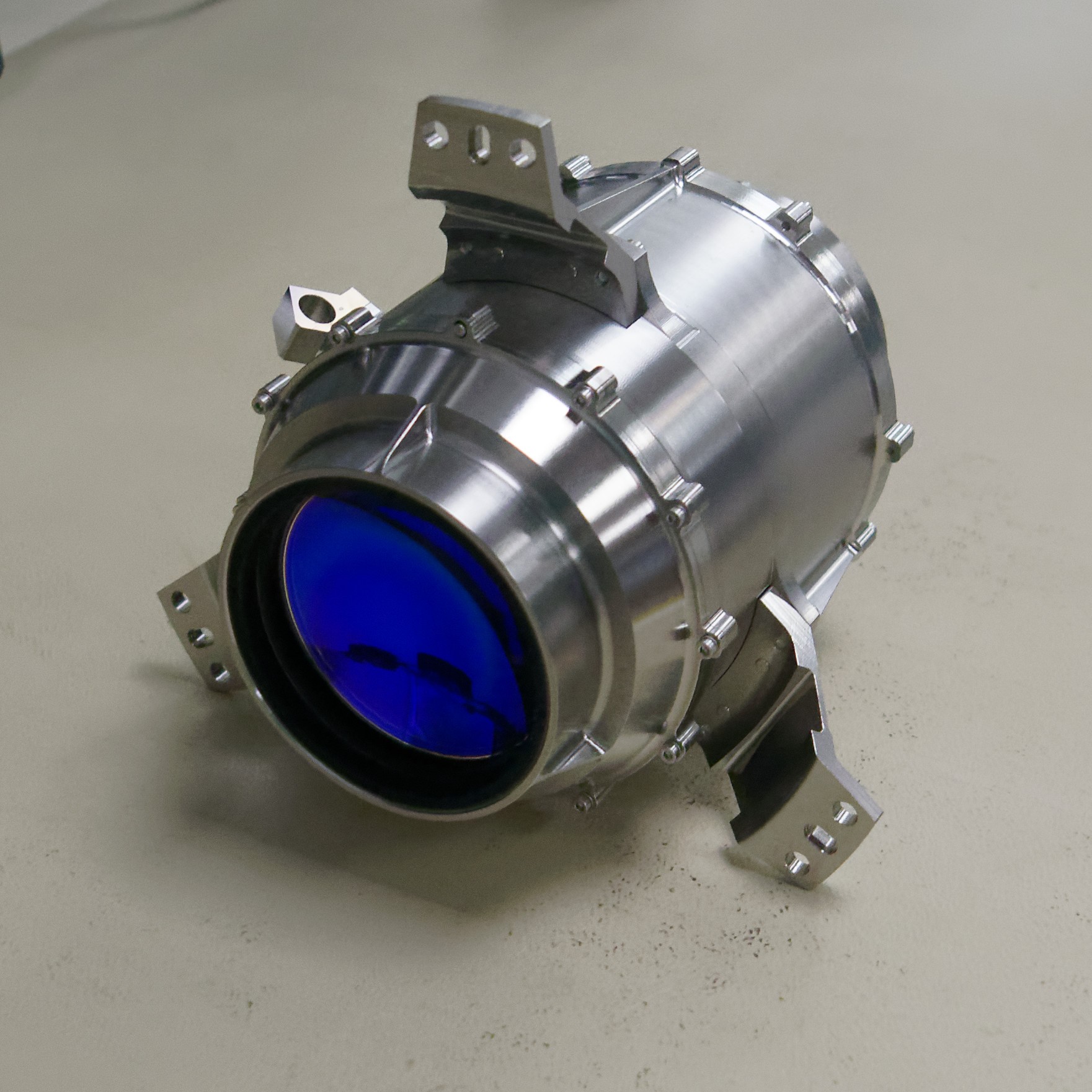}
    \caption{Picture of the Filtergraph oven.}
    \label{Fig. FG_Photo }
\end{figure}


The etalon for \sophi\ was produced by CSIRO in Australia. \citet{gensemer14} report on the fabrication and metrology of the etalons, which were tuned by ion deposition to absolute optical thickness, a novelty, which was  necessary due to the strong temperature restrictions within the instrument.  The etalon performances were then calibrated at the Institut d'Astrophysique Spatiale (IAS) facility using a dye laser.  The wavelength sensitivity of the FM etalon on the applied voltage was calibrated to $(351.1\pm 1.0)$\,m\AA\,kV$^{-1}$.
The wavelength sensitivity of the FM etalon on temperature was calibrated to be $(37.9\pm 4.9)$\,m\AA\,K$^{-1}$.
The free spectral range (FSR) of the etalon was calibrated to be 0.301\,nm with a mean full-width-at-half-maximun (FWHM) of $(106\pm 5)$\,m\AA , resulting in an effective finesse of 30. All values are averaged over the illuminated area of the etalon in the FG. 

The two prefilters were produced by Materion, USA.  The FWHM bandwidths of the 2 prefilters are 0.27\,nm and 10\,nm, respectively.  The homogeneity of the bandwidth of the narrow prefilter is better than 10\,\%\ peak-to-peak over the used optical area.  The prefilters are wedged and mounted at two different angles for ghost image control. 

The etalon is thermally stabilised by  conductive and radiative insulation.  For radiative insulation, the lenses are coated with a low-emissivity coating and act as thermal IR shields between the prefilters and the instrument cavity. Further radiative decoupling is ensured by a multi-layer insulation wrapping the etalon and its mount.  For conductive insulation, the etalon is mounted in a glass fibre-reinforced plastic (GFRP) structure with low thermal conductivity to the rest of the oven.  The oven is made of titanium, which has a lower thermal conductivity than aluminum, while having the same strength.  With this design, it is possible to stabilise the temperature of the etalon at $66\,^\circ$C for any outside temperature between $-20\,^\circ$C and $+65\,^\circ$C with less than 1.5\,W of heating power.  The typical time scale of the thermal inertia seen by the etalon is about 4 hours. The thermal stability of the etalon was measured at 0.3\,mK rms (corresponding to a wavelength stability of $1.03\times 10^{-5}$\,\AA\ rms, which amounts to an error of 50\,cm\,s$^{-1}$ rms in the obtained Doppler velocities).

The High Voltage Power Supply (HVPS) of the etalon provides a range from $-2.6$\,kV to $3.9$\,kV.  The HVPS was designed by the IAS and built by the EREMS company.  The stability of the HVPS is calibrated to be 1.3\,V rms for time scales shorter than 1\,s (corresponding to a wavelength shift of 0.45\,m\AA\ or a Doppler velocity of 22\,m\,s$^{-1}$ rms).  The overall combined stability including the temperature and high voltage stabilities lead to the noise induced velocity of the FG to be less than 30\,m\,s$^{-1}$ rms, sufficient for detecting solar $p$-modes with a degree larger than 10. The power supply is located in the E-Unit of \sophi\ (Section~\ref{e-unit}) and connected to the FG by two high-voltage cables.

\subsubsection{Focal Plane Array and camera optics}

The focal plane array (FPA) is based on a $2048 \times 2048$\,pixel APS sensor from CMOSIS in Belgium (now AMS), custom made for \sophi. 
The pixel pitch is $10\,\mu$m. The frame rate is 11 images per second. 

\begin{figure}[h]
    \centering
    \includegraphics[width=\columnwidth]{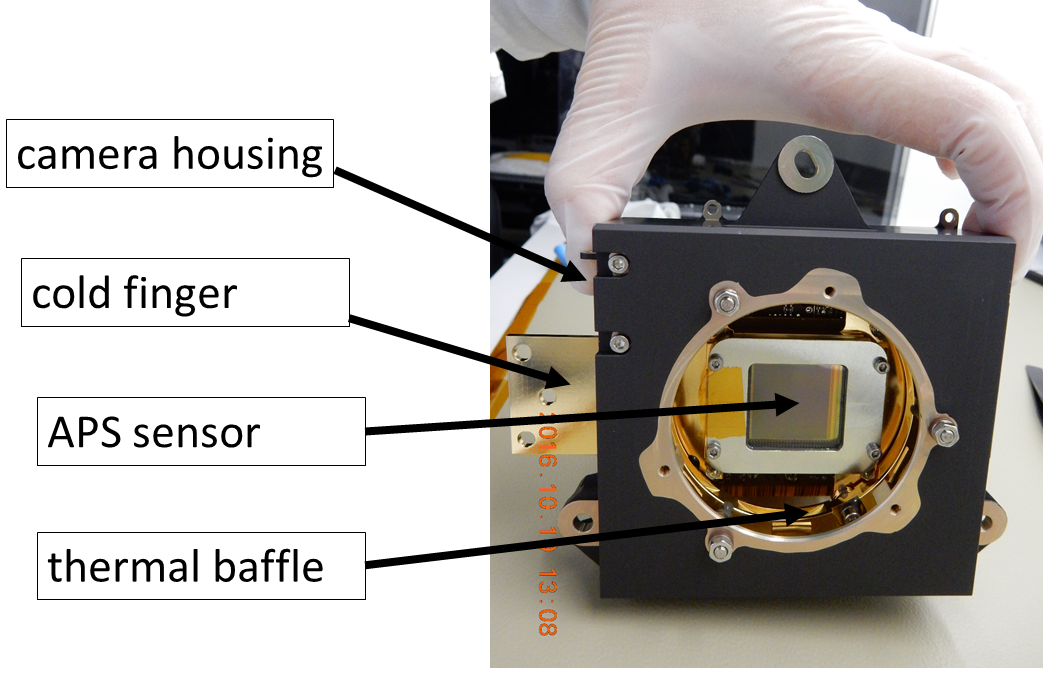}
    \caption{Focal Plane Array with APS sensor and cold finger during assembly. The optical baffle in front of the detector is not yet mounted, such that the gold plated thermal baffling can be seen.}
    \label{Fig. FPA }
\end{figure}

The camera is mechanically built around a large aluminum beryllium alloy plate that carries the sensor, attached with thermal glue, and protrudes from the housing to form the camera cold finger, which is connected to a flexible cold strap, draining energy from the sensor to a dedicated cold radiator on the side of the spacecraft (see Fig.~\ref{Fig. FPA }). This cold element interface is controlled by the spacecraft to be $-10\,^\circ$C.  

The sensor is connected, via pins passing through the cold finger, to a custom-made sensor socket, soldered to the front-end electronics printed circuit board (PCB). Two fast 14-bit analogue-to-digital converters (ADCs) are directly coupled to the sensor outputs. A second PCB contains the components for the command interface to the \sophi\  power and processing unit, the control signals to operate the detector
and its support electronics, and to acquire and transfer the image
data. 

The camera housing is made of aluminum
parts with conductive conversion coating; all external surfaces are black coated. Back-shells and baffles are used in order to
ensure the light tightness of the sub-assembly. A labyrinth-type
structure is built around the sensor to minimise potential contamination from the electronics while avoiding contact with the mounting interface, thus ensuring thermal isolation.


The camera lens optics consists of three lenses in one group, which re-image and de-magnify the etalon focus onto the APS. The lens group is mechanically attached to the camera and forms part of the light tight baffling system, which is needed to separate the light path after the FG from the bright open parts of the telescope. It also contains a pupil stop, which is used as a Lyot stop for ghost suppression of the etalon/prefilter ghosts (see Section~\ref{sec:OPT_Baffling}). 

\subsubsection{Image Stabilisation System}\label{sec:ISS}

\begin{figure}
    \centering
    \includegraphics[width=\columnwidth]{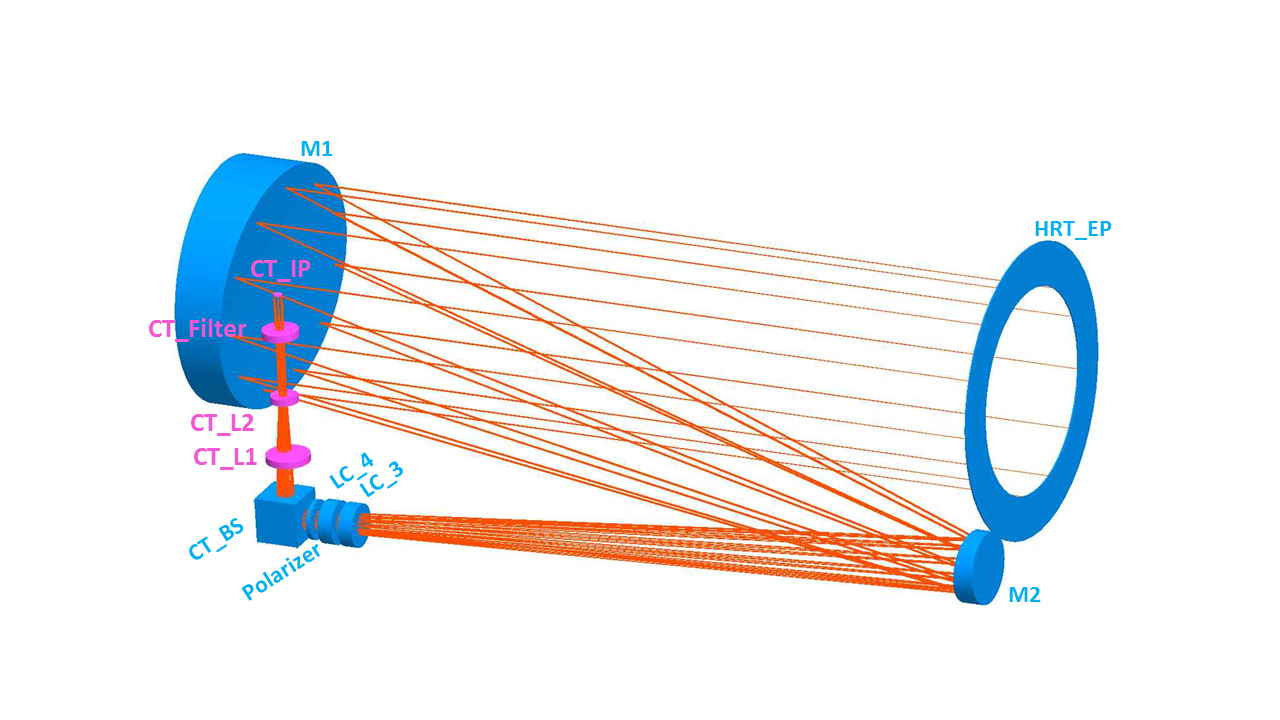}
    \caption{Optical scheme of the ISS path. M2 is used as the tip/tilt compensator; CT\_BS is the beam splitter which feeds a small fraction of the light to the image plane (CT\_IP) in the Correlation Tracker Camera.}
    \label{Fig. iss_optical_path}
\end{figure}

The Image Stabilisation System \citep[ISS; see][]{volkmer12} reduces the residual image motion on the detector, which is due to pointing inaccuracies of the spacecraft. This is mandatory, since the differential imaging needed for the polarimetry is sensitive to relative image shifts between the individual exposures. In order to bring the differential errors below the noise limit, a relative shift from one exposure to the next must be smaller than $1/20$ of a pixel. For the HRT channel this is incompatible with the  residual pointing error that is expected from the spacecraft. 
To this end, an adaptive, real-time  image motion compensator (correlation tracker) has been implemented in the \sophi\ HRT channel. 
A dedicated fast camera images the solar scene. From the correlation of these images, an error signal is computed and sent to a movable mirror (tip/tilt mirror). The light for this camera is taken out of the HRT path by a beam-splitter cube, which reflects 2.8\,\%\ of the intensity (cf. Fig.~\ref{Fig. iss_optical_path}). Since this beam-splitter is placed behind the active mirror, the Correlation Tracker Camera (CTC) sees the "corrected" scene in closed loop. The ISS optical path is equipped with an independent Correlation tracker Refocus Mechanism (CRM).


The sensor of the CTC is a Star1000 from ON Semiconductor and allows  integration times between 0.02\,ms and 3\,ms, frame rates up to 600\,fps and a FOV from $64\times 64$ to $128\times 128$ pixels. The data from the CTC are linked by a fast interface to the \sophi\ Data Processing Unit (DPU, see Section~\ref{DPU}).
The ISS control firmware runs in one of the two Reconfigurable Field Programmable Gate Arrays (RFPGAs) of the DPU \citep[][]{carmona14}. It compares the position of the real time image with a reference image. The reference image must be updated approximately every 60\,s due to the evolution of the solar surface pattern. The displacements of the images are calculated by an absolute differences algorithm with sub pixel resolution in a field of $7\times 7$ pixels \citep[][]{casas16}. The detected shifts are converted to a tilt angle and the corresponding signal is sent to the fast tip-tilt mirror M2 in closed loop. The resulting band width of the ISS is above 30\,Hz.

The CTC is equipped with its own pre-filter and re-imaging optics, which adapts the plate scale, and which can be shifted in order to act as an in-flight focus compensator. 

The HRT secondary mirror was chosen as the active optical component. The actuator is an optimised device of the series S-340 manufactured for \sophi\ by P.I. Systems, Germany. The low voltage PICMA\textsuperscript{\textregistered} piezo stacks with a few hundred layers are encapsulated to be insensitive to vacuum and environmental impact. The housing of the S-340 device is made of titanium to achieve a stiff and light-weight structure. 


\begin{figure}[h]
    \centering
    \includegraphics[width=\columnwidth]{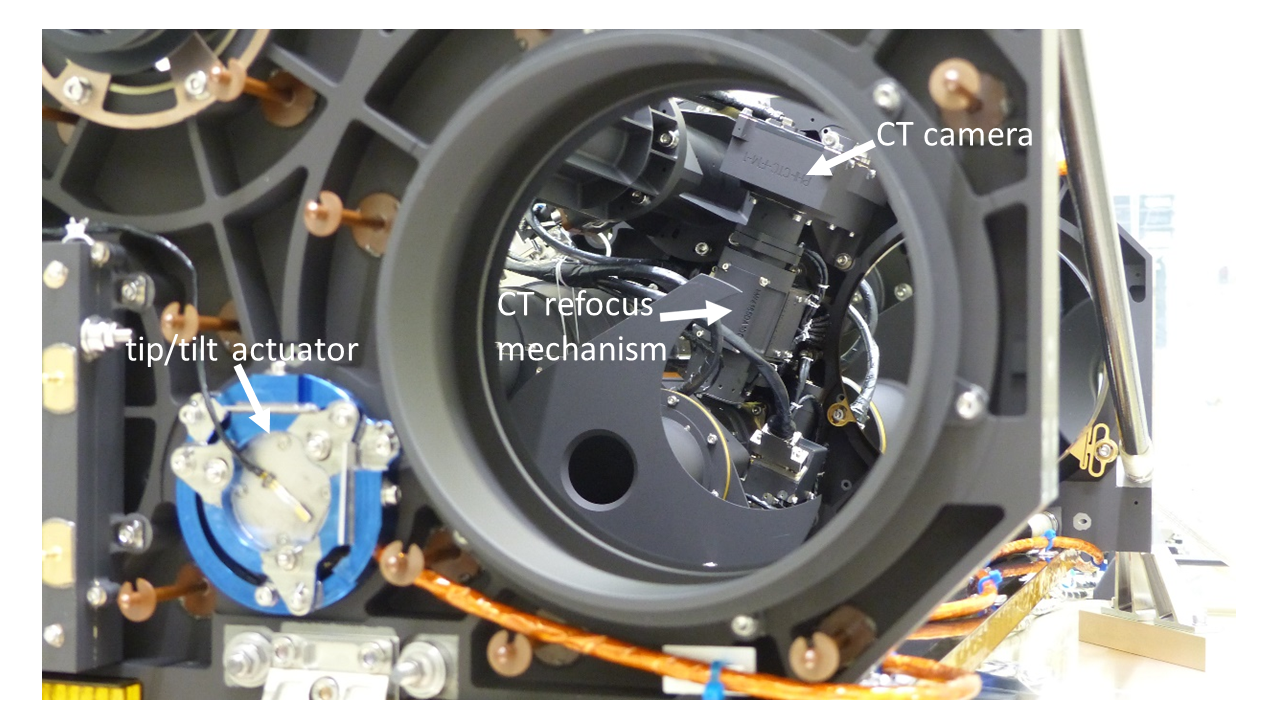}
    \caption{Main components of the Image Stabilisation System as seen through the HRT aperture.}
    \label{Fig. ISS }
\end{figure}



\begin{figure}[h]
    \centering
    \includegraphics[width=\columnwidth]{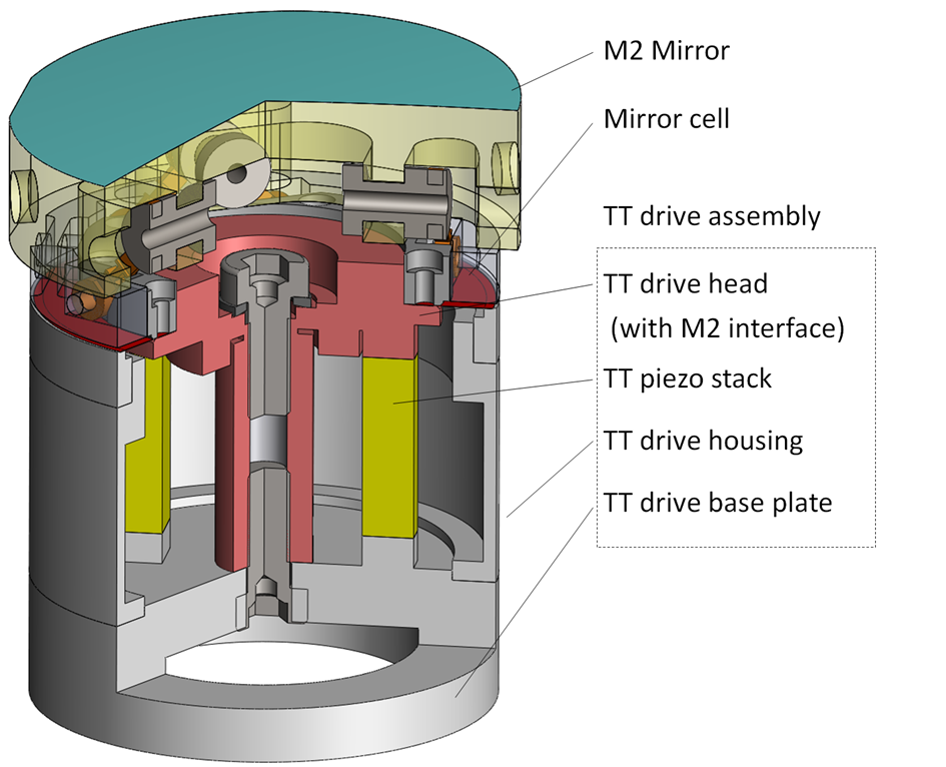}
    \caption{Schematic  overview of the tip/tilt drive assembly with mounted M2 mirror}
    \label{Fig. TT }
\end{figure}


\subsection{Thermal Aspects}\label{thermomechanical} 


\begin{figure}[h]
    \centering
    \includegraphics[width=\columnwidth]{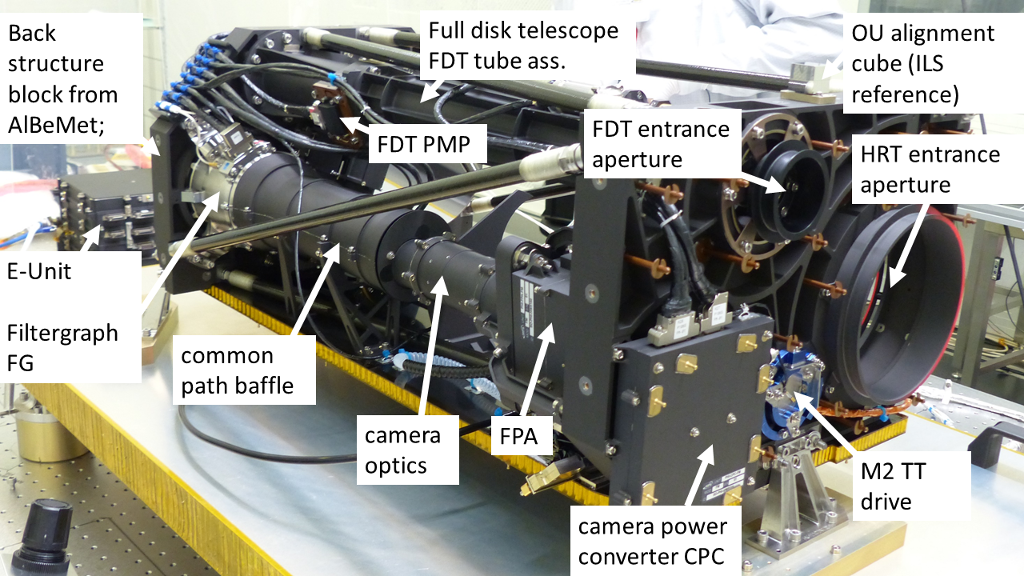}
    \includegraphics[width=\columnwidth]{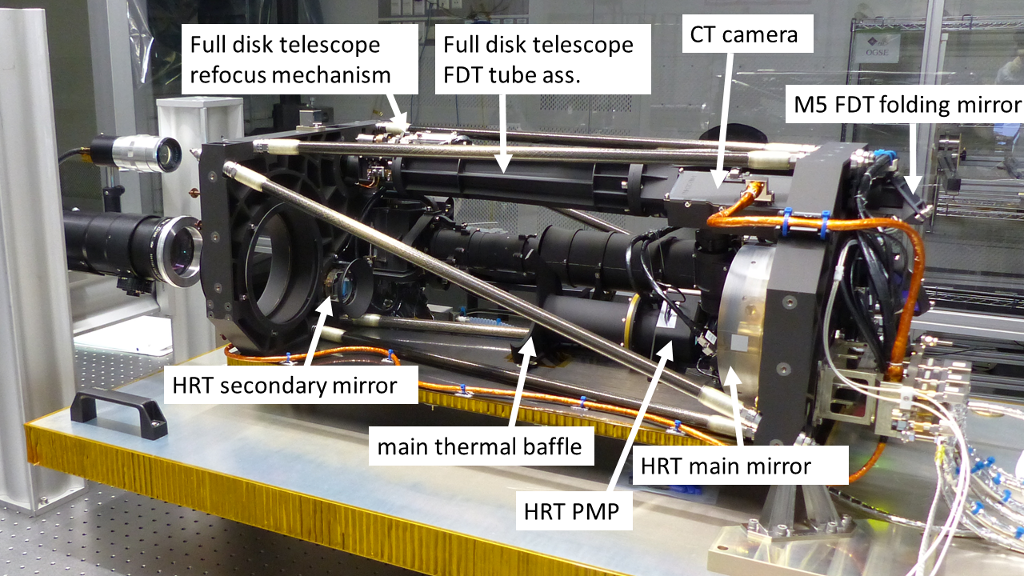}
   \caption{Main subsystems of the \sophi\ O-Unit as seen on the right (upper figure) and from the left side of the instrument (lower figure).}
    \label{Fig. PHI_OU_right}
\end{figure}


\subsubsection{Thermal design}\label{thermal}

\paragraph{SO/PHI O-Unit thermal loads:}

The thermal behaviour of the \sophi\ O-Unit is driven by the external thermal loads (solar light and IR radiation) entering the instrument through the apertures, the internal dissipation of the electronics, and the thermal conditions at the boundaries of the instrument, both radiative and conductive. The solar irradiation constitutes the main source of energy. The maximum external load is found during closest perihelion passage at 0.28\,AU, where the spacecraft sees practically 13 solar constants (17.5\,kW\,m$^{-2}$). As already explained above, only a narrow wavelength band is of interest for science purposes. Therefore, the HREWs filter this band, allowing only 3.2\,\% of the incoming solar radiation to enter the instrument, which is about 10\,W at closest perihelion. The Sun irradiation heats up the entrance feedthrough system, resulting in an IR load of  about 8\,W entering the instrument. The unit's internal power dissipation is 6\,W in hot conditions, and additional 4\,W of heater power is available for the thermal stabilisation of the etalon and the LCVRs.

The unit sits within the spacecraft cavity, where the environmental temperatures (both conductive and radiative) are guaranteed by the spacecraft management system. In this way, during the science phase, the environment of \sophi\ will be kept between $+10\,^\circ$C and $+50\,^\circ$C, whereas in the non-operational phase it will be between $-30\,^\circ$C and $+60\,^\circ$C.

\paragraph{SO/PHI O-Unit thermal concept:}

\begin{figure}[h]
    \centering
    \includegraphics[width=\columnwidth]{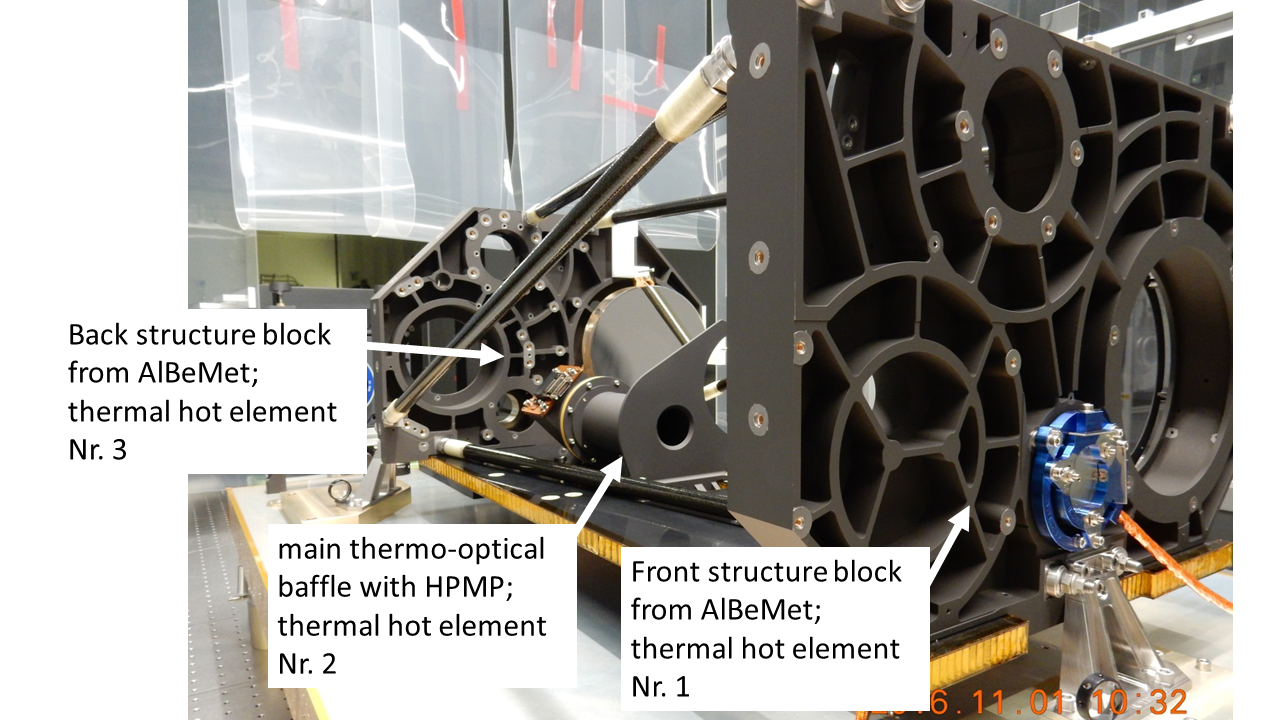}
    \caption{\sophi\ O-Unit structure with the three hot element interfaces: structural blocks, and the main thermo-optical baffle.}
    \label{Fig. ifs }
\end{figure}

\begin{figure}[h]
    \centering
    \includegraphics[width=\columnwidth]{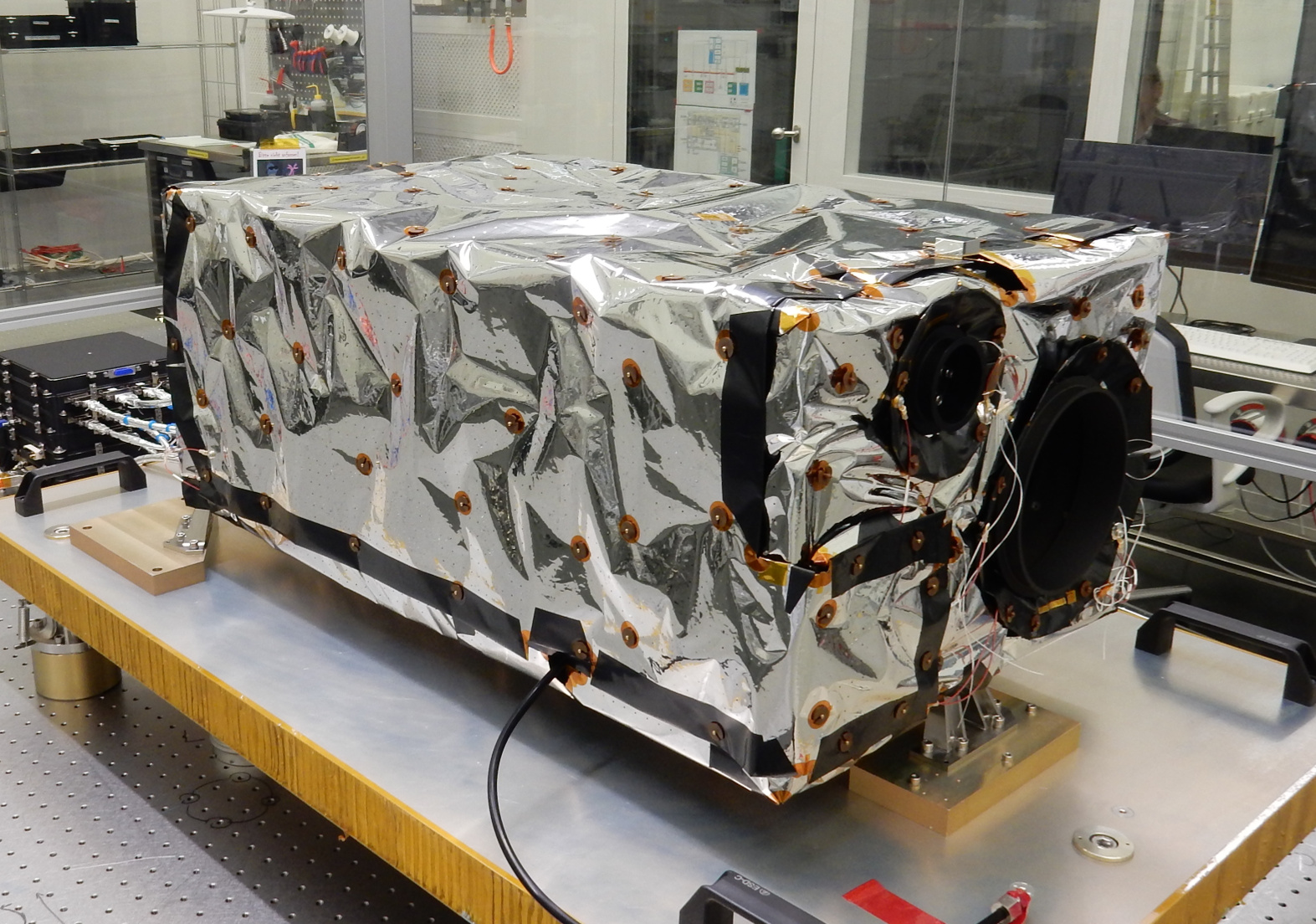}
    \caption{\sophi\ O-Unit with the Multi Layer Insulation installed.}
    \label{Fig. PHI_OU_MLI }
\end{figure}

The instrument thermal design has been carried out following the guidelines defined for Solar Orbiter and the applicable standards listed in European Cooperation for Space Standardization. It is based on the following points:

\begin{itemize}[noitemsep]

\item The HREWs limit the energy entering the instrument to 3.2\,\% of the total solar energy seen by the instrument apertures (see Section~\ref{hrew_design}).

\item The \sophi\ O-Unit is a so-called thermally ‘insulated unit’, a design requirement imposed by ESA. This means that the O-Unit is allowed to transfer to the spacecraft cavity no more than 1\,W of thermal radiation and 1\,W conductively through the mounting feet. The heat dissipated by the equipment, and the solar and infrared loads entering the instrument have to be rejected to space through dedicated radiators connected to the instrument by means of 4 specific conductive interfaces, three hot element interfaces (HE) to keep the instrument practically isothermal at room temperature, and one cold element interface (CE) to keep the detector below the maximum allowable temperature of $-25\,^\circ$C during all science phases (see Figure~\ref{Fig. ifs }). 

\item The instrument's opto-mechanical design (see Section~\ref{sec:struct_design}) 
keeps the O-Unit practically isothermal. 

\item In order to achieve the insulation level required and to  minimise the leakage to the spacecraft, the unit is wrapped in 12-layer multi-layer insulation (MLI). 
The surface of the MLI facing the instrument is black Kapton to avoid stray light within the unit, and the outer layer is Vacuum Deposited Aluminum (VDA) embossed perforated Kapton, highly reflective to decouple radiatively the unit from the spacecraft. 
A photograph of the \sophi\ O-Unit with the MLI is shown in Fig.~\ref{Fig. PHI_OU_MLI }.

\item All the internal surfaces are  coated black, not only for optical reasons but also to reduce hot spots within the instrument.  

\item Stabilisation heaters are used to control the temperature of those elements that require stable temperature during the science acquisition periods: the etalon and the prefilters in the FG and the LCVRs in the PMPs (see Sections.~\ref{pmp_design} and~\ref{fg_design} for further explanations of the fine thermal control of these elements). 

\end{itemize}


\paragraph{Thermal Analysis.} 

\begin{figure}[htb]
    \centering
    \includegraphics[width=\columnwidth]{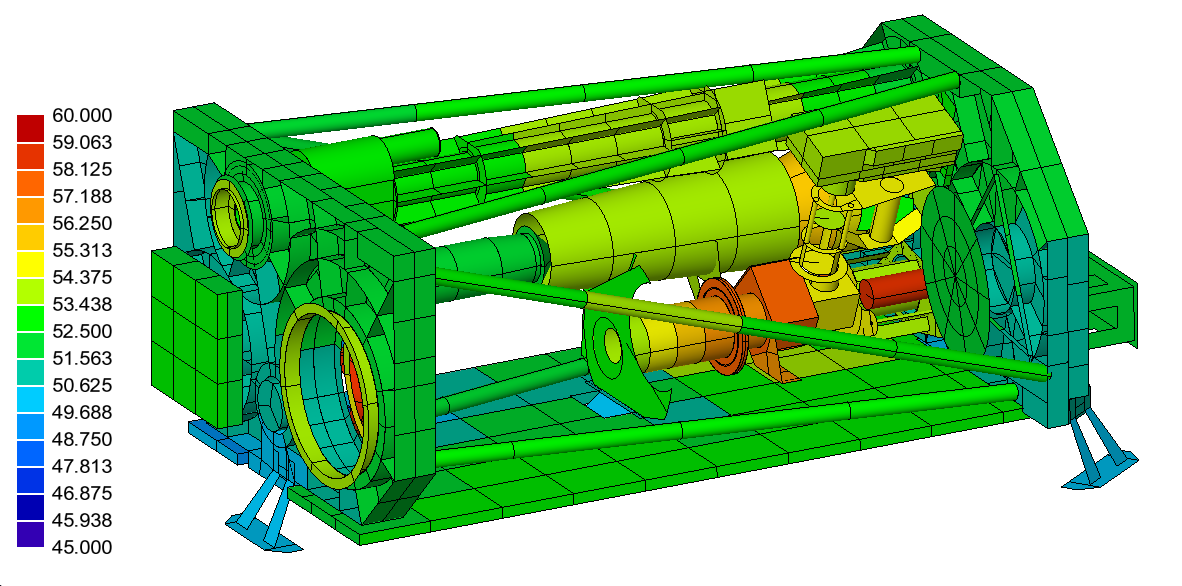}
    \includegraphics[width=\columnwidth]{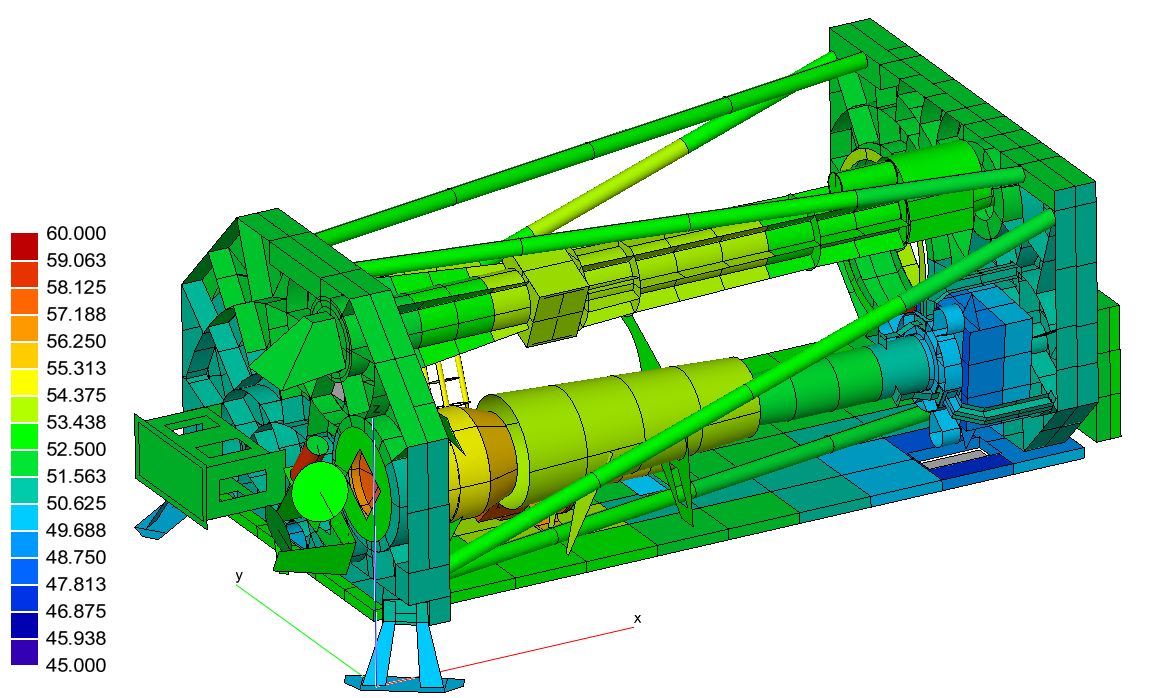}
    \caption{\sophi\ O-Unit thermal analysis results for the hot operational case. The colour bars shows temperatures given in $^\circ$C.}
    \label{Fig. hot}
\end{figure}

A detailed Thermal Mathematical Model (TMM) along with the corresponding Geometrical Mathematical Model (GMM) was set up to size the thermal control pieces of hardware and to analyse the performance of the unit. The tool used was ESATAN-TMS. With these models, the worst thermal dimensioning cases and the flight performance cases were analysed. 
The model was iterated with the design and correlated with thermal vacuum test data.
Results for all the cases were obtained. They include temperatures, heat fluxes through the instrument interfaces with the spacecraft, and temperature gradients, which were used for both structural and optical analysis. The results were satisfactory, proving that the instrument is expected to be within the thermal limits during all mission phases. For simplicity, only the hot operational case results are presented in Fig.~\ref{Fig. hot} (MLI is not displayed). Detailed information about the \sophi\ thermal model and the results obtained can be found in \citet{perezgrande16}.

\subsection{Testing}

\subsubsection{Mechanical testing}

One cornerstone of the \sophi\ environmental test campaign was the vibration test of the O-Unit conducted at  Industrieanlagen-Betriebsgesellschaft mbH (IABG) in Ottobrunn, Germany. Due to the high vibration levels and the delicate specimen, it was decided to follow a force-limited vibration approach. This allowed close monitoring of the forces at the unit interfaces and significantly eased the notching negotiations with the agency.

In order to get immediate information about the unit's integrity, functional tests were  conducted after vibration testing along each axis. The functional tests used a fibre optic light source connected to  lens barrels adapted to the focal length of HRT and FDT. The 
simplistic mobile setup was sufficient for assessing the fundamental health status of the instrument.

\subsubsection{Thermal testing}

\begin{figure}[h]
    \centering
    \includegraphics[width=\columnwidth]{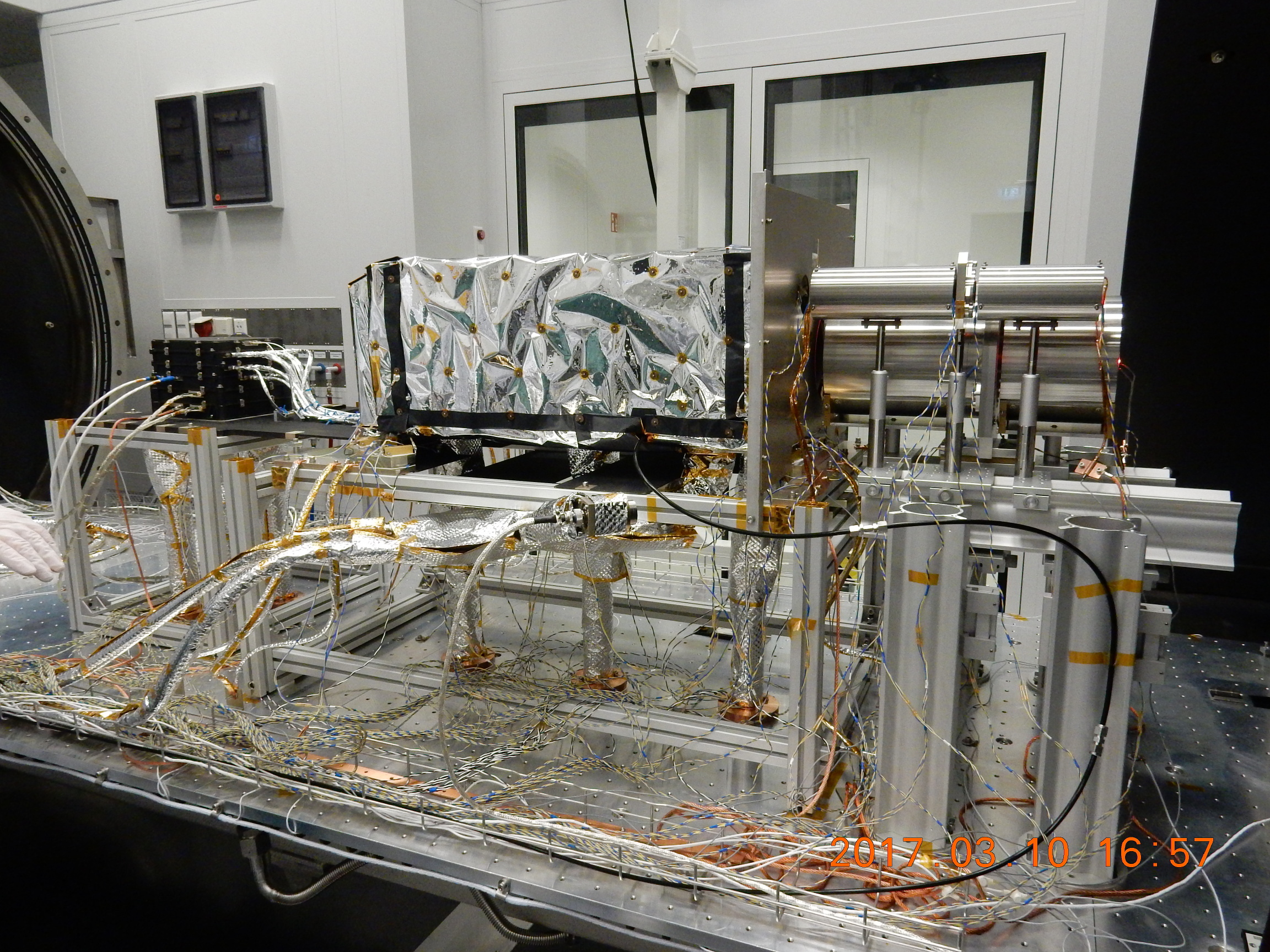}
    \caption{\sophi\ Flight Model thermal test setup prior to insertion into the large thermal vacuum chamber at MPS.}
    \label{Fig. ThermalTest}
\end{figure}

As part of the verification test campaign, a thermal vacuum test was carried out on the fully integrated \sophi\ instrument. The thermal test had a threefold objective: the acceptance test of both, O- and E-Units, the correlation of the thermal mathematical model of the Optics Unit, and the spectral calibration of the instrument (see Section~\ref{calibration}). The test facility chosen was the large thermal vacuum chamber ("Big Mac") at the Max Planck Institute for Solar System Research, in G\"ottingen, Germany. The spectral calibration of the instrument was made possible by feeding sunlight into the thermal vacuum chamber. The sunlight was driven from the facility building roof down to the vacuum tank with a coelostat and several feed mirrors located inside the test facility building. The instrument (including O-Unit, E-Unit, harness and HREWs) with all the auxiliary equipment,  prior to  insertion into the thermal vacuum chamber, is depicted in Figure~\ref{Fig. ThermalTest}. For more information on the \sophi\ thermal test campaign, we refer to \citet{fernandez18}.


\section{Electronics Unit}\label{e-unit}

\begin{figure*}
    \centering
    \includegraphics[width=2\columnwidth]{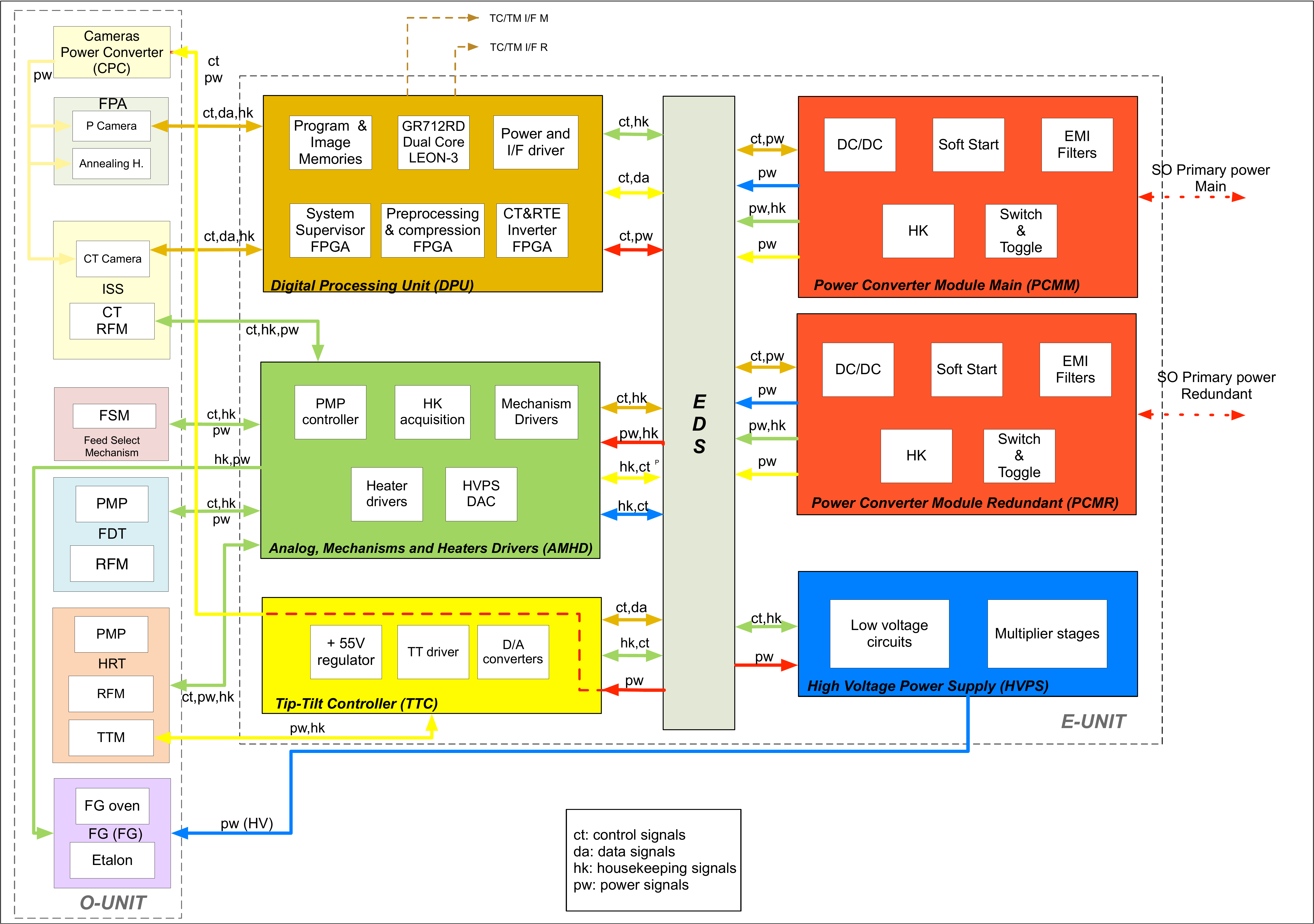}
    \caption{E-Unit block diagram.
    \label{Fig. BLOCK-E-UNIT}}
\end{figure*}

\begin{figure}
    \centering
    \includegraphics[width=\columnwidth]{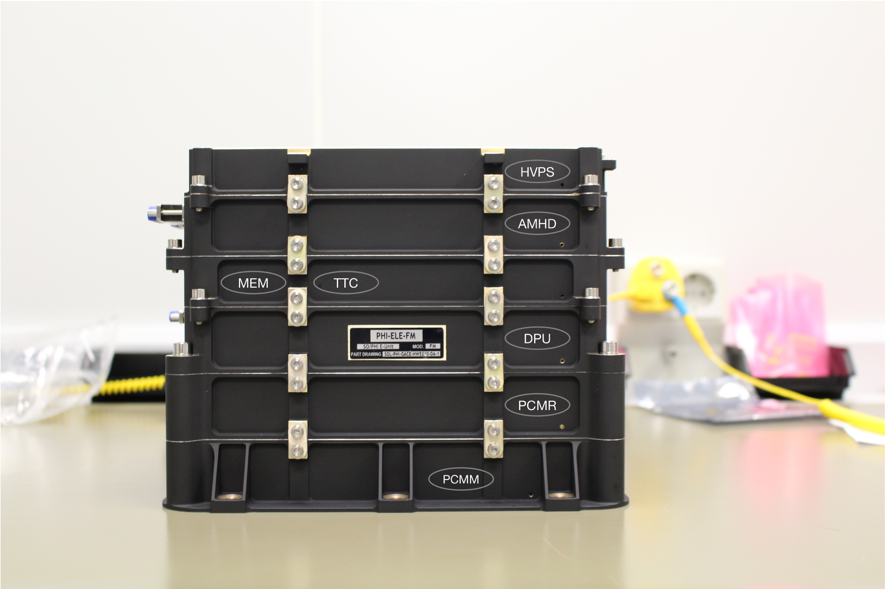}
    \caption{Flight model of the E-Unit. The various subsystems are labelled. From top to bottom we have the high-voltage power supply; the Analogue, Motor, and Heater Driver; the mezzanine memory board of the data processing unit and the tip-tilt controller; the main data processing unit board; the redundant power converter module; and the main power converter module.}
    \label{Fig. FM-E_UNIT}
\end{figure}

The Electronics Unit (E-Unit) controls the whole instrument and provides communication and power interfaces to the spacecraft. It is  a modular system with individual boards stacked on top of each other and interconnected by means of a motherboard, the Electrical Distribution System (EDS; cf. Section~\ref{sec:EDS}). Its conceptual design is shown in Fig.~\ref{Fig. BLOCK-E-UNIT}, where the various boards are distinguished by colours. The interconnections between boards are shown by means of lines where the type of signal is indicated. A picture of the E-Unit flight model is shown in Figure~\ref{Fig. FM-E_UNIT}. A short description of each board follows.


\subsection{The Power Converter Module}\label{sec:PCM}

The Power Converter Module (PCM)  handles the instrument power interface with the spacecraft and  provides the different input voltages to the various \sophi\ subsystems \citep[][]{sanchis14}. It is composed of one printed circuit board (PCB) with more than 1200 components on both sides. To enhance reliability, the PCM is doubled, with one acting as the main and the other as the redundant PCM, configured in cold redundancy. Fulfilling the very stringent electromagnetic cleanliness (EMC) requirements of Solar Orbiter posed critical challanges to the design \citep[][]{sanchis16}. 

As the interface with the spacecraft, the PCM manages the bus voltage, the on-off commands via the High Voltage High Power Pulse Commands (HV-HPC), and the status signal via the Bi-Level Switch Monitor (BSM).

The PCM has three main blocks, the input section, the DC/DC converter and the power distribution. The input section handles the power interface with the satellite and the DC/DC converter is custom made to better fulfill the subsystem needs. The power distribution provides power to all subsystems, directly to the Data Processing Unit (DPU) and the Analogue Motor and Heater Driver (AMHD) boards and, via an on/off switch controlled by the DPU, to all other subsystems in the E-Unit and O-Unit.

To improve the reliability of the PCM, the module contains undervoltage and overvoltage lockouts, overcurrent detectors on the primary and secondary side as well as an overvoltage detector on the most critical secondary side voltage. The PCM boards are placed in the two lower modules of the E-Unit box and labelled PCMM (main) and PCMR (redundant) in Fig.~\ref{Fig. FM-E_UNIT}.

\begin{figure*}
    \centering
    \includegraphics[width=2\columnwidth]{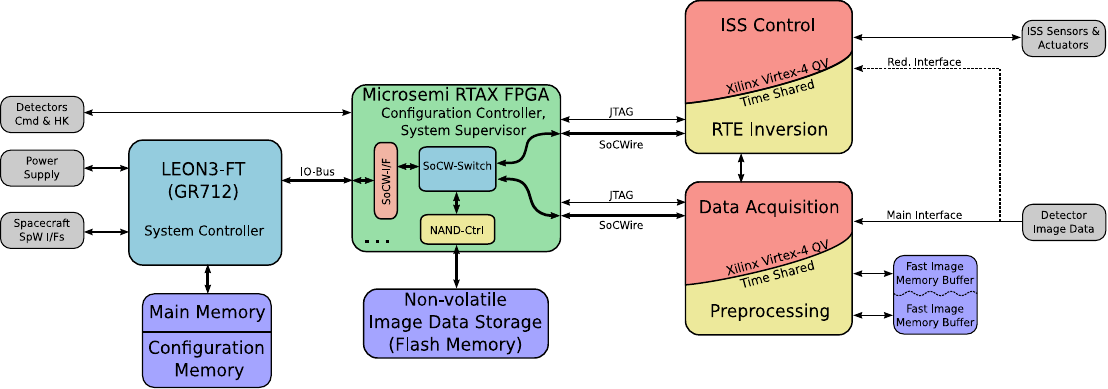}
    \caption{Block diagram of the Data Processing Unit}
    \label{Fig. DPU}
\end{figure*}

\subsection{The Data Processing Unit}\label{DPU}

A basic block diagram of the Data Processing Unit (DPU) is shown in Figure~\ref{Fig. DPU}. The design \citep{fiethe12} is based on a flexible approach, like a System on Chip (SoC). The limited telemetry rate combined with the large amount of scientific information retrieved from the FPA demands sophisticated on-board functionalities. Thus, the DPU utilises a combination of a processor Application Specific Integrated Circuit (ASIC) together with a radiation hardened and Triple Modular Redundancy (TMR) by design, one-time programmable Field Programmable Gate Array (FPGA; Microsemi RTAX) and a set of dedicated processing cores implemented within two in-flight reconfigurable Xilinx Virtex-4 FPGAs \citep[][]{lange15b,lange17}. The in-flight reconfigurable processing cores are attached to the running system by a flexible on-chip communication architecture, named System-on-Chip Wire (SoCWire) and based on the well-established SpaceWire standard \citep{osterloh09}. 

The Cobham Gaisler GR712RC processor ASIC, containing a LEON-3FT based system, is employed as main system controller for high-level instrument control and for communication with the platform implementing the SpaceWire protocol. All static interfaces and instrument control/monitoring functions needed for basic operations are integrated in the Microsemi FPGA to achieve the highest reliability. Additionally, this FPGA acts as a system supervisor to achieve the configuration control and required Single Event Effect (SEE) radiation tolerance of the reconfigurable FPGAs \citep{michel13}, supported by a high-reliability  configuration memory (NOR flash) for firmware storage. A combination of a small amount of volatile Fast Image Memory Buffer (1\,GiByte SDRAM) and a large non-volatile Image Data Storage (512\,GiByte NAND-flash)  provides significant storage capacity and fulfills all needs for intermediate data storage at very low resources \citep{lange15a}. For high data rates, dedicated memory controllers directly control the memories within the FPGAs, including complete error correction and taking into account the NAND-flash handling.

The image acquisition and processing functionalities are split into the two reconfigurable Virtex-4 FPGAs. The dynamic reconfigurability of these FPGAs enables multiple use of the FPGA resources during different modes of operation. This will be used for both simple update capability of hardware functions during the long mission and repetitive in-flight reconfiguration for sharing of complex algorithms on limited FPGA resources. The operational and processing requirements of \sophi\ facilitates very well the Time-Space Partitioning (TSP) of different modules in two main operation modes. During the image acquisition period, FPGA\#1 is used to run the correlation tracker algorithm for controlling the tip/tilt mirror, while FPGA\#2 is in charge of image data accumulation and stores the results in the NAND-flash memory. When observations are stopped, FPGA\#2 is reconfigured several times to perform pre-processing (calibration and polarimetric demodulation) of the stored image data, while FPGA\#1 executes the RTE inversion of the observed Stokes parameters to get the properties of the observed portion of the solar atmosphere. After that, FPGA\#2 is reconfigured again to perform reordering and bit truncation of the data, which are then sent to FPGA\#1 where the processed data are finally compressed. 

The main aspects of the data pre-processing and the inversion of the thus obtained Stokes parameters are described in Sections~\ref{sec:preprocessing} and~\ref{sec:RTE}. 
The goals and basic properties of the correlation tracker operation are described in Section~\ref{sec:ISS}. 

\begin{figure}
    \centering
    \includegraphics[width=0.8\columnwidth]{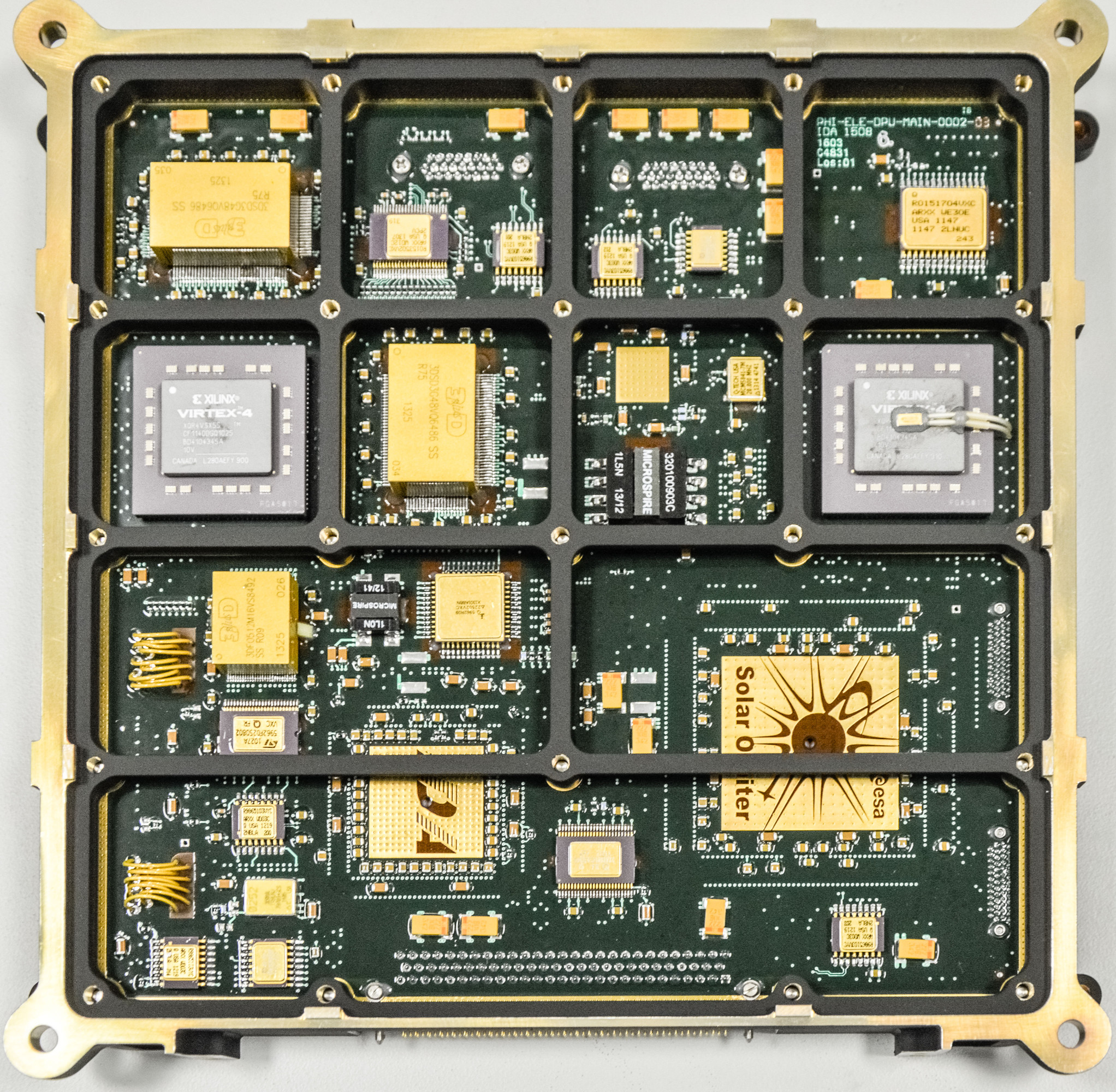}
    \caption{Flight model of the Data Processing Unit.}
    \label{F:DPUborad}
\end{figure}

Physically, the DPU subsystem consists of two boards, one main (cf Fig.~\ref{F:DPUborad}) and one mezzanine board. The latter contains the non-volatile memory devices and shares the frame with the tip/tilt controller. 

\subsection{The Analogue, Motor and Heater Driver}\label{sec:AMHD}

The Analogue, Motor and Heater Driver (AMHD) acquires the instrument housekeeping (HK) data and controls  mechanisms and heaters. In order to perform these tasks, it has interfaces with all the E-Unit as well as with some of the O-Unit subsystems.

Regarding HK, the AMHD acquires and adapts the signals from the different sensors (voltages, currents and temperatures) located either in the E-Unit or in the O-Unit, which provide the necessary information about the health of the whole instrument. 

\begin{figure}
    \centering
    \includegraphics[width=\columnwidth]{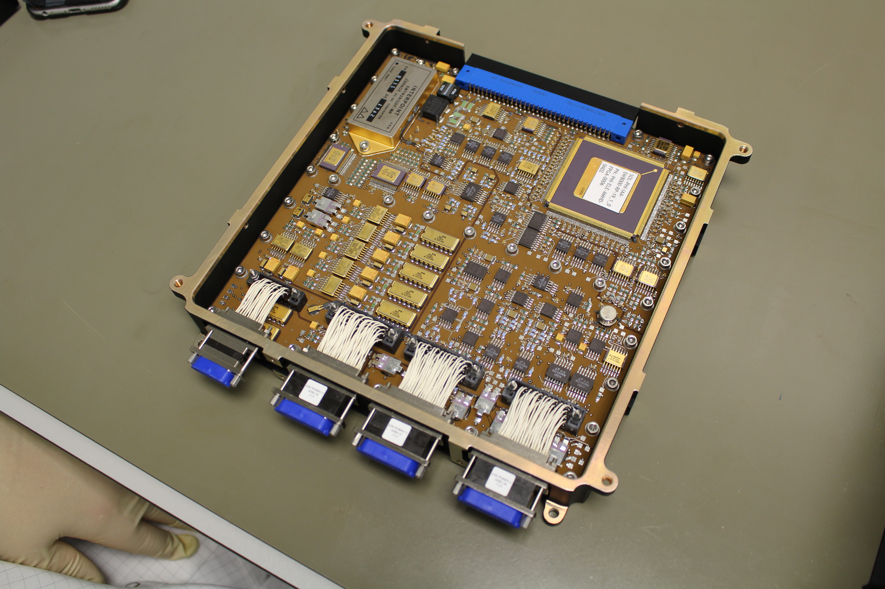}
    \caption{Flight model of the Analogue, Motor, and Heater Driver board.}
    \label{Fig. AMHD}
\end{figure}

The AMHD also generates and synchronises the accurate square modulated signals needed to control the LCVRs included in the PMPs. Regarding the instrument mechanisms control, the AMHD generates the phase signals to command the four stepper motors included in the O-Unit mechanisms that correspond to the FSM, the HRM, the FRM and the CRM. The AMHD also performs the control of the fail-safe mechanism.

The AMHD controls the instrument heaters. On the one hand, it controls the instrument heaters of both PMPs and the FG oven. This is achieved by means of proprtional-inegral-derivative (PID) algorithms implemented on an FPGA for temperature control. On the other hand, the AMHD provides the power and enable signals of the Camera Power Converter (CPC) annealing heaters. Finally, the AMHD also generates the signals to control the HVPS. A picture of the AMHD flight model is shown in Fig.~\ref{Fig. AMHD}.

\subsection{The High Voltage Power Supply}\label{sec:HVPS}

The High Voltage Power Supply (HVPS) provides a controlled high voltage to tune the LiNbO$_3$ etalon. The differential voltage covering the range from $-5$\,kV to $+5$\,kV, with a high short-term stability, is generated in this board. Under flight conditions the HVPS will only be used over a reduced range from $-1.3$\,kV to $2.0$\,kV.

The differential voltage provided to the etalon is obtained from the combination of two separated positive and negative branches based on multiplier stages that produce two voltages: $+2.5$\,kV and $-2.5$\,kV. The HVPS design includes a specific block with the capability of reversing the voltage polarity from $-5$\,kV to $+5$\,kV. The core of this block is based on four high-voltage optocouplers in a specific configuration that allows the controlled electric charging and discharging of the etalon.

For safety reasons, the HVPS is equipped with specific circuitry to protect the etalon. The protections are implemented through a voltage limiter and a set of filters that limit the high voltage variation applied to the etalon, maintain the voltage set point for a given wavelength,  and protect the etalon if a power shut down occurs. 

To achieve the fastest tuning performance of the etalon, the HVPS also provides an accurate voltage variation slope of $300$\,V\,s$^{-1}$, sufficiently smaller than the maximum allowed that avoids potential stresses in the etalon. 

The HVPS provides the instrument with three HK parameters, one for each high-voltage output and the readout of one temperature sensor. The sensor is placed near the most thermally critical components: the optocouplers.

\subsection{The Tip/Tilt Controller}\label{sec:TTC}

The Tip/Tilt Controller (TTC) board generates the analogue signals that drive the piezoelectric actuators of the HRT tip-tilt mirror. It consists of three main blocks: the $+55$\,V voltage regulator, the $+5$\,V voltage regulator and the piezo driver amplifier. The $+55$\,V voltage regulator provides a very accurate output in a Quasi LDO (Low Dropout) topology from a 60\,V input generated by the PCM board. To achieve the stringent requirements, this block design includes a filter, an accurate reference voltage, an error amplifier, a feedback divider, and a loop compensation. 

The $+5$\,V voltage regulator is based on the well-known, space-qualified RHFL4913 circuit.  The piezo driver amplifier consists of two serial Digital-to-Analogue Converter (DAC) devices that provide two analogue signals to be amplified by two transconductance amplifiers in closed loop configuration. All the regulator output (voltages and currents) are read out by the AMHD to check the integrity of these signals. During the data acquisition phase, the TTC receives digital data from the DPU, which performs the image stabilisation function (two-dimensional correlation of the live and reference images). The data are then converted to analogue signals with a maximum amplitude of $+55$\,V. 

\subsection{The Electric Distribution System}\label{sec:EDS}

The Electrical Distribution System (EDS) consists of a rigid printed circuit board that allows the interconnection of the previously described E-Unit subsystems. It works as a motherboard, distributing the interface signals (data, control, power and HK) between the boards inside the E-Unit. It also contains the ground star point of \sophi\ that joins the secondary grounds corresponding to the different supplies and connects the ground star point to the structure ground.

\subsection{The E-Unit housing}\label{sec:housing}

The E-Unit housing consists of an assembly of six modules made of aluminium alloy 7075 that host the \sophi\ electronics subsystems. It is painted with PUK low-outgassing, black polyurethane, thermal control coating (PUK black conductive paint), except for the interface areas, which remain coated with Alodine\textsuperscript{\textregistered}.

The chosen modular design has allowed an easy integration by just stacking the modules and screwing them together. The layout of the modules in the E-Unit housing follows thermal and mechanical guidelines and improves its electronic functionality. This distribution allows keeping the E-Unit centre of mass centred and near the mechanical interface and enhancing its thermal behaviour; in addition, this distribution fits the optimal solution to interconnect the different E-Unit electronic subsystems (power and signals).

The E-Unit module distribution is displayed in Fig.~\ref{Fig. FM-E_UNIT}. The hottest and heaviest modules (DPU, PCMR and PCMM) have been placed at the bottom part, as they need to enhance the heat conduction to the interface. The power and data interface connectors with the spacecraft have been placed on one lateral side of the E-Unit. A second side of the unit is used for the connection to the O-Unit elements and a third side is used to place the EDS board.

\subsection{The \sophi\ harness}\label{sec:harness}

The \sophi\ harness consists of 10 bundles of different cable types that connect the various subsystems placed in the two instrument units. 
Two of these harnesses are high voltage cables that connect the HVPS with the etalon. All harnesses have been designed and manufactured using ESA space qualified parts and procedures. Regarding connectors, non-magnetic micro-D connectors with non-magnetic backshells with stress relief have been used, except for the two high voltage cables that are equipped with high voltage connectors.

In order to fulfill the strict Solar Orbiter EMC requirements, special attention has been paid to some design parameters such as cable twisting and cable shielding. Each harness is protected by an external overshielding made of silver-plated copper and an additional tape of aluminised Kapton.

\subsection{Thermal analysis of the E-Unit}

The thermal design of the E-Unit has been performed by setting up the E-Unit TMM together with the corresponding E-Unit GMM. The software tool used for the analysis is ESATAN-TMS. The E-Unit GMM and TMM contain a representation of the main parts of the unit from the thermal point of view, which are the E-Unit structure, divided in six stacked frames, the PCBs with the main dissipating components represented and the EDS. For these dissipating components both cases and junctions nodes have been modelled.

\begin{figure}
    \centering
    \includegraphics[width=0.75\columnwidth]{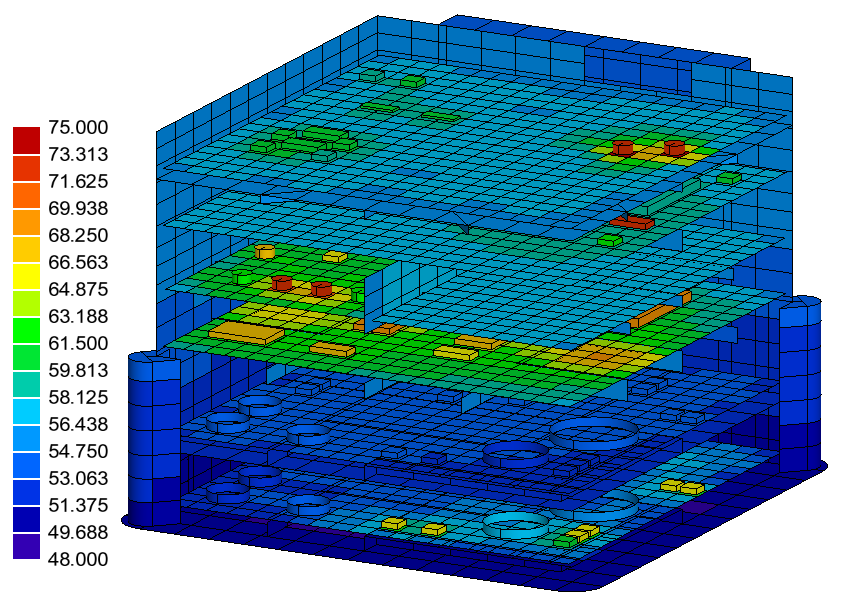}
    \caption{E-Unit thermal analysis results corresponding to hot operational conditions, data acquisition mode, main Power Converter Module active. The colour bar shows temperatures given in $^\circ$C.}
    \label{Fig. E-Unit_TMM}
\end{figure}

For each operational mode, two different cases have been analysed. The first corresponds to the nominal case, the E-Unit working with the PCMM, while the second refers to the redundant case, the E-Unit working with the PCMR. Temperatures and heat fluxes have been calculated for the different load cases under steady-state conditions. The E-Unit thermal model has been iterated with the design and it was correlated with the Structural Thermal Model (STM) thermal test  results. Figure~\ref{Fig. E-Unit_TMM} shows the temperatures corresponding to hot operational conditions.

\section{Instrument characterisation and calibration}\label{calibration}

\subsection{On-ground optical verification and ground calibration}

The optical verification of the instrument and the ground calibration are strongly related and were done in different stages of the instrument assembly. 
All subsystems were characterised individually first before integration into the instrument. Especially the FG subsystem was spectrally calibrated before integration into the O-Unit. 
Also, the PMPs were characterised on subsystem level. 
During the alignment and the assembly of the instrument a number of optical tests were run in order to check the proper performance of the system. Interferometric testing of the full light path in distinct parts of the field-of-view was used for optical functional characterisation. 
Once the FPA was mounted, an interferometric test was no longer possible. The instrument was then stimulated by two counter-telescopes, or stimulus telescopes, which were adapted in aperture and FOV to the telescope under test, that is the FDT and the HRT channels.  

\begin{figure}[]
  \centering
  \includegraphics[width=\columnwidth]{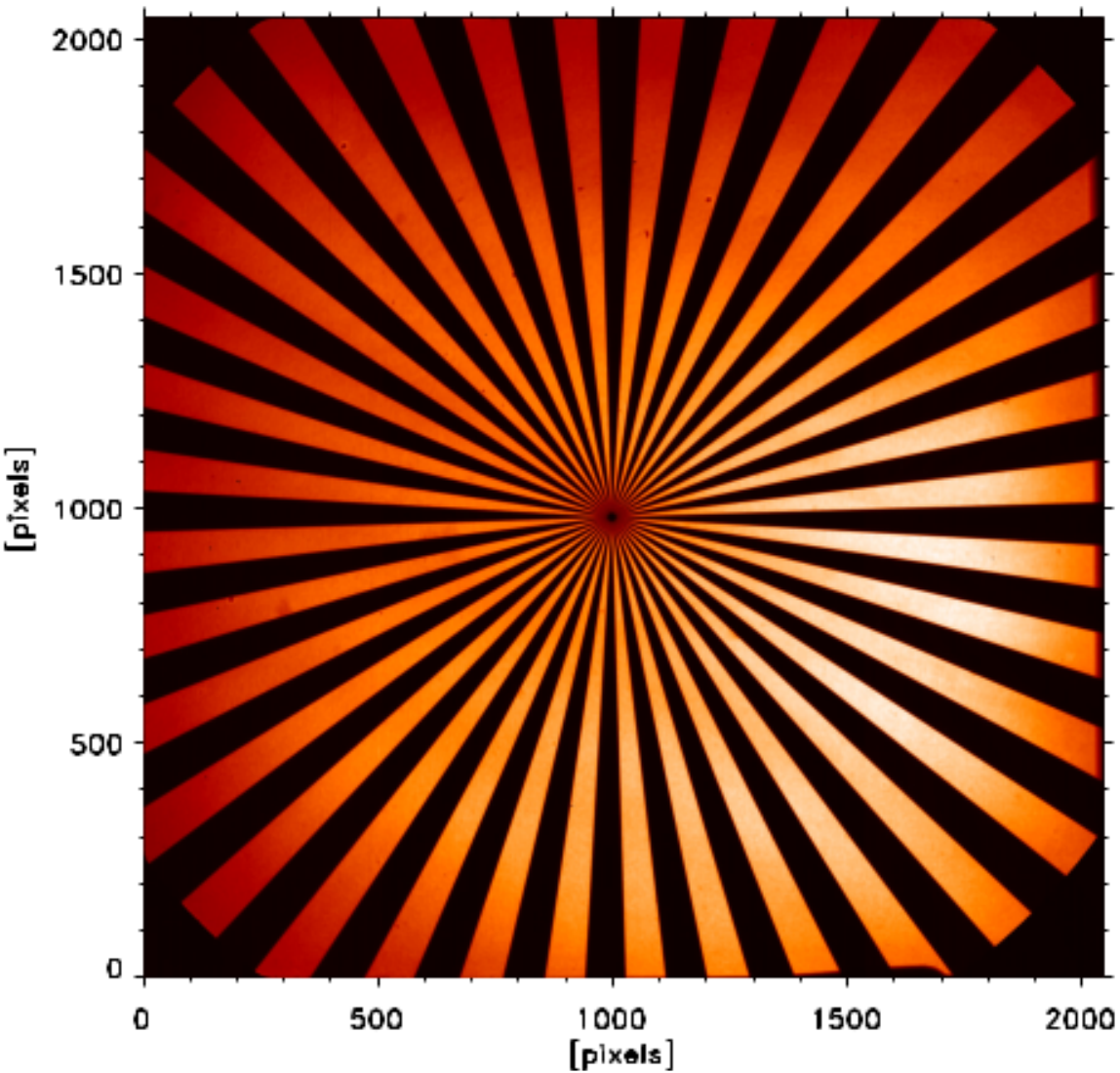}
  \caption[]{HRT image of a Siemens star target obtained after all ground testing and qualification campaigns had been completed. \label{F:siemens} %
  }
\end{figure}

These stimulus telescopes allow projecting different optical test targets to infinity.  
Targets included: A Siemens star (for estimating the modulation transfer functions,  MTFs), grids (for image geometry), scale (plate scale), pinhole (ghosts and false-light), pinhole arrays, hole array, and polka dot array (electronic ghosting). In Fig.~\ref{F:siemens} an image of a Siemens star target mounted in the HRT stimulus telescope, obtained after vibration and thermal vacuum tests is depicted. The derived MTF is in agreement with the diffraction limit of a 14-cm telescope at 617\,nm. The FOV is limited in the extreme corners by an optical baffle inside the FG, that is by the mechanical size of the etalon (see Section~\ref{telescope_implications}). This limitation, in combination with all alignment errors, amounts to 3.4\,\%\ of the detector area.

\begin{figure}[]
  \centering
  \includegraphics[width=\columnwidth]{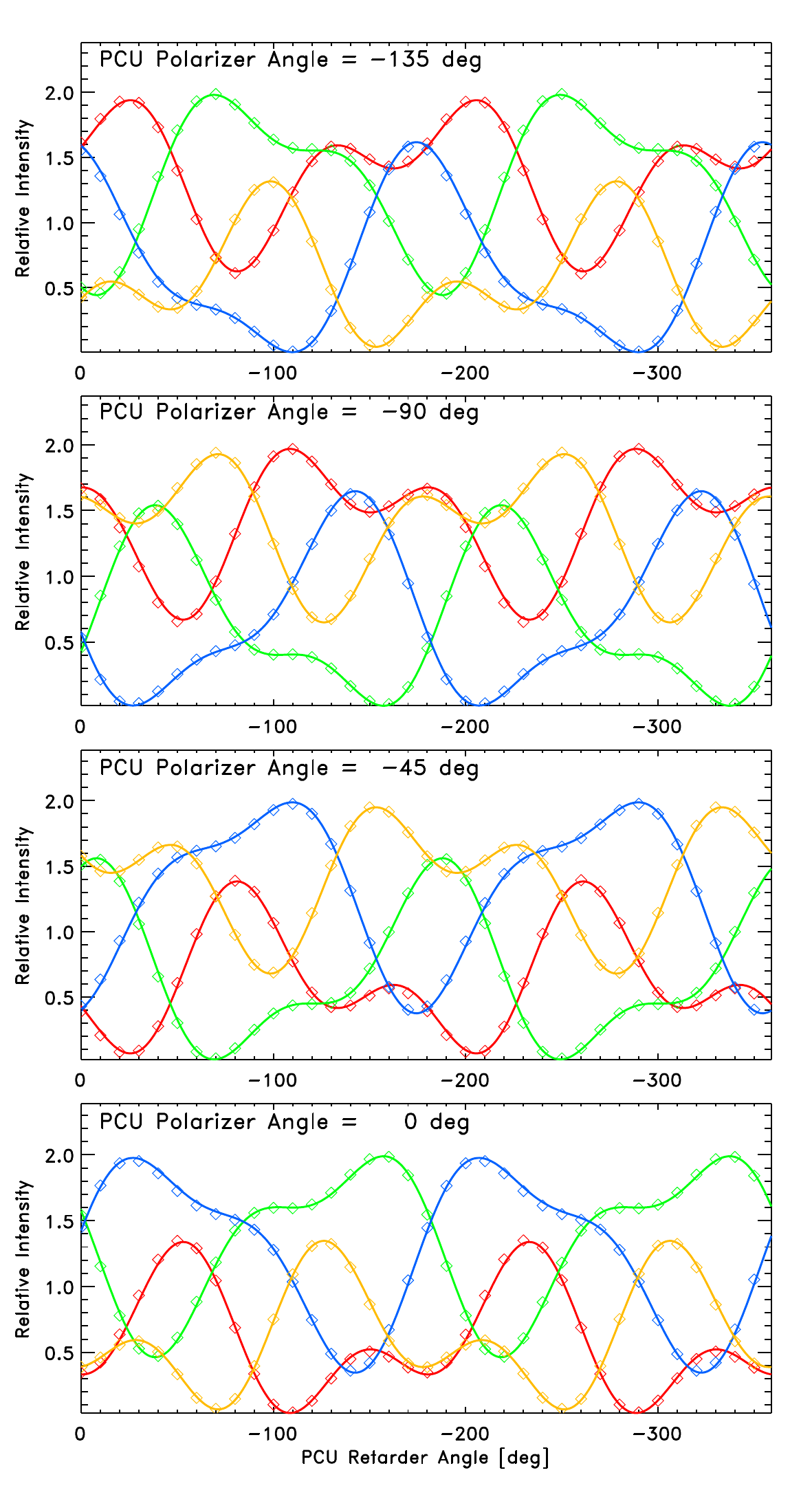}
  \caption[]{Polarimetric calibration of the FDT for $45\,^\circ$C liquid crystal temperature. The diamonds denote mean intensities of the images taken at the 4 polarimetric states of the \sophi\ modulation scheme (see Tab.~\ref{T:lcvr_retardances}) at each input polarimetric state; solid lines represent the fitted model. \label{F:polcal_modelFDT40} %
  }
\end{figure}

Polarimetric calibration of the full instrument was done in ambient conditions only, but for different set temperatures of the liquid crystals following the thorough calibration of the PMPs on subsystem level \citep[][]{alvarez18,silvalopez17}. For this calibration we used the Polarimetric Calibration Unit (PCU), which was  developed for the HMI instrument on board SDO \citep[described in][]{schou12a}. This unit contains a linear polariser and a quarter-wave retarder, which can be independently moved into the beam and which can be freely rotated, such that many different polarimetric input states can be sent to the instrument under test. We used, typically, $4\times 36$ input states and fitted the observed data to a model, which yields the response matrix of the instrument and the alignment and retardation errors of the test equipment. A typical example is shown in Fig.~\ref{F:polcal_modelFDT40}. The corresponding FOV averaged polarimetric efficiencies are $\epsilon =[0.9917,0.5697,0.5666,0.5745]$, which is close to the theoretical optimum of 1.0 for Stokes $I$ and 0.5774 for Stokes $Q$, $U$ and $V$.   

\begin{figure}[]
  \centering
  \includegraphics[width=\columnwidth]{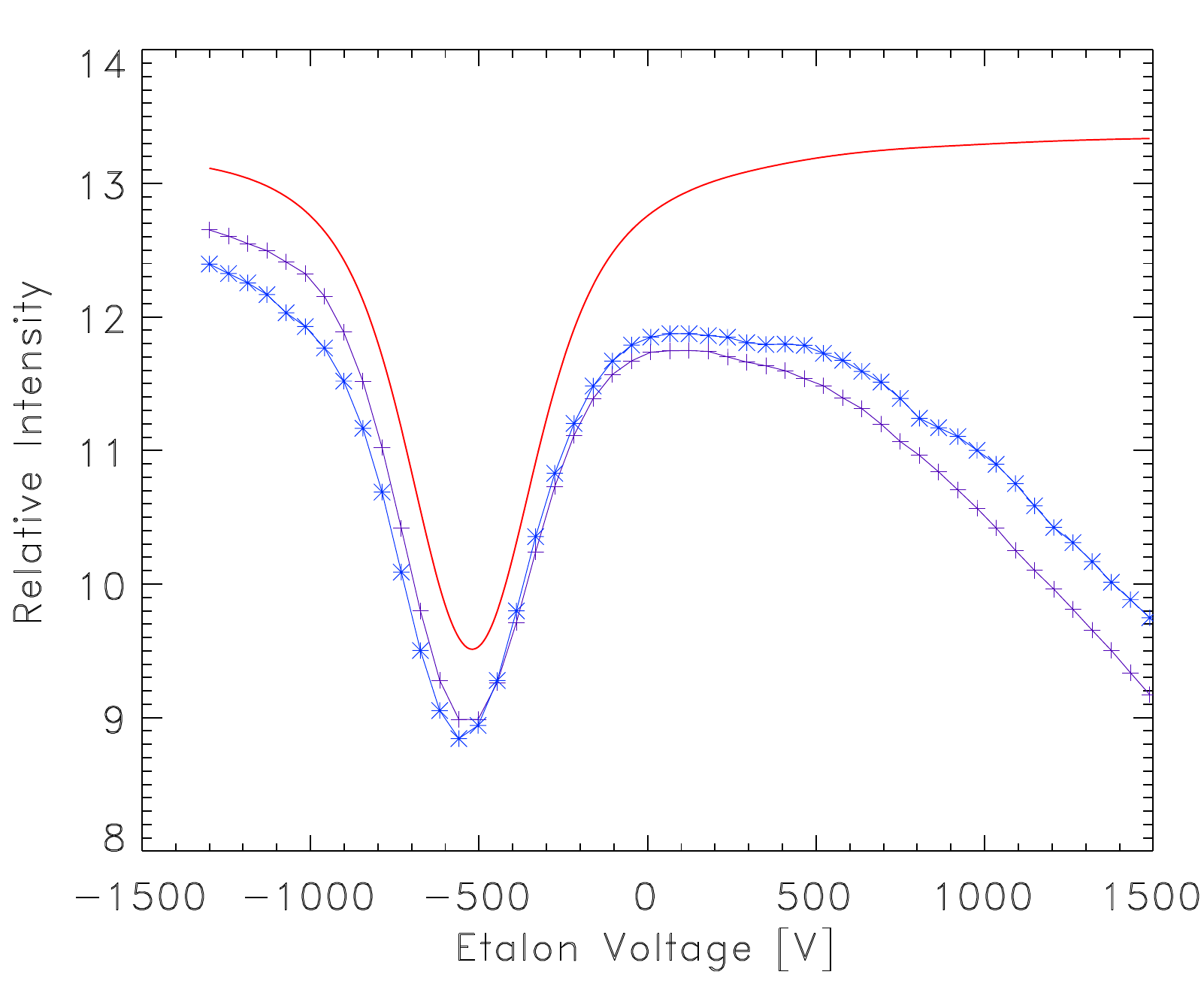}
  \caption[]{Spectral scans across the \ion{Fe}{i}\,6173\,\AA\ spectral line (crosses and asterisks) obtained from different sectors of the science detector during HRT ground calibration. The solid line denotes a convolution of the FTS spectrum with an Airy function of 106\,m\AA\ FWHM. \label{F:HRT_spectralscan} %
  }
\end{figure}

\begin{figure}[]
  \centering
  \includegraphics[width=\columnwidth]{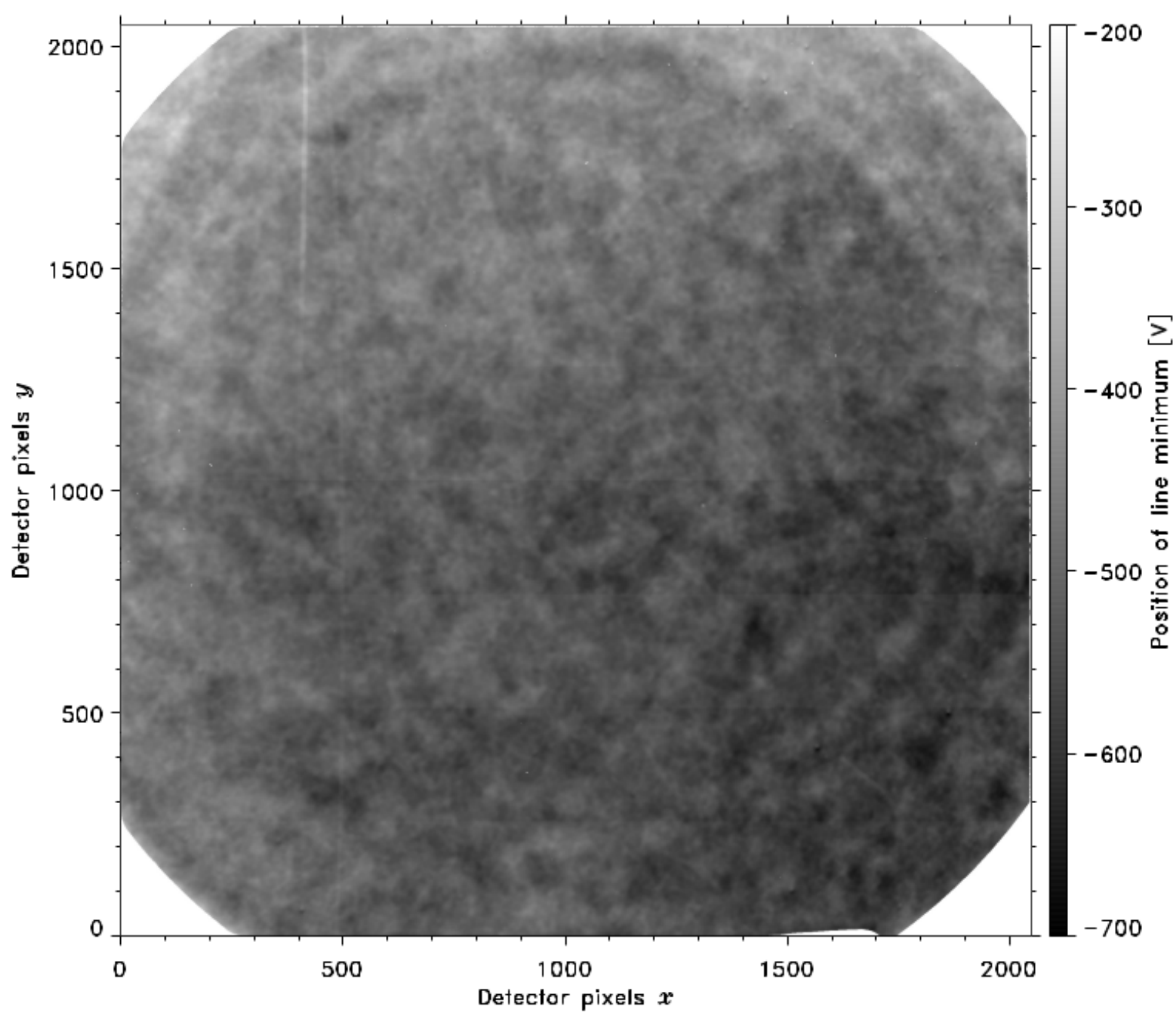}
  \caption[]{Etalon cavity map obtained from Gaussian fits to the centre of the \ion{Fe}{i}\,6173\,\AA\ spectral line during HRT ground calibration. The large-scale trend and the slight shift of the two scans shown in Fig.~\ref{F:HRT_spectralscan} arise from the solar rotation pattern imprinted in the incoming light, since the HRT FOV covers almost the entire solar disc at 1\,AU distance, even if the feed optics does not provide a focused solar image. \label{F:HRT_cavitymap} %
  }
\end{figure}

Spectral calibration of the instrument requires tuning of the etalon, which is not possible in ambient conditions due to electric arcing in the high voltage chain. 
Therefore the spectral calibration was exclusively done in vacuum conditions during the thermal-vacuum acceptance and thermal balance testing of the instrument at MPS. To this end, the instrument was illuminated with sunlight, provided by a 53-cm coelostat at MPS premises. Figures~\ref{F:HRT_spectralscan} and~\ref{F:HRT_cavitymap} show spatially averaged \ion{Fe}{i}\,6173\,\AA\ line profiles and a cavity thickness map of the etalon obtained by applying Gaussian fits to the calibration data. For spectrally flat illumination (over spectral scales of 1\,nm) we used a high power LED source. With this, the transmission band-pass over the FOV, as well as the prefilter curve could be measured.

\subsection{In-flight calibration}

\sophi , in its standard operation mode, provides maps of physical quantities obtained by means of a scientific on-board data analysis. It is, therefore, required that the obtained data undergo an autonomous on-board calibration procedure that is schematically displayed in Figure~\ref{F:pre_processing_pipeline}. The individual modules of this processing pipeline and the modules required to obtain the on-board calibration data are briefly  described in Section~\ref{S:dataprocessing} and the details of the software implementation concept are given in \citep{albert18}. 

More details on the ground and on-board calibration procedures and the ultimate instrument characteristics will be presented in a dedicated publication after in-flight instrument commissioning.


\section{Scientific operations}\label{sc_operations}

As Solar Orbiter is a deep space mission in an orbit with highly variable distance from Earth, \sophi\ operations have to counterbalance the difficulties this produces. In addition, \sophi\ must perform coordinated observations with the other instruments aboard Solar Orbiter as well as with other observatories in deep space \citep[e.g. Parker Solar Probe; see][]{Velli2019a}, in Near-Earth Orbit (NEO) and on ground \citep[see][for the overall mission Science Activity Plan]{Zouganelis2019a}. The major obstacles for the operations are the limited data rate, the often high latency of the data return and the lack of knowledge of the solar scene when the mission is far from Earth \citep[][]{Sanchez2019}. Therefore, \sophi\ operations have to be planned and commanded a long time in advance and almost all operations have to be carried out with a high degree of on-board autonomy.

\sophi\  has available a telemetry volume of 20\,kbits/s within only $3\times 10$ days in each science orbit, hence the total scientific data return will amount to approximately 6.5\,GBytes per orbit. On the other hand, the \sophi\ detector runs at a frame rate of 11 frames per second with a size of 6.3\,MBytes of each image (12\,bits digital depth). \sophi, therefore, has to adopt elaborate on-board data processing and compression procedures.

\subsection{Software operations concept}

\begin{figure*}[ht]
  \centering
  \includegraphics[width=1.0\textwidth]{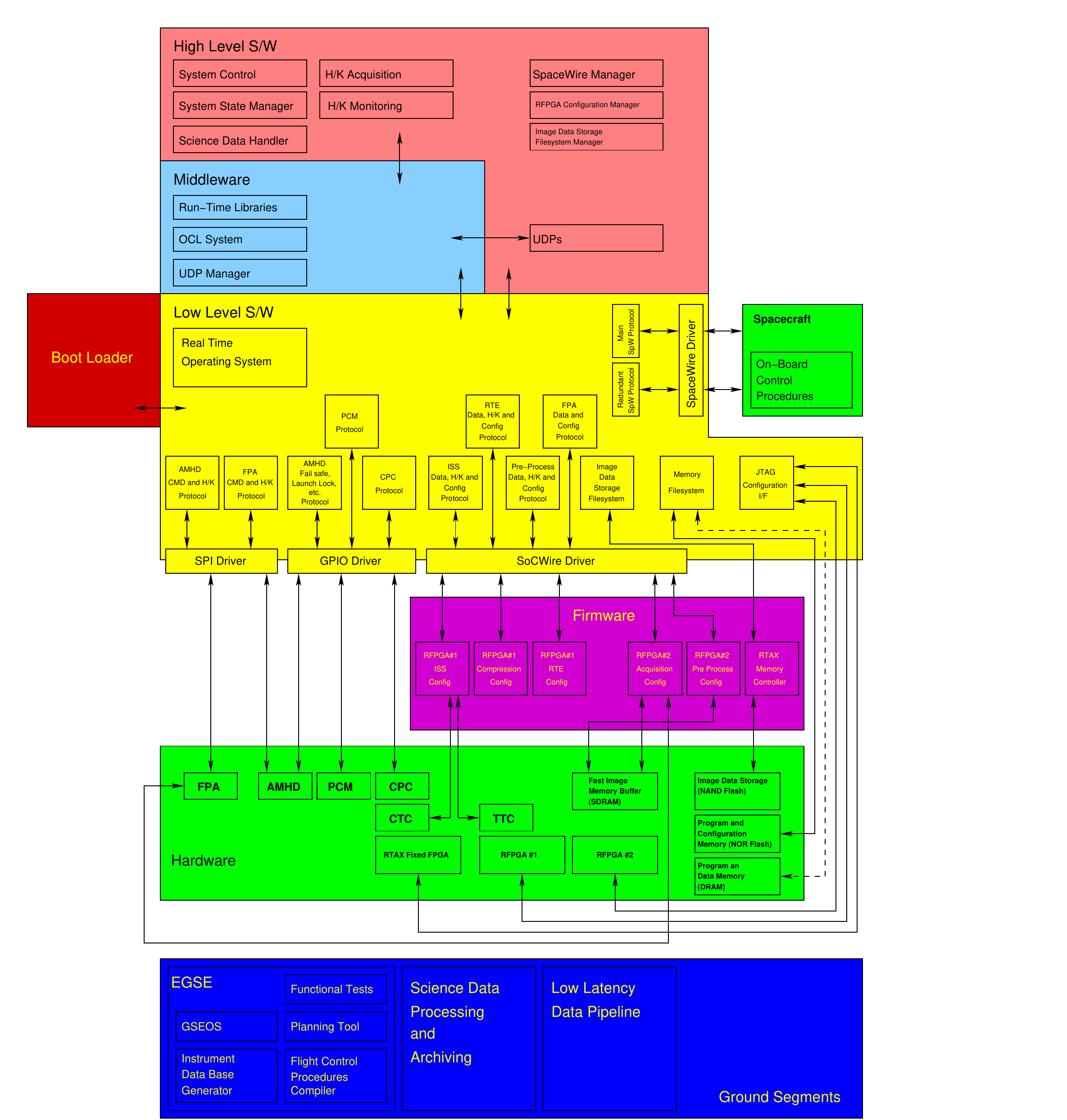}
  \caption[]{\label{F:sw_block_diagram} %
  \sophi\ software block diagram (including on-board software and ground segment) and interfaces to the instrument hardware and the Solar Orbiter spacecraft.}
\end{figure*}

The complexity of the \sophi\ instrument and the challenging operational conditions require a software concept that is able to cope with the autonomy and timing requirements and the expected precision of the produced science data. The \sophi\ software block diagram is displayed in Figure~\ref{F:sw_block_diagram}. The on-board software is composed of mainly three layers: the low-level code, the middleware and the high-level routines. In addition, \sophi\ utilises a comprehensive firmware concept, which consists of several FPGA configurations that are used to perform the on-board data processing as well as a fixed configuration of the system controller (see Section~\ref{e-unit}).    

The low-level software comprises a boot loader and an Real-Time Operating System for Multiprocessor Systems (RTEMS) as well as the hardware drivers and protocols.The middleware  consists of C runtime libraries and the On-board Command Language \citep[OCL, see][]{wittrock03}, which is a dedicated programming language for space instruments. The high-level software is is written in OCL. It comprises, besides the system control, instrument HK and the science data control, also a set of User Defined Programmes (UDPs) which contain the science and calibration operation procedures. 

\sophi\ is supported by a comprehensive software ground segment partly composed of a data processing and archiving system as well as a pipeline for the autonomous processing of low-latency data products. The instrument control is carried out with  GSEOS~V\footnote{The GSEOS Ground Support Equipment Operating System is a commercial software system provided by the Institut für Datentechnik und Kommunikationsnetze at the Technical University of Braunschweig.}, but also with fully autonomous software tools to set up the instrument data base (i.e. the parameter and command interfaces to the mission control) and the flight control procedures. In order to perform a reliable and traceable operations planning, the \sophi\ team developed a dedicated planning tool that is attached to a data base storing  all parameters and operational constraints of each flight control procedure. With this tool the entire operation planning can be displayed graphically. The tool also performs automatic checks of compliance to instrument and mission constraints (e.g., telemetry limits, power restrictions, spacecraft attitude stability, etc.).

Because of the complexity of the instrument and the limited commanding volume available (only 150 telecommands per day) the observation and on-board processing procedures have to run as standardised routines with a certain set of input parameters launched by commanding from ground. These UDPs can be launched by a single telecommand and will carry out, e.g., the acquisition of a single data set. Input parameters to this specific UDP will be the size of the FOV, that is start and end addresses of the detector read-out, or the number of frames to be accumulated at each wavelength position and polarisation state. The inputs to the UDPs can be provided either by ground commanding or by reading parameters from the configuration memory, which is a non-volatile file system located in the NOR-flash (see Section~\ref{e-unit}).
Typical configuration parameters are the states of the polarimetric modulation scheme and the corresponding voltages to be applied to the LCVRs. Parts of the data stored in the configuration memory will be updated in flight after each instrument re-calibration.

\subsubsection{\sophi\ system states}

\sophi\ operations are organised in 12 system states:
\begin{enumerate}
\item {\sf off:} \sophi\ is entirely off.
\item {\sf boot:} \sophi\ is booting.
\item {\sf safe:} \sophi\ is safe and can be switched off any time, all optical subsystems, HVPS and RFPGAs are off. 
\item {\sf idle:} same as {\sf safe}, but the UDP manager is active and the housekeeping telemetry rate is increased. 
\item {\sf observational idle:} PMPs and FG are thermally stabilised; all other optical subsystems, HVPS and RFPGAs  are off.
\item {\sf observation:} \sophi\ is ready to acquire science data.
\item {\sf process\_sci:} \sophi\ performs on-board data processing; all optical subsystems are off.
\item {\sf process\_heater:} \sophi\ performs on-board data processing; PMPs and FG are thermally stabilised; all optical subsystems are off.
\item {\sf process\_cal:} \sophi\ carries out on-board calibration data processing; all subsystems are on.
\item {\sf process\_anneal:} \sophi\ performs on-board data processing and detector annealing; all other optical subsystems are off.
\item {\sf annealing:} \sophi\ performs detector annealing; all other optical subsystems, HVPS and the RPFGAs are off.
\item {\sf debug:} all systems can be commanded individually; low level commands are allowed.
\end{enumerate}

This organisation allows to, first,  keep \sophi\ always in a defined state, so that power consumption is known and, secondly, avoid unintended or hazardous actions by erroneous commanding. If any hazardous instrument anomaly is detected, \sophi\  automatically enters the {\sf safe} system sate and sends out a flag that asks for it to be switched off.  

Scientific operations are carried out in a rather flexible way with a list of operation modes (see Table~\ref{tab:sc_ops_modes}) differing mainly by the number of physical parameters provided for telemetry downlink. Within each of these modes the cadence and image sizes are tunable within rather wide ranges, depending on the scientific goal and the available telemetry. 

\begin{table*}
\caption{\sophi\ science operating modes. The digital depth (total number of bits/pixel) is given prior to compression; the compression ratio is based on estimates obtained from simulations with test data.\label{tab:sc_ops_modes}}
\begin{tabular}{lllcrrrr}
\hline
Mode & Name & Description & Telescope & Image Size & Total size   & Compr. & Cadence \\
     &      &             &           & [pixel]    & [bits/pixel] & ratio      & [min$^{-1}$] \\  
\hline
\hline
0    & PHI\_nominal\_FDT   & $I_{\rm c}, v_{\rm LOS}, B, \gamma, \phi$ & FDT & $512 - 2048$ & 46 & 2 & $1-1/60$ \\ 
     & PHI\_nominal\_HRT   & $I_{\rm c}, v_{\rm LOS}, B, \gamma, \phi$ & HRT & $512 - 2048$ & 46 & 2 & $1-1/60$ \\ 
\hline
1    & PHI\_vector\_FDT    &                 $B, \gamma, \phi$ & FDT & $512 - 2048$ & 26 & 2 & $1-1/60$ \\ 
     & PHI\_vector\_HRT    &                 $B, \gamma, \phi$ & HRT & $512 - 2048$ & 26 & 2 & $1-1/60$ \\ 
\hline
2    & PHI\_magnetograph\_FDT    & $I_{\rm c}, v_{\rm LOS}, B_{\rm LOS}$ & FDT & $512 - 2048$ & 30 & 2 & $1-1/60$ \\ 
     & PHI\_magnetograph\_HRT    & $I_{\rm c}, v_{\rm LOS}, B_{\rm LOS}$ & HRT & $512 - 2048$ & 30 & 2 & $1-1/60$ \\ 
\hline
3    & PHI\_global\_helioseismology & $I_{\rm c}, v_{\rm LOS}$ & FDT & $128 - 512$ & 16 & 2 & 1 \\
     & PHI\_local\_helioseismology  & $I_{\rm c}, v_{\rm LOS}$ & HRT & $128 - 512$ & 16 & 2 & 1 \\
\hline
4    & PHI\_synoptic                & $I_{\rm c}, v_{\rm LOS}, B, \gamma, \phi$ & FDT & 1024 & 40 & 2 & $1/240 - 1/1440$ \\
\hline
5    & PHI\_burst$^\ast$            & $I_{\rm c}$                               & HRT & 2048 & 10 & 2 & 60 \\
\hline
6    & PHI\_raw\_data\_FDT & 24 images & FDT & $512 - 2048$ & 384 & 5 & $1 - 1/60$ \\
     & PHI\_raw\_data\_HRT & 24 images & HRT & $512 - 2048$ & 384 & 5 & $1 - 1/60$ \\
\hline
     & low latency data    & $I_{\rm c}, B_{\rm LOS}$ & FDT & 1024 & 16 & 2 & 1 per day \\
\hline     
\multicolumn{8}{l}{} \\
\multicolumn{8}{l}{$^\ast$ PHI\_burst will be operated for a few minutes only and will be interlaced with magnetograms} \\
\end{tabular}
\end{table*}

\subsection{Data acquisition}

\sophi\ aims to acquire high signal-to-noise (SNR) image data at high cadence while minimising spurious polarisation signals introduced by solar evolution and residual spacecraft jitter. This can be achieved by employing a semi-fast polarimetric modulation scheme, based on 4 modulation states, combined with a slow wavelength tuning mode, which samples the \ion{Fe}{i}\,6173\,\AA\ photospheric absorption line at 6 wavelength positions. Of these, five lie within the spectral line, at $[-140, -70, 0, 70, 140]$\,m\AA\ around the line centre, while the sixth is a continuum point located either at $-300$\,m\AA\ or at $+300$\,m\AA , depending on orbital velocity relative to the solar surface, which is in the range of $\pm 23.6$\,km\,s$^{-1}$ and results in a Doppler shift of $\pm 486.9$\,m\AA . The entire required tuning range is, therefore, $\pm 626.9$\,m\AA , which is less than one half of the FSR of the etalon. \sophi\ uses the same modulation scheme for FDT and HRT and the corresponding LCVR retardances as well as the LCVR switching times, $t_P$, are given in Table~\ref{T:lcvr_retardances}. By $t_P$ we understand the longest switching time among those for the two LCVRs of each PSA. These times are measured with a null ellipsometer for each single cell. Specifically, $t_P$ is the interval needed to go from 10\,\% to 90\,\% of the maximum intensity level.

\begin{table}

\caption[]{Nominal polarimetric modulation scheme for \sophi . The LCVR retardances are given in degrees. The switching times assume a cyclic sampling from left to right.
\label{T:lcvr_retardances}}

\vspace*{0mm}

\centerline{
\begin{tabular}{l@{\qquad}r@{\qquad}r@{\qquad}r@{\qquad}r}
\hline
\hline
pol. state           & 0 &  1  &  2    & 3    \\
\hline
\hline 
LCVR 1               & 315.00     & 315.00      & 225.00        & 225.00 \\
\hline
LCVR 2               & 234.74     & 125.26      &  54.74        & 305.26 \\ 
\hline
\hline
$t_P$ [ms] (FDT)          & 57.7  &     19.0  &   83.2    &   95.4 \\
$t_P$ [ms] (HRT)          & 56.2  &     18.3  &   82.7    &   95.1 \\
\hline
\end{tabular}
}
\end{table}

The total optical efficiency of the \sophi\ instrument is estimated to be 7.5\% for the HRT and 7.8\% for the FDT 
and the product of quantum efficiency and fill factor of the detector at science wavelength is $Q\cdot f=59$\,\%. To stay securely within the linear range of the detector, for each individual exposure the detector will be filled to 65\% of its full-well capacity of $10^5$ electrons. Each raw image thus comprises a SNR of SNR$_{\rm single}=255$ and the exposure times for an individual frame obtained at disc centre and in the continuum of 6173\,\AA\ amounts to approximately 32\,ms for the FDT and 24\,ms for the HRT. As \sophi\ needs to provide data with a polarimetric sensitivity of $10^{-3}$, corresponding to an SNR$\, = 10^3$ in the Stokes $I$ continuum, and as the instrument provides a polarimetric efficieny, $\epsilon$, greater than 0.5, at each polarimetric state and spectral position,   
\begin{equation}\label{E:accumulations}
N_{\rm acc} = \frac{1}{4}\left(\frac{\rm SNR}{\bar\epsilon\, \rm SNR_{\rm single}}\right)^2 \simeq 16
\end{equation}
frames have to be accumulated, where $\bar\epsilon$ is the average of the $Q$, $U$, and $V$ polarimetric efficiencies \citep{deltoroiniesta00}. 

In order to reduce polarimetric artefacts produced by spacecraft jitter, polarimetric modulation will be, usually, carried out in a fast modulation scheme, that means by switching the polarisation state after each single exposure. In addition to the exposure time and the waiting time required to carry out the LCVR switches, also the wavelength tuning time, which is restricted to a speed of 300\,V\,s$^{-1}$ at a tuning constant of $d\lambda/dV = 0.3511$\,m\AA\,V$^{-1}$ in order not to jeopardise the LiNbO$_3$ etalon, has to be considered. Consequently, a compromise between polarimetric noise and total acquisition cycle length has to be found. \sophi\ can alter the modulation mode, that is the number of exposures before switching to another modulation state, in flight. Theoretically calculated total cycle times are given in Table~\ref{T:cycle_times}. $N_P$ denotes the number of polarisation cycles while accumulating 16 frames. It shows clearly that the optimum solution of $N_P = 16$ will result in a cycle time of $t_{\rm cycle}=76.43\,s$, which is longer than the required maximum cadence of 60\,s. For certain science goals, which require a high polarimetric sensitivity at lower spatial and temporal resolution, this mode is, however, well suited.  

\begin{table}

\caption[]{Theoretically calculated total cycle times, $t_{\rm cycle}$, for acquiring an entire wavelength scan, listed as a function of the number of polarisation cycles, $N_P$. According to Eq.~\ref{E:accumulations} 16 frames per wavelength position and polarisation state have to be accumulated to get the required SNR.}\label{T:cycle_times}
\centerline{
\begin{tabular}{rrrrrr}
\hline
\hline
$N_P$                 & 1       &  2    &  4    & 8     & 16  \\
\hline
$t_{\rm cycle}$  [s] (FDT) & 45.51   & 47.81 & 52.41 & 61.61 & 80.01 \\ 
$t_{\rm cycle}$  [s] (HRT) & 45.50   & 47.78 & 52.35 & 61.48 & 79.76 \\ 
\hline
\end{tabular}
}
\end{table}

Data acquisition requires synchronisation of the FPA, the PMPs and the FG. This is controlled by absolute timing using software. After setting up all subsystems, the image data acquisition is launched by triggering the acquisition firmware, which has been loaded to reconfigurable FPGA\#2 (see Fig.~\ref{Fig. DPU}) to read out a number, $N$, of images from the FPA. The speed of the FPA is constant at approximately 11 frames per second and the corresponding images are accumulated to a certain address in the fast image data memory connected to FPGA\#2. After these $N$ images are recorded, the read out is paused and the voltages applied to the PMPs and/or the FG are changed. As the LCVR switching times and the etalon tuning times are known, the image read-out continues after a predefined delay and the next set of $N$ frames is accumulated to another address in the fast image data memory. If $N_P$ is greater than 1, the memory addresses to which the frames are accumulated are cycled $N_P$ times. 

Science data acquisition is performed only if the instrument is in the {\sf observation} system state. Upon entering into this mode, the firmware to control the ISS is loaded automatically into the reconfigurable FPGA\#1. If the HRT is used, then it runs in closed loop, while the FDT runs in open loop and the resulting image shifts are sent out via inter-instrument communication. Open-loop operation provides true image shifts (mean, maximum and rms values obtained within a configurable period) caused by spacecraft jitter, while closed-loop operations  provide only residuals after internal stabilisation. In order to provide useful pointing accuracy information to the other instruments on board Solar Orbiter, \sophi\  also provides a flag if the tip/tilt mirror has reached its maximum range during this period.  

\subsection{On-board data processing}\label{S:dataprocessing}

As Solar Orbiter is a deep space mission, its remote-sensing instruments, which provide high resolution data at high cadence, are always telemetry limited. In order to achieve the science goals under these restrictions, the downlink volume required from \sophi\ is reduced by processing the acquired data on board. The optimum way to reduce the telemetry volume is to perform an on-board scientific data analysis, which implies, in addition, an accurate calibration of instrumental and systematic effects. Because space-grade general purpose processors such as the employed GR712RC have limited processing performance, on-board data processing makes use of two in-flight reconfigurable FPGAs. While the inversion of the RTE and the image compression are implemented as dedicated FPGA designs (see section \ref{sec:RTE}), other parts of the scientific data processing have to be kept flexible to adapt to changes during the mission. This includes the calibration of the acquired images during pre-processing as well as the generation of flat-field data. Therefore, a flexible image processing framework has been implemented. This framework mainly consists of multiple processing modules implemented in a set of FPGA configurations. Furthermore, it includes a software part that is in charge of parameterisation, execution and monitoring of the processing modules. During processing, the set of loaded processing modules can be reconfigured without loss of data inside the buffer memory. The hardware-accelerated on-board processing allows a speedup of about a factor of 20 to 600 compared to a fixed-point implementation of the processing functions on the GR712RC \citep{lange18}.

\subsubsection{Pre-processing}\label{sec:preprocessing}

\begin{figure}[ht]
\centerline{
  \includegraphics[width=0.45\textwidth]{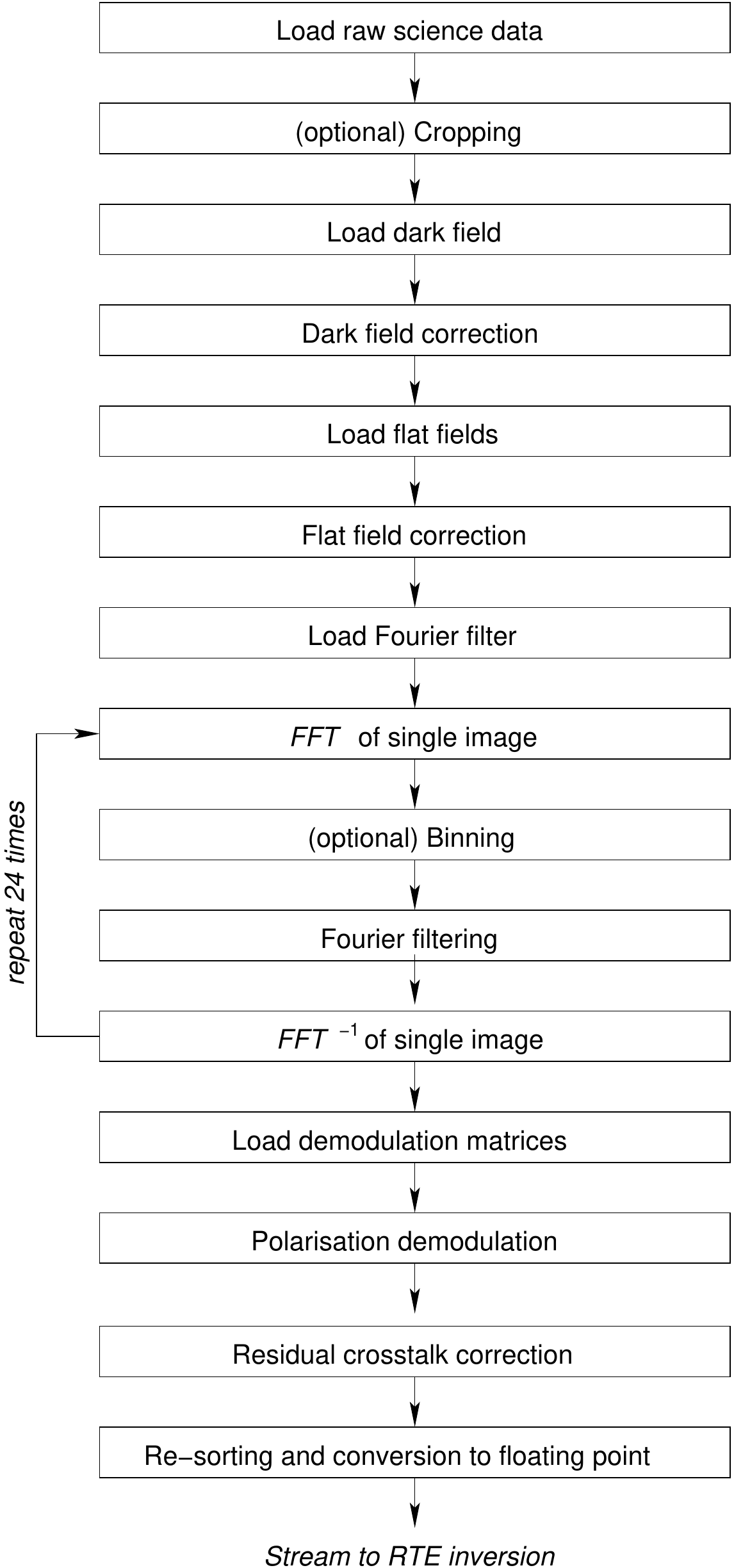}
}
\caption{Pre-processing sequence.\label{F:pre_processing_pipeline}}
\end{figure}

During on-board data pre-processing a data calibration pipeline is applied to the accumulated raw science images. The result of this pipeline is a set of fully calibrated images of the four Stokes parameters at each of the 6 spectral positions. The pre-processing pipeline \citep[see][]{albert18} is based on UDPs, which sequentially call basic mathematical functions implemented in dedicated firmware configurations to be loaded to FPGA\#2, that means on-board data calibration is controlled by software whereas the actual processing is carried out in hardware. The overall processing framework includes several basic functions (i.e. pixel-wise addition, subtraction, multiplication and division of images) as well as median filtering, pixel reordering, averaging, min/max determination and many more. Because many filtering operations can be carried out in the frequency domain, the framework also includes a 2D Fast Fourier Transform. We note that many complex processing operations \citep[e.g. the Hough transform and flat-field calculation based on the technique developed by][]{kuhn91} can be broken down into a sequence of these hardwired functions. As a backup solution and for test purposes all the processing functions are also implemented in software, with particular attention being paid to coding these functions such that their performance is as similar as possible to the corresponding hardwired functions. Nominal pre-processing is performed by loading the science data sets and the required calibration data into the fast image memory buffer (SDRAM) which is attached to FPGA\#2. A complete overview of the pre-processing pipeline is given in Figure~\ref{F:pre_processing_pipeline}. This pipeline is divided into linear parts where all processing steps will be applied to complete images (e.g. flat or dark fielding) and parallel parts where single pixels from more than one image have to be processed (e.g. polarimetric demodulation, re-sorting for RTE streaming). All calibration data loaded by the pre-processing pipeline are assumed to be fully corrected in separate pipelines. Basic checks, such as for correct exposure times, are carried out and warnings or errors are indicated if these checks fail. 
%

In the first pipeline module, the images are optionally clipped to a sub-region of the detector area, which helps to save telemetry. This module will be used only for high-resolution science cases that require a high cadence but do not necessarily require a large FOV. If cropping will be carried out, all subsequently used calibration data (e.g. flat and dark fields) have to be clipped accordingly.   

Dark and flat field correction pipeline blocks are straightforward procedures, however. Within these blocks bad pixels (e.g. bad detector areas and cosmic ray traces) and regions outside the solar disc have to be marked. This procedure contains an elaborate management of over- and underflows, which is done by a system of binary image masks that contains this information. Subsequent modules, such as Fourier filtering, require a detailed treatment of these regions. Therefore, the flat-fielding module also contains an algorithm, based on a $5\times 5$\,pixels median filtering, to interpolate bad detector areas.

As \sophi\ is the first-ever solar imaging spectropolarimeter on a highly elliptic orbit around the Sun, and as the realistic environmental conditions could not be simulated to their whole extent during ground testing, we have to be prepared that the extremes of the thermal environment on Solar Orbiter trajectory might produce variations of the image quality with orbital position. These effects might range from changes of the telescope point spread functions due to the variable temperature gradients on the HREWs to optical fringes produced by multiple reflections caused by thermo-elastic deformations of the instrument optics. Therefore, the data pre-processing pipeline includes several modules that will be able to correct for such potential degradation effects by applying Fourier filtering techniques. In addition, optional data binning for those science cases that can afford low spatial resolution will be carried out in the Fourier domain. All these corrections require a deconvolution, which is carried out by applying a Fourier filter that has to be produced on ground.

\begin{figure}[ht]
\centerline{
  \hspace*{-4mm}\includegraphics[width=\columnwidth]{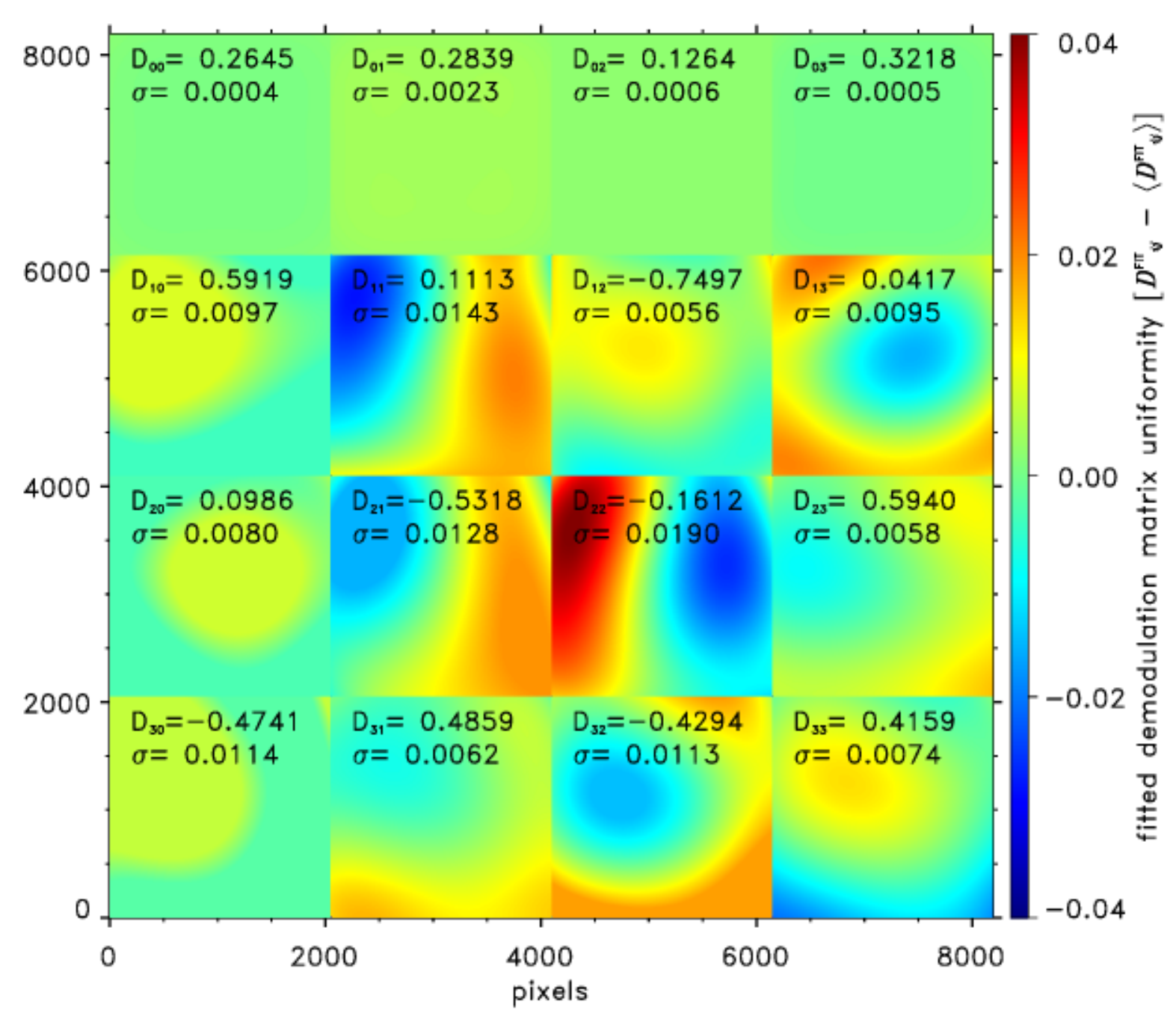}
}
\caption{2D fitted demodulation matrix for the HRT and a PMP temperature set point of $40\,^\circ$C. Each subimage shows the variation of one matrix element across the entire $2048\times 2048$\,pixels area of the HRT FOV. The mean value of each matrix element, $\left< D_{ij}\right>$, is subtracted and annotated (together with the standard deviation, $\sigma_{ij}$, across the FOV within the illuminated area) to each element. \label{F:demod_fit}}
\end{figure}

After the obtained images have been cleaned of instrumental effects, the data sets can be demodulated. The LCVRs induce optical retardance variations across the FOV. This and the limited electronic power for the temperature stabilisation of the modulators require applying different temperature set points along the orbit (we assume in a range between $+40\,^\circ$C and $+70\,^\circ$C). In addition, changes of the demodulation matrices with orbital position have to be considered. Therefore, the pre-processing pipeline contains a demodulation module that applies field and temperature dependent demodulation matrices. These matrices are stored by a set of fit parameters in the configuration memory prior to launch. As the FOV-dependence of the single elements of the demodulation matrices are smooth, these matrices can be easily produced on board by computing the  $4^{\rm th}$ order polynomials from the stored fit parameters and it is, moreover, simply possible to update the fit parameter by ground commanding, if it turns out that, for instance, systematic cross-talk effects will be detected. Figure~\ref{F:demod_fit} shows the FOV variation of the demodulation matrix for the HRT for the PMP temperature of $40\,^\circ$C as calculated from the 15 stored parameters of the $4^{\rm th}$ order fit applied to the corresponding ground calibration data. In order to use the same colour scale for each matrix element, the mean value of each element was subtracted.  


It is well known from comparable instruments that demodulation matrices result in residual crosstalk between the Stokes parameters. Consequently, the \sophi\ pre-processing pipeline will apply an ad-hoc correction \citep{SanchezAlmeida,Schlichenmaier} to the retrieved Stokes maps. Stokes $I\rightarrow Q, U, V$ can be obtained from the mean offsets of the $Q, U, V$ maps at the continuum position where no intrinsic signal is expected. It is assumed that the area fraction of highly Doppler-shifted signals is low, so that they do not significantly influence the obtained offsets. The $V\rightarrow Q, U$ crosstalk will be obtained by linear fits of the measured signals $Q_{\rm measured}$ and $U_{\rm measured}$ to  $V_{\rm measured}$: 
\parbox{7.2cm}{
\begin{eqnarray*}
Q_{\rm measured} & = & Q_{\rm corr} + aV_{\rm measured}, \\
U_{\rm measured} & = & U_{\rm corr} + bV_{\rm measured}, \\
\end{eqnarray*}}\hfill\parbox{1cm}{\begin{equation}\end{equation}}
from which the crosstalk terms $aV_{\rm measured}$ and $bV_{\rm measured}$ will be retrieved and the corrected signals $Q_{\rm corr}$ and $U_{\rm corr}$ can be calculated.

Prior to inverting the thus calibrated  data, the Stokes maps have to be transformed into 32-bit floating point values and re-sorted to individual Stokes profiles for each pixel of the spatial FOV. The inversion of the Stokes profiles carried out after that (see Section~\ref{sec:RTE}) is configured to provide the magnetic field vector only for pixels containing Stokes $Q, U, V$ signals above the noise level which is typically of the order $10^{-3}$ of the continuum intensity. The photon-induced centre-to-limb variation of the noise level can be counterbalanced by increasing the number of accumulations (see Eq.~\ref{E:accumulations}). Since in most of the cases the Stokes profiles will not be available on ground it is intended to set the magnetic field strength (as well as the inclination and the azimuth) to zero in those regions where the magnetic signal is below the noise level and to provide only the Doppler velocity and the continuum intensity. 

\subsubsection{Inversion of the Stokes profiles}\label{sec:RTE}

The last step before the lossy or lossless compression in the data processing pipeline is the inversion of the radiative transfer equation (RTE). As commented on earlier (see e.g. Sects. \ref{sec:products}, \ref{DPU}) this task is carried out on board by means of a specific firmware design working on one of the FPGAs in the DPU \citep{cobos14,cobos15,cobos16a,cobos16b,cobost}. We call this an  electronic RTE inverter. After pre-processing, one set of 24 images, corresponding to 4 polarisation states times 6 wavelengths, is translated into five solar physical quantities, namely the plasma line-of-sight velocity, $v_\mathrm{LOS}$, the three components of the vector magnetic field (strength, $B$, inclination relative to the line of sight, $\gamma$, and azimuth, $\phi$) and the continuum intensity, $I_{\rm c}$. These quantities are taken from the parameters describing a Milne-Eddington atmosphere, which is assumed to prevail. A full description of inversion procedures can be found in \cite{deltoroiniesta16}. Here we follow the implementation by \cite{orozco}. Additionally, it also takes into account the contribution from the spectral transmission of \sophi. The RTE inverter assumes a single homogeneous magnetic component in any given pixel. This step results in a data compression ratio of approximately a factor of five. 

The RTE inverter is composed of two main blocks: the RTE inversion core and the communications block. The latter is based on SoCWire and implements the communications with the other parts of the DPU. The inversion core is itself divided into two blocks. The first one is implemented through a SIMD (Single Instruction Multiple Data) multi-processor with several computation cores. The second block is in charge of some few operations that may be used just a few times. To carry out the inversion of the Stokes profiles two main tasks can be distinguished: The spectral synthesis of the Stokes profiles and the evaluation of the analytical response functions \citep[][]{orozco} on the one side, and a singular value decomposition (SVD) of a correlation matrix on the other side. While the synthesis and response functions are calculated by the multi-processor, SVD is carried out in the common operations block. Any inversion is an iterative process. The number of iterations is programmable within an internal register that allows for a maximum of 128 iterations, with 15 being the default number.  
There are five modes in which the inverter can be executed: 
\begin{itemize}
\item[-] \emph{RTE inversion}: The inverter infers the components of the magnetic field vector, $B$, $\gamma$, $\phi$, and the line-of-sight velocity, $v_\mathrm{LOS}$, using the Milne-Eddington approximation and an ad-hoc initial model atmosphere. 
\item[-] \emph{Classical estimations}: The inverter infers $B_\mathrm{LOS}$ and $v_\mathrm{LOS}$ using the centre of gravity technique as introduced by \citet{semel} \citep[see also][]{rees}. Then, $B$, $\gamma$, $\phi$ are obtained through the weak-field approximation \citep{landi}. This mode is not iterative.
\item[-] \emph{RTE inversion with classical estimations}: The classical estimations are used to get initial estimates of the free parameters that are then fed into the RTE inversion algorithm. This is the default mode.
\item[-] \emph{Longitudinal}: The inverter  delivers $v_\mathrm{LOS}$ and  $B_\mathrm{LOS}$ using the centre of gravity technique.
\item[-] \emph{No polarisation modulation}: The RTE inverter delivers $v_\mathrm{LOS}$ only.
\end{itemize}
The RTE inverter is able to process the four Stokes profiles recorded at each spatial pixel of the $2048\times 2048$\,pixel camera sensor at a rate of 3500 Stokes profiles per second.

\subsubsection{Post-processing, data compression and telemetry packing}

The on-board RTE inversion will stream the resulting parameters pixelwise as 32-bit floating point numbers to the image data memory (NAND flash). For post-processing, these data have to be reordered into parameter maps that can be compressed and delivered to the telemetry packeting module. Therefore, after the RTE inversion is finished, the Xilinx FPGAs have to be reconfigured again to carry out these tasks. FPGA\#2 will be configured to reload the inverted data from the image data memory and perform the resorting task. In addition, a scaling of the output parameters will be applied and all data will be converted to 16-bit signed integer. As the continuum intensity, $I_{\rm c}$, is an input parameter to the RTE the corresponding parameter map has to be separately loaded from the image data memory, scaled  and converted to  integer without reordering.

The final data compression is also based on firmware  
that will be loaded to FPGA\#1 \citep[see][]{hernandez18} 
and the reordered parameter maps will be streamed from the fast memory attached to FPGA\#2 to this compression core. The compression will be mainly applied to 16-bits integer fully-processed images (based on the output of the RTE inverter). Exceptionally, it will also be applied to raw or partly processed images either from nominal observations (to the extent that telemetry allows), or from the commissioning and instrumental check-out windows. We note that a single data set contains five images in nominal mode while it contains 24 in the case of raw data. The selected compression scheme is based on the CCSDS Recommended Standard for Image Data Compression, CCSDS~122.0-B-1\footnote{Consultative Committee for Space Data Systems, ``IMAGE DATA COMPRESSION Recommendation for Space Data System Standards 122.0-B-1,'' 2005.}. This algorithm is able to perform lossless and lossy compression, yet its limited computational complexity makes it suitable for hardware implementations. The CCSDS 122.0-B-1 compression algorithm was implemented with a multi-processor architecture that relies on a modified version of the RTE inverter. This FPGA implementation of the CCSDS~122.0-B-1 stands out over other solutions because it does not require any external memory for compressing 16~bits-per-pixel $2048\times 2048$\,pixel images and performs an acceleration of $\sim\,$30 times with respect to a software implementation running on space-qualified LEON3. All on-board data compression will be carried out in the lossy mode for which either a fixed compression ratio or a fixed image quality can be specified. Compression of processed data will allow for a compression ratio of 3-4, resulting in 4-5 bits per pixel \citep[see however][who discuss the implementation of an even higher compression ratio]{loeptien16a}, raw data can be compressed by a factor of 5 without violating the signal-to-noise requirements.

After compression, the data will be transferred to the RAM file system (main memory, cf. Fig.~\ref{Fig. DPU}) where they will be packed and provided to the spacecraft through the SpaceWire interface. Telemetry packing requires incorporating metadata to the header of the telemetry packets. 

\subsubsection{Additional on-board processing pipelines}

In addition to the standard scientific on-board data processing pipeline, \sophi\ requires several additional processing pipelines to produce the  data (e.g. flat fields or dark fields) required as input for the science data processing pipeline and to carry out preparatory tasks, such as re-focusing prior to the observations. 

\paragraph{Dark field calibration}

As \sophi\ is not equipped with an internal mechanical shutter and since it is assumed that the detector temperature might change significantly when the instrument doors are closed, dark field data have to be adopted from illuminated images. This can be achieved by dividing the dark field into its components. The fixed pattern noise is constant throughout each column on the detector and can be acquired from two masked detector rows. The mean dark current will be estimated from the detector areas which are obstructed by the baffle inside the FG (see Fig.~\ref{F:siemens}) and the dark current non-uniformity will be obtained from dedicated images which will be taken at the minimum possible exposure time ($86.191\,\mu$s). Ground tests and simulations show that the required accuracy of the dark field can be easily achieved.

\paragraph{FDT flat field data generation}

The FDT flat field data will be produced by applying the iterative method developed by \citet{kuhn91}. This procedure requires observational data taken at the same conditions as the science data and by off-pointing the telescope to several positions which should be, ideally, linearly independent. As \sophi\ does not have the possibility to off-point the FDT, the corresponding data acquisition requires off-pointing of the entire spacecraft. In addition to disc-centre pointing, 8 off-point positions between 0.08 and 0.43 solar radii (both in pitch and yaw) are envisaged. 

\paragraph{HRT flat field data generation}

HRT flat fields will be obtained by long-term accumulation of images while moving the tip/tilt mirror along a Lissajous-type trajectory. As the maximum off-pointing amplitude of the ISS is limited to $\pm 20\varcsec$, the required flatness is only achievable by also taking into account solar surface evolution (by averaging exposures covering several granular turn-over time scales). A backup procedure based on the method of \citet{kuhn91} can be also be used to generate the HRT flat fields (this choice will be made autonomously if, for instance,. a sunspot is in the FOV). 

\paragraph{ISS flat-field generation}

The ISS flat field generation procedure is similar to that of the HRT flat field generation procedure. However, the ISS cannot acquire images while moving the tip-tilt mirror. Flat field image accumulation is, therefore, interrupted by short tip-tilt movements which in sum result in the same Lissajous-type trajectory as for the HRT flat fielding. 

As a backup procedure the flat fielding method of \citet{kuhn91} can also be adopted for ISS flat-field generation. As the granulation contrast of the CTC images is expected to change significantly with orbital position, a decision about which of the two methods to use will be based on results obtained during instrument commissioning and cruise phase instrument check-outs. 

\paragraph{Re-focusing operations}

As the \sophi\ optics is not completely athermal, both the science focus and the correlation tracker focus change with orbital position. The main contribution is a variation of the temperature distribution on the HREWs. 

Re-focusing is carried out following a two-step automated procedure: 

\begin{enumerate}
\item If the focus position is completely unknown, the entire focus range is scanned and only a coarse focus position is estimated;
\item Afterwards, a second scan within a small range around the coarse focus position is carried out.
\end{enumerate}
If only a small change of the focus position is assumed to be required, then only the second step is carried out. This procedure is particularly relevant during long-term observations, for example when a solar feature is tracked across the solar disc by the HRT. Feature tracking requires periodic re-focusing since the temperature distribution on the HREWs changes with pointing position. The HREWs are, however, designed such, that the temperature distribution always has a parabolic shape which changes with solar distance and pointing position, thus it causes only a focus term in the resulting wavefront errors \citep[][]{garranzo17}.

The HRT is refocused by stepping through the corresponding mechanism range and accumulating a continuum image at each position. These images are dark-fielded and the rms contrast of each image is calculated. A parabolic fit around the maximum contrast then gives the focus position.

As the image contrast of the FDT images is dominated by the solar limb, refocusing of the FDT requires a more sophisticated procedure. 
The limb position is determined and a so-called masked gradient, $\delta I$, 
\begin{equation}
\delta I = \frac{1}{\left<I\right> \sum\limits_{i,j}M_{i,j}}\sum\limits_{i,j}\left[\left\{\left(\frac{\partial I(x,y)}{\partial x}\right)^2 + \left(\frac{\partial I(x,y)}{\partial y}\right)^2\right\} M_{i,j}\right]\,\,\,,
\end{equation}
within a narrow annular binary mask containing the defocused solar limb is computed.  $M_{i,j}$ is one pixel of the annular mask and $I$ and $\left<I\right>$ are the image intensity and its spatial mean. As for the HRT, the focus position is computed by fitting a parabola around the obtained maximum contrast values.

To refocus the correlation tracker, the CRM is moved through its entire range. After each step the ISS is put into closed-loop mode and the reference image is updated. The corresponding rms contrast is calculated inside the ISS control and provided to the DPU via HK from which the maximum contrast position is then computed. 

\paragraph{Additional calibration data generation}

All additional calibration data required for on-board processing (e.g. point spread functions or polarimetric modulation matrices) are generated on ground and uploaded to the spacecraft. The corresponding data acquisition and re-calibration procedures will be described in a dedicated publication.


\section{Conclusions/summary}\label{summary}

\sophi, the Polarimetric and Helioseismic Imager on board the Solar Orbiter spacecraft simultaneously provides magnetic field and velocity measurements. It will be the first such instrument to observe the Sun from outside the Sun-Earth line. This opens up many unresolved problems in solar and heliospheric physics for \sophi\ to tackle either on its own, or in conjunction with other instruments on board Solar Orbiter, on ground, or on other platforms. 

With its two telescopes, a high-resolution telescope that will reach a spatial resolution corresponding to 200\,km on the Sun at a distance of 0.28\,AU, and a full-disc telescope that will observe the full solar disc at all phases of the orbit, \sophi\ is a very versatile instrument. It will provide maps of the full magnetic field vector, the line-of-sight velocity and continuum intensity. These will enable addressing a variety of important science questions. These include questions related to the nature and functioning of the solar dynamo, and the structure of the magnetic  fields and flows for which \sophi\ will be the main instrument on Solar Orbiter. Thus it will play the main role in answering one of the top-level science questions of Solar Orbiter,
``How does the solar dynamo work and drive connections between the Sun and the heliosphere?''  

In addition, \sophi\ will provide the magnetic field data at the solar surface from which the coronal and heliospheric magnetic field will be extrapolated. Such extrapolations will play a key role in answering the remaining top-level science questions to be addressed by Solar Orbiter.

Beyond that, \sophi\ will enable an extremely rich set of science investigations whose importance has become evident in recent years, after Solar Orbiter was selected. 

To achieve its goals and to be accommodated within the tight mass, volume, power and telemetry constraints set by the platform and orbit, \sophi\ builds on a number of novel concepts that were verified through technology development efforts for first-time application in space.

These include the Fabry-Perot interferometers made from LiNbO$_3$ 
as electrically tunable narrow-band filters and the electro-optic polarisation analyser based on Liquid Crystal Variable Retarders. 
The newly developed camera system uses an APS sensor for space-based solar magnetometry for the first time.
Sophisticated on-board data processing will extract the final \sophi\ data products from the primary observables. This includes the on-board inversion of the radiative transfer equation. 
For the Heat-Rejecting Entrance Windows protecting the telescopes from the intense solar flux, a novel coating design has been qualified for space application.
Besides these ‘firsts’, specific challenges were imposed on the opto-mechanical design of the instrument, which has to guarantee stringent alignment of the telescopes for all orbital conditions.    

%
%

\begin{acknowledgements}
We are grateful to the ESA payload and mission support (SOC and MOC) teams for their cooperative support also during difficult phases of the Solar Orbiter project. We also thank the competent members of the spacecraft team. The dedication and hard work of the technical and administrative staff of the participating institutes is also gratefully acknowledged. Finally, we thank an anonymous referee for helpful comments.
The German contribution to \sophi\ is funded by the Bundesministerium f{\"u}r Wirtschaft und Technologie through Deutsches Zentrum f{\"u}r Luft- und Raumfahrt e.V. (DLR), Grants No. 50 OT 1001/1201/1901 as well as 50 OT 0801/1003/1203/1703, and by
the President of the Max Planck Society (MPG).
The Spanish contribution has been partially funded by Ministerio de Ciencia, Innovaci\'on y Universidades through projects ESP2014-56169-C6 and ESP2016-77548-C5. 
IAA-CSIC acknowledges financial support from the Spanish Research Agency (AEI/MCIU) through the ``Center of Excellence Severo Ochoa'' award for the Instituto de Astrof\'\i sica de Andaluc\'\i a (SEV-2017-0709). DOS acknowledges support from a {\em Ram\'on y Cajal} fellowship. The French contribution is funded by the Centre National d’Etudes Spatiales. 
\end{acknowledgements}

\bibliographystyle{aa}
\bibliography{bibfile_phi_paper.bib,SO_Book_cross_references.bib}

\begin{thebibliography}{218}
\expandafter\ifx\csname natexlab\endcsname\relax\def\natexlab#1{#1}\fi

\bibitem[{{Albert} {et~al.}(2018){Albert}, {Hirzberger}, {Busse}, {Lange},
  {Kolleck}, {Fiethe}, {Orozco Su\'arez}, {Woch}, {Schou}, {Blanco Rodr\'\i
  guez}, {Gandorfer}, {Cobos Carrascosa}, {Hern\'andez Exp\'osito}, {del Toro
  Iniesta}, {Solanki}, \& {Michalik}}]{albert18}
{Albert}, K., {Hirzberger}, J., {Busse}, D., {et~al.} 2018, in \procspie, Vol.
  10707, Software and Cyberinfrastructure for Astronomy V, 107070O

\bibitem[{{Alvarez-Herrero} {et~al.}(2015){Alvarez-Herrero}, {Garc{\'{\i}}a
  Parejo}, {Laguna}, {Villanueva}, {Barandiar{\'a}n}, {Bastide}, {Reina},
  {S{\'a}nchez}, {Gonzalo}, {Navarro}, {Vera}, \& {Royo}}]{alvarez15}
{Alvarez-Herrero}, A., {Garc{\'{\i}}a Parejo}, P., {Laguna}, H., {et~al.} 2015,
  in \procspie, Vol. 9613, Polarization Science and Remote Sensing VII, 96130I

\bibitem[{{Alvarez-Herrero} {et~al.}(2011){Alvarez-Herrero}, {Uribe-Patarroyo},
  {Garc{\'{\i}}a Parejo}, {Vargas}, {Heredero}, {Restrepo},
  {Mart{\'{\i}}nez-Pillet}, {Del Toro Iniesta}, {L{\'o}pez}, {Fineschi},
  {Capobianco}, {Georges}, {L{\'o}pez}, {Boer}, \& {Manolis}}]{alvarez11}
{Alvarez-Herrero}, A., {Uribe-Patarroyo}, N., {Garc{\'{\i}}a Parejo}, P.,
  {et~al.} 2011, in \procspie, Vol. 8160, Polarization Science and Remote
  Sensing V, 81600Y

\bibitem[{{\' Alvarez-Herrero} {et~al.}(2018){\' Alvarez-Herrero}, {Garc\'\i a
  Parejo}, \& {Silva-L\'opez}}]{alvarez18}
{\' Alvarez-Herrero}, A., {Garc\'\i a Parejo}, P., \& {Silva-L\'opez}, M. 2018,
  Optics Express, 26, 12038

\bibitem[{{Antonucci} {et~al.}(2019){Antonucci}, {Romoli}, {Andretta}, \&
  et~al.}]{Antonucci2019a}
{Antonucci}, E., {Romoli}, M., {Andretta}, V., \& et~al. 2019, \aap, this
  volume

\bibitem[{{Auch{\` e}re} {et~al.}(2019){Auch{\` e}re}, {Andretta}, {Antonucci},
  \& et~al.}]{Auchere2019a}
{Auch{\` e}re}, F., {Andretta}, V., {Antonucci}, E., \& et~al. 2019, \aap, this
  volume

\bibitem[{{Babcock}(1961)}]{babcock61}
{Babcock}, H.~W. 1961, \apj, 133, 572

\bibitem[{{Baldner} \& {Schou}(2012)}]{Baldner12}
{Baldner}, C.~S. \& {Schou}, J. 2012, \apjl, 760, L1

\bibitem[{{Barandiaran} {et~al.}(2017){Barandiaran}, {Zuluaga}, {Fernandez},
  {Vera}, {Garranzo}, {Nu{\~n}ez}, {Bastide}, {Royo}, \&
  {Alvarez}}]{barandiaran17}
{Barandiaran}, J., {Zuluaga}, P., {Fernandez}, A.~B., {et~al.} 2017, in Society
  of Photo-Optical Instrumentation Engineers (SPIE) Conference Series, Vol.
  10563, Society of Photo-Optical Instrumentation Engineers (SPIE) Conference
  Series, 1056319

\bibitem[{{Barthol} {et~al.}(2011){Barthol}, {Gandorfer}, {Solanki},
  {Sch{\"u}ssler}, {Chares}, {Curdt}, {Deutsch}, {Feller}, {Germerott},
  {Grauf}, {Heerlein}, {Hirzberger}, {Kolleck}, {Meller}, {M{\"u}ller},
  {Riethm{\"u}ller}, {Tomasch}, {Kn{\"o}lker}, {Lites}, {Card}, {Elmore},
  {Fox}, {Lecinski}, {Nelson}, {Summers}, {Watt}, {Mart{\'{\i}}nez Pillet},
  {Bonet}, {Schmidt}, {Berkefeld}, {Title}, {Domingo}, {Gasent Blesa}, {Del
  Toro Iniesta}, {L{\'o}pez Jim{\'e}nez}, {{\'A}lvarez-Herrero},
  {Sabau-Graziati}, {Widani}, {Haberler}, {H{\"a}rtel}, {Kampf}, {Levin},
  {P{\'e}rez Grande}, {Sanz-Andr{\'e}s}, \& {Schmidt}}]{barthol11}
{Barthol}, P., {Gandorfer}, A., {Solanki}, S.~K., {et~al.} 2011, \solphys, 268,
  1

\bibitem[{{Basu}(2016)}]{basu16}
{Basu}, S. 2016, Living Reviews in Solar Physics, 13, 2

\bibitem[{{Benz}(2017)}]{Benz17}
{Benz}, A.~O. 2017, Living Reviews in Solar Physics, 14, 2

\bibitem[{{Berkefeld} {et~al.}(2011){Berkefeld}, {Schmidt}, {Soltau}, {Bell},
  {Doerr}, {Feger}, {Friedlein}, {Gerber}, {Heidecke}, {Kentischer},
  {v.~D.~L{\"u}he}, {Sigwarth}, {W{\"a}lde}, {Barthol}, {Deutsch}, {Gandorfer},
  {Germerott}, {Grauf}, {Meller}, {{\'A}lvarez-Herrero}, {Kn{\"o}lker},
  {Mart{\'{\i}}nez Pillet}, {Solanki}, \& {Title}}]{berkefeld11}
{Berkefeld}, T., {Schmidt}, W., {Soltau}, D., {et~al.} 2011, \solphys, 268, 103

\bibitem[{{Bernasconi} {et~al.}(2000){Bernasconi}, {Rust}, {Eaton}, \&
  {Murphy}}]{bernasconi00}
{Bernasconi}, P.~N., {Rust}, D.~M., {Eaton}, H.~A., \& {Murphy}, G.~A. 2000, in
  \procspie, Vol. 4014, Airborne Telescope Systems, ed. R.~K. {Melugin} \&
  H.-P. {R{\"o}ser}, 214--225

\bibitem[{{Bischoff} {et~al.}(2014){Bischoff}, {Grauf}, {Staub}, {Gandorfer},
  {Woch}, {Clark}, {Zimmermann}, {Kolb}, {Metz}, \& {Rucks}}]{bischoff14}
{Bischoff}, J., {Grauf}, B., {Staub}, J., {et~al.} 2014, in Ilmenau Scientific
  Colloquium. Technische Universität Ilmenau, Vol.~58, Shaping the Future by
  Engineering, Technische Universität Ilmenau (ilmedia)

\bibitem[{{B{\"o}ning} {et~al.}(2017){B{\"o}ning}, {Roth}, {Jackiewicz}, \&
  {Kholikov}}]{2017ApJ...845....2B}
{B{\"o}ning}, V.~G.~A., {Roth}, M., {Jackiewicz}, J., \& {Kholikov}, S. 2017,
  \apj, 845, 2

\bibitem[{{Borrero} {et~al.}(2017){Borrero}, {Jafarzadeh}, {Sch{\"u}ssler}, \&
  {Solanki}}]{borrero17}
{Borrero}, J.~M., {Jafarzadeh}, S., {Sch{\"u}ssler}, M., \& {Solanki}, S.~K.
  2017, \ssr, 210, 275

\bibitem[{{Borucki} {et~al.}(2010){Borucki}, {Koch}, {Basri}, {Batalha},
  {Brown}, {Caldwell}, {Caldwell}, {Christensen-Dalsgaard}, {Cochran},
  {DeVore}, {Dunham}, {Dupree}, {Gautier}, {Geary}, {Gilliland}, {Gould},
  {Howell}, {Jenkins}, {Kondo}, {Latham}, {Marcy}, {Meibom}, {Kjeldsen},
  {Lissauer}, {Monet}, {Morrison}, {Sasselov}, {Tarter}, {Boss}, {Brownlee},
  {Owen}, {Buzasi}, {Charbonneau}, {Doyle}, {Fortney}, {Ford}, {Holman},
  {Seager}, {Steffen}, {Welsh}, {Rowe}, {Anderson}, {Buchhave}, {Ciardi},
  {Walkowicz}, {Sherry}, {Horch}, {Isaacson}, {Everett}, {Fischer}, {Torres},
  {Johnson}, {Endl}, {MacQueen}, {Bryson}, {Dotson}, {Haas}, {Kolodziejczak},
  {Van Cleve}, {Chandrasekaran}, {Twicken}, {Quintana}, {Clarke}, {Allen},
  {Li}, {Wu}, {Tenenbaum}, {Verner}, {Bruhweiler}, {Barnes}, \&
  {Prsa}}]{borucki10}
{Borucki}, W.~J., {Koch}, D., {Basri}, G., {et~al.} 2010, Science, 327, 977

\bibitem[{{Brandenburg}(2005)}]{brandenburg05}
{Brandenburg}, A. 2005, \apj, 625, 539

\bibitem[{{Brun} {et~al.}(2004){Brun}, {Miesch}, \& {Toomre}}]{brun04}
{Brun}, A.~S., {Miesch}, M.~S., \& {Toomre}, J. 2004, \apj, 614, 1073

\bibitem[{{Buehler} {et~al.}(2013){Buehler}, {Lagg}, \& {Solanki}}]{buehler13}
{Buehler}, D., {Lagg}, A., \& {Solanki}, S.~K. 2013, \aap, 555, A33

\bibitem[{{Cacciani} {et~al.}(1990){Cacciani}, {Varsik}, \&
  {Zirin}}]{cacciani90}
{Cacciani}, A., {Varsik}, J., \& {Zirin}, H. 1990, \solphys, 125, 173

\bibitem[{{Cameron} \& {Sch{\"u}ssler}(2015)}]{cameron15}
{Cameron}, R. \& {Sch{\"u}ssler}, M. 2015, Science, 347, 1333

\bibitem[{{Cameron} {et~al.}(2017){Cameron}, {Dikpati}, \&
  {Brandenburg}}]{cameron17}
{Cameron}, R.~H., {Dikpati}, M., \& {Brandenburg}, A. 2017, \ssr, 210, 367

\bibitem[{{Cao} {et~al.}(2010){Cao}, {Gorceix}, {Coulter}, {Ahn}, {Rimmele}, \&
  {Goode}}]{cao10}
{Cao}, W., {Gorceix}, N., {Coulter}, R., {et~al.} 2010, Astronomische
  Nachrichten, 331, 636

\bibitem[{{Carlsson} {et~al.}(2004){Carlsson}, {Stein}, {Nordlund}, \&
  {Scharmer}}]{carlsson04}
{Carlsson}, M., {Stein}, R.~F., {Nordlund}, {\AA}., \& {Scharmer}, G.~B. 2004,
  \apjl, 610, L137

\bibitem[{{Carmona} {et~al.}(2014){Carmona}, {G{\'o}mez}, {Roma}, {Casas},
  {L{\'o}pez}, {Bosch}, {Herms}, {Sabater}, {Volkmer}, {Heidecke}, {Maue},
  {Nakai}, \& {Schmidt}}]{carmona14}
{Carmona}, M., {G{\'o}mez}, J.~M., {Roma}, D., {et~al.} 2014, in \procspie,
  Vol. 9150, Modeling, Systems Engineering, and Project Management for
  Astronomy VI, 91501U

\bibitem[{{Casas} {et~al.}(2016){Casas}, {G{\'o}mez}, {Roma}, {Carmona},
  {L{\'o}pez}, {Bosch}, {Herms}, {Sabater}, {Volkmer}, {Heidecke}, {Maue},
  {Nakai}, {Baumgartner}, \& {Schmidt}}]{casas16}
{Casas}, A., {G{\'o}mez}, J.~M., {Roma}, D., {et~al.} 2016, in \procspie, Vol.
  9911, Modeling, Systems Engineering, and Project Management for Astronomy VI,
  991123

\bibitem[{{Charbonneau}(2010)}]{charbonneau10}
{Charbonneau}, P. 2010, Living Reviews in Solar Physics, 7, 3

\bibitem[{{Charbonneau}(2013)}]{charbonneau13}
{Charbonneau}, P. 2013, in Journal of Physics Conference Series, Vol. 440,
  Journal of Physics Conference Series, 012014

\bibitem[{{Charbonneau}(2014)}]{charbonneau14}
{Charbonneau}, P. 2014, \araa, 52, 251

\bibitem[{{Chen}(2011)}]{chen11}
{Chen}, P.~F. 2011, Living Reviews in Solar Physics, 8, 1

\bibitem[{{Chen} \& {Zhao}(2017)}]{2017ApJ...849..144C}
{Chen}, R. \& {Zhao}, J. 2017, \apj, 849, 144

\bibitem[{{Chitta} {et~al.}(2018){Chitta}, {Peter}, \& {Solanki}}]{chitta18}
{Chitta}, L.~P., {Peter}, H., \& {Solanki}, S.~K. 2018, \aap, 615, L9

\bibitem[{{Chitta} {et~al.}(2017){Chitta}, {Peter}, {Solanki}, {Barthol},
  {Gandorfer}, {Gizon}, {Hirzberger}, {Riethm{\"u}ller}, {van Noort}, {Blanco
  Rodr{\'{\i}}guez}, {Del Toro Iniesta}, {Orozco Su{\'a}rez}, {Schmidt},
  {Mart{\'{\i}}nez Pillet}, \& {Kn{\"o}lker}}]{chitta17}
{Chitta}, L.~P., {Peter}, H., {Solanki}, S.~K., {et~al.} 2017, \apjs, 229, 4

\bibitem[{{Christensen-Dalsgaard}(2002)}]{christensen02}
{Christensen-Dalsgaard}, J. 2002, Reviews of Modern Physics, 74, 1073

\bibitem[{{Cobos Carrascosa}(2016)}]{cobost}
{Cobos Carrascosa}, J.~P. 2016, in Doctoral Thesis Dissertation, Universidad de
  Granada

\bibitem[{{Cobos Carrascosa} {et~al.}(2015){Cobos Carrascosa}, {Aparicio del
  Moral}, {Ramos Mas}, {Balaguer}, {L{\'{o}}pez Jim{\'{e}}nez}, \& {del Toro
  Iniesta}}]{cobos15}
{Cobos Carrascosa}, J.~P., {Aparicio del Moral}, B., {Ramos Mas}, J.~L.,
  {et~al.} 2015, in 2015 {NASA/ESA} Conference on Adaptive Hardware and
  Systems, {AHS} 2015, Montreal, QC, Canada, June 15-18, 2015, 1--8

\bibitem[{{Cobos Carrascosa} {et~al.}(2016{\natexlab{a}}){Cobos Carrascosa},
  {Aparicio del Moral}, {Ramos Mas}, {Balaguer}, {L{\'o}pez Jim{\'e}nez}, \&
  {Del Toro Iniesta}}]{cobos16a}
{Cobos Carrascosa}, J.~P., {Aparicio del Moral}, B., {Ramos Mas}, J.~L.,
  {et~al.} 2016{\natexlab{a}}, in \procspie, Vol. 9913, Software and
  Cyberinfrastructure for Astronomy IV, 991342

\bibitem[{{Cobos Carrascosa} {et~al.}(2014){Cobos Carrascosa}, {Aparicio del
  Moral}, {Ramos Mas}, {L{\'{o}}pez Jim{\'{e}}nez}, \& {Del Toro
  Iniesta}}]{cobos14}
{Cobos Carrascosa}, J.~P., {Aparicio del Moral}, B., {Ramos Mas}, J.~L.,
  {L{\'{o}}pez Jim{\'{e}}nez}, A.~C., \& {Del Toro Iniesta}, J.~C. 2014, in
  {IEEE} 8th International Symposium on Embedded Multicore/Manycore SoCs, MCSoC
  2014, Aizu-Wakamatsu, Japan, September 23-25, 2014, 223--228

\bibitem[{{Cobos Carrascosa} {et~al.}(2016{\natexlab{b}}){Cobos Carrascosa},
  {Ramos Mas}, {Aparicio del Moral}, {Balaguer}, {L{\'{o}}pez Jim{\'{e}}nez},
  \& {Del Toro Iniesta}}]{cobos16b}
{Cobos Carrascosa}, J.~P., {Ramos Mas}, J.~L., {Aparicio del Moral}, B.,
  {et~al.} 2016{\natexlab{b}}, Journal of Systems Architecture - Embedded
  Systems Design, 62, 1

\bibitem[{{Corbard} {et~al.}(2013){Corbard}, {Salabert}, {Boumier},
  {Appourchaux}, {Hauchecorne}, {Journoud}, {Nunge}, {Gelly}, {Hochedez},
  {Irbah}, {Meftah}, {Renaud}, \& {Turck-Chi{\`e}ze}}]{corbard13}
{Corbard}, T., {Salabert}, D., {Boumier}, P., {et~al.} 2013, in Journal of
  Physics Conference Series, Vol. 440, Journal of Physics Conference Series,
  012025

\bibitem[{{Cranmer}(2009)}]{Cranmer09}
{Cranmer}, S.~R. 2009, Living Reviews in Solar Physics, 6, 3

\bibitem[{{Cranmer} {et~al.}(2007){Cranmer}, {van Ballegooijen}, \&
  {Edgar}}]{cranmer07}
{Cranmer}, S.~R., {van Ballegooijen}, A.~A., \& {Edgar}, R.~J. 2007, \apjs,
  171, 520

\bibitem[{{Danilovic} {et~al.}(2016){Danilovic}, {Rempel}, {van Noort}, \&
  {Cameron}}]{danilovic16}
{Danilovic}, S., {Rempel}, M., {van Noort}, M., \& {Cameron}, R. 2016, \aap,
  594, A103

\bibitem[{{Danilovic} {et~al.}(2010){Danilovic}, {Sch{\"u}ssler}, \&
  {Solanki}}]{danilovic10}
{Danilovic}, S., {Sch{\"u}ssler}, M., \& {Solanki}, S.~K. 2010, \aap, 513, A1

\bibitem[{{Del Toro Iniesta}(2003)}]{deltoroiniesta03}
{Del Toro Iniesta}, J.~C. 2003, {Introduction to Spectropolarimetry}
  (Cambridge, UK: Cambridge University Press)

\bibitem[{{Del Toro Iniesta} \& {Collados}(2000)}]{deltoroiniesta00}
{Del Toro Iniesta}, J.~C. \& {Collados}, M. 2000, Appl. Opt., 39, 1637

\bibitem[{{Del Toro Iniesta} \& {Ruiz Cobo}(2016)}]{deltoroiniesta16}
{Del Toro Iniesta}, J.~C. \& {Ruiz Cobo}, B. 2016, Living Reviews in Solar
  Physics, 13, 4

\bibitem[{{Desai} \& {Giacalone}(2016)}]{desai16}
{Desai}, M. \& {Giacalone}, J. 2016, Living Reviews in Solar Physics, 13, 3

\bibitem[{{Duvall} {et~al.}(1993){Duvall}, {Jefferies}, {Harvey}, \&
  {Pomerantz}}]{duvall93}
{Duvall}, Jr., T.~L., {Jefferies}, S.~M., {Harvey}, J.~W., \& {Pomerantz},
  M.~A. 1993, \nat, 362, 430

\bibitem[{{Ermolli} {et~al.}(2013){Ermolli}, {Matthes}, {Dudok de Wit},
  {Krivova}, {Tourpali}, {Weber}, {Unruh}, {Gray}, {Langematz}, {Pilewskie},
  {Rozanov}, {Schmutz}, {Shapiro}, {Solanki}, \& {Woods}}]{ermolli13}
{Ermolli}, I., {Matthes}, K., {Dudok de Wit}, T., {et~al.} 2013, Atmospheric
  Chemistry \& Physics, 13, 3945

\bibitem[{Fernandez-Rico {et~al.}(2018)Fernandez-Rico, Alvarez-Copano, Deutsch,
  Gandorfer, Ramanath, Staub, Bambach, \& Torralbo}]{fernandez18}
Fernandez-Rico, G., Alvarez-Copano, M., Deutsch, W., {et~al.} 2018, in
  48$^{th}$ International Conference on Environmental Systems, Albuquerque, NM

\bibitem[{Fiethe {et~al.}(2012)Fiethe, Bubenhagen, Lange, Michalik, Michel,
  Woch, \& Hirzberger}]{fiethe12}
Fiethe, B., Bubenhagen, F., Lange, T., {et~al.} 2012, in 2012 {NASA/ESA}
  Conference on Adaptive Hardware and Systems, {AHS} 2012, Erlangen, Germany,
  June 25-28, 2012, 31--37

\bibitem[{{Fisk} {et~al.}(2003){Fisk}, {Gloeckler}, {Zurbuchen}, {Geiss}, \&
  {Schwadron}}]{fisk03}
{Fisk}, L.~A., {Gloeckler}, G., {Zurbuchen}, T.~H., {Geiss}, J., \&
  {Schwadron}, N.~A. 2003, in American Institute of Physics Conference Series,
  Vol. 679, Solar Wind Ten, ed. M.~{Velli}, R.~{Bruno}, F.~{Malara}, \&
  B.~{Bucci}, 287--292

\bibitem[{{Fisk} {et~al.}(1999){Fisk}, {Schwadron}, \& {Zurbuchen}}]{fisk99}
{Fisk}, L.~A., {Schwadron}, N.~A., \& {Zurbuchen}, T.~H. 1999, \jgr, 104, 19765

\bibitem[{{Fligge} {et~al.}(2000){Fligge}, {Solanki}, \& {Unruh}}]{fligge00}
{Fligge}, M., {Solanki}, S.~K., \& {Unruh}, Y.~C. 2000, \ssr, 94, 139

\bibitem[{{Foukal} {et~al.}(2006){Foukal}, {Fr{\"o}hlich}, {Spruit}, \&
  {Wigley}}]{Foukal06}
{Foukal}, P., {Fr{\"o}hlich}, C., {Spruit}, H., \& {Wigley}, T.~M.~L. 2006,
  \nat, 443, 161

\bibitem[{{Fr{\"o}hlich} {et~al.}(1995){Fr{\"o}hlich}, {Romero}, {Roth},
  {Wehrli}, {Andersen}, {Appourchaux}, {Domingo}, {Telljohann}, {Berthomieu},
  {Delache}, {Provost}, {Toutain}, {Crommelynck}, {Chevalier}, {Fichot},
  {D{\"a}ppen}, {Gough}, {Hoeksema}, {Jim{\'e}nez}, {G{\'o}mez}, {Herreros},
  {Cort{\'e}s}, {Jones}, {Pap}, \& {Willson}}]{froehlich95}
{Fr{\"o}hlich}, C., {Romero}, J., {Roth}, H., {et~al.} 1995, \solphys, 162, 101

\bibitem[{{Gabriel} {et~al.}(1995){Gabriel}, {Grec}, {Charra}, {Robillot},
  {Roca Cort{\'e}s}, {Turck-Chi{\`e}ze}, {Bocchia}, {Boumier}, {Cantin},
  {Cesp{\'e}des}, {Cougrand}, {Cr{\'e}tolle}, {Dam{\'e}}, {Decaudin},
  {Delache}, {Denis}, {Duc}, {Dzitko}, {Fossat}, {Fourmond}, {Garc{\'{\i}}a},
  {Gough}, {Grivel}, {Herreros}, {Lagard{\`e}re}, {Moalic}, {Pall{\'e}},
  {P{\'e}trou}, {Sanchez}, {Ulrich}, \& {van der Raay}}]{gabriel95}
{Gabriel}, A.~H., {Grec}, G., {Charra}, J., {et~al.} 1995, \solphys, 162, 61

\bibitem[{{Gandorfer} {et~al.}(2018){Gandorfer}, {Grauf}, {Staub}, {Bischoff},
  {Woch}, {Hirzberger}, {Solanki}, {{\' A}lvarez-Herrero}, {Garc{\'\i}a
  Parejo}, {Volkmer}, {Appourchaux}, \& {del Toro Iniesta}}]{gandorfer18}
{Gandorfer}, A., {Grauf}, B., {Staub}, J., {et~al.} 2018, in \procspie, Vol.
  10698, Space Telescopes and Instrumentation 2018: Optical, Infrared, and
  Millimeter Wave, 106984N

\bibitem[{{Gandorfer} {et~al.}(2011){Gandorfer}, {Solanki}, {Woch},
  {Mart{\'{\i}}nez Pillet}, {{\'A}lvarez Herrero}, \&
  {Appourchaux}}]{gandorfer11}
{Gandorfer}, A., {Solanki}, S.~K., {Woch}, J., {et~al.} 2011, GONG-SoHO 24: A
  New Era of Seismology of the Sun and Solar-Like Stars, 271, 012086

\bibitem[{{Garcia-Marirrodriga}(2019)}]{Garcia2019}
{Garcia-Marirrodriga}, C. e.~a. 2019, \aap, this volume

\bibitem[{{Garranzo} {et~al.}(2017){Garranzo}, {N{\'u}{\~n}ez},
  {Zuluaga-Ram{\'{\i}}rez}, {Barandiar{\'a}n}, {Fern{\'a}ndez-Medina},
  {Belenguer}, \& {{\'A}lvarez-Herrero}}]{garranzo17}
{Garranzo}, D., {N{\'u}{\~n}ez}, A., {Zuluaga-Ram{\'{\i}}rez}, P., {et~al.}
  2017, in Society of Photo-Optical Instrumentation Engineers (SPIE) Conference
  Series, Vol. 10563, Society of Photo-Optical Instrumentation Engineers (SPIE)
  Conference Series, 105634T

\bibitem[{{Gensemer} \& {Farrant}(2014)}]{gensemer14}
{Gensemer}, S.~D. \& {Farrant}, D. 2014, Advanced Optical Technologies, 3, 309

\bibitem[{{Gilbert} {et~al.}(2007){Gilbert}, {Zurbuchen}, \&
  {Fisk}}]{gilbert07}
{Gilbert}, J.~A., {Zurbuchen}, T.~H., \& {Fisk}, L.~A. 2007, \apj, 663, 583

\bibitem[{{Gizon} \& {Birch}(2005)}]{gizon05}
{Gizon}, L. \& {Birch}, A.~C. 2005, Living Reviews in Solar Physics, 2, 6

\bibitem[{{Gizon} {et~al.}(2018){Gizon}, {Fournier}, {Yang}, {Birch}, \&
  {Barucq}}]{2018arXiv181000402G}
{Gizon}, L., {Fournier}, D., {Yang}, D., {Birch}, A.~C., \& {Barucq}, H. 2018,
  ArXiv e-prints

\bibitem[{{Haigh}(2007)}]{haigh07}
{Haigh}, J.~D. 2007, Living Reviews in Solar Physics, 4, 2

\bibitem[{{Harvey} {et~al.}(1996){Harvey}, {Hill}, {Hubbard}, {Kennedy},
  {Leibacher}, {Pintar}, {Gilman}, {Noyes}, {Title}, {Toomre}, {Ulrich},
  {Bhatnagar}, {Kennewell}, {Marquette}, {Patron}, {Saa}, \&
  {Yasukawa}}]{harvey96}
{Harvey}, J.~W., {Hill}, F., {Hubbard}, R.~P., {et~al.} 1996, Science, 272,
  1284

\bibitem[{{Hathaway}(2010)}]{hathaway10}
{Hathaway}, D.~H. 2010, Living Reviews in Solar Physics, 7, 1

\bibitem[{{Hern{\'a}ndez Exp{\'o}sito} {et~al.}(2018){Hern{\'a}ndez
  Exp{\'o}sito}, {Cobos Carrascosa}, {Ramos Mas}, {Rodr{\'{\i}}guez Valido},
  {Orozco Su{\'a}rez}, {Hirzberger}, {Woch}, {Solanki}, \& {del Toro
  Iniesta}}]{hernandez18}
{Hern{\'a}ndez Exp{\'o}sito}, D., {Cobos Carrascosa}, J.~P., {Ramos Mas},
  J.~L., {et~al.} 2018, in Society of Photo-Optical Instrumentation Engineers
  (SPIE) Conference Series, Vol. 10707, Software and Cyberinfrastructure for
  Astronomy V, 107072F

\bibitem[{{Hirzberger} \& {Wiehr}(2005)}]{hirzberger05}
{Hirzberger}, J. \& {Wiehr}, E. 2005, \aap, 438, 1059

\bibitem[{{Horbury} {et~al.}(2019){Horbury}, {Owen}, {Maksimovic}, \&
  et~al.}]{Horbury2019b}
{Horbury}, T., {Owen}, C., {Maksimovic}, M., \& et~al. 2019, \aap, this volume

\bibitem[{{Horbury}(2019)}]{Horbury2019a}
{Horbury}, T. e.~a. 2019, \aap, this volume

\bibitem[{{Howard} {et~al.}(2008){Howard}, {Moses}, {Vourlidas}, {Newmark},
  {Socker}, {Plunkett}, {Korendyke}, {Cook}, {Hurley}, {Davila}, {Thompson},
  {St Cyr}, {Mentzell}, {Mehalick}, {Lemen}, {Wuelser}, {Duncan}, {Tarbell},
  {Wolfson}, {Moore}, {Harrison}, {Waltham}, {Lang}, {Davis}, {Eyles},
  {Mapson-Menard}, {Simnett}, {Halain}, {Defise}, {Mazy}, {Rochus}, {Mercier},
  {Ravet}, {Delmotte}, {Auchere}, {Delaboudiniere}, {Bothmer}, {Deutsch},
  {Wang}, {Rich}, {Cooper}, {Stephens}, {Maahs}, {Baugh}, {McMullin}, \&
  {Carter}}]{howard08}
{Howard}, R.~A., {Moses}, J.~D., {Vourlidas}, A., {et~al.} 2008, \ssr, 136, 67

\bibitem[{{Howard} {et~al.}(2019){Howard}, {Vourlidas}, {Colaninno}, \&
  et~al.}]{Howard2019a}
{Howard}, R.~A., {Vourlidas}, A., {Colaninno}, R.~C., \& et~al. 2019, \aap,
  this volume

\bibitem[{{Howe}(2009)}]{howe09}
{Howe}, R. 2009, Living Reviews in Solar Physics, 6, 1

\bibitem[{{Howe} {et~al.}(2018){Howe}, {Hill}, {Komm}, {Chaplin}, {Elsworth},
  {Davies}, {Schou}, \& {Thompson}}]{2018ApJ...862L...5H}
{Howe}, R., {Hill}, F., {Komm}, R., {et~al.} 2018, \apjl, 862, L5

\bibitem[{{Jiang} {et~al.}(2014){Jiang}, {Hathaway}, {Cameron}, {Solanki},
  {Gizon}, \& {Upton}}]{jiang14}
{Jiang}, J., {Hathaway}, D.~H., {Cameron}, R.~H., {et~al.} 2014, \ssr, 186, 491

\bibitem[{{Kaiser} {et~al.}(2008){Kaiser}, {Kucera}, {Davila}, {St.~Cyr},
  {Guhathakurta}, \& {Christian}}]{kaiser08}
{Kaiser}, M.~L., {Kucera}, T.~A., {Davila}, J.~M., {et~al.} 2008, \ssr, 136, 5

\bibitem[{{Keller} {et~al.}(2004){Keller}, {Sch{\"u}ssler}, {V{\"o}gler}, \&
  {Zakharov}}]{keller04}
{Keller}, C.~U., {Sch{\"u}ssler}, M., {V{\"o}gler}, A., \& {Zakharov}, V. 2004,
  \apjl, 607, L59

\bibitem[{{Kilpua} {et~al.}(2017){Kilpua}, {Koskinen}, \&
  {Pulkkinen}}]{kilpua17}
{Kilpua}, E., {Koskinen}, H.~E.~J., \& {Pulkkinen}, T.~I. 2017, Living Reviews
  in Solar Physics, 14, 5

\bibitem[{{Klimchuk}(2001)}]{klimchuk01}
{Klimchuk}, J.~A. 2001, Washington DC American Geophysical Union Geophysical
  Monograph Series, 125

\bibitem[{{Klimchuk}(2006)}]{klimchuk06}
{Klimchuk}, J.~A. 2006, \solphys, 234, 41

\bibitem[{{Knaack} {et~al.}(2001){Knaack}, {Fligge}, {Solanki}, \&
  {Unruh}}]{knaack01}
{Knaack}, R., {Fligge}, M., {Solanki}, S.~K., \& {Unruh}, Y.~C. 2001, \aap,
  376, 1080

\bibitem[{{Komm} {et~al.}(2018){Komm}, {Howe}, \& {Hill}}]{2018SoPh..293..145K}
{Komm}, R., {Howe}, R., \& {Hill}, F. 2018, \solphys, 293, 145

\bibitem[{{Kosovichev}(2014)}]{kosovichev14}
{Kosovichev}, A.~G. 2014, ArXiv e-prints

\bibitem[{{Kosovichev} \& {Zharkova}(1998)}]{kosovichev98}
{Kosovichev}, A.~G. \& {Zharkova}, V.~V. 1998, \nat, 393, 317

\bibitem[{{Krivova} {et~al.}(2003){Krivova}, {Solanki}, {Fligge}, \&
  {Unruh}}]{krivova03}
{Krivova}, N.~A., {Solanki}, S.~K., {Fligge}, M., \& {Unruh}, Y.~C. 2003, \aap,
  399, L1

\bibitem[{{Krucker}(2019)}]{Krucker2019a}
{Krucker}, S., e.~a. 2019, \aap, this volume

\bibitem[{{Kuhn} {et~al.}(1991){Kuhn}, {Lin}, \& {Loranz}}]{kuhn91}
{Kuhn}, J.~R., {Lin}, H., \& {Loranz}, D. 1991, \pasp, 103, 1097

\bibitem[{{Landi Degl'Innocenti} \& {Landolfi}(2004)}]{landi}
{Landi Degl'Innocenti}, E. \& {Landolfi}, M., eds. 2004, Astrophysics and Space
  Science Library, Vol. 307, {Polarization in Spectral Lines}

\bibitem[{Lange {et~al.}(2018)Lange, Albert, Busse, Fiethe, Guan, Hirzberger,
  \& Michalik}]{lange18}
Lange, T., Albert, K., Busse, D., {et~al.} 2018, in Data Systems in Aerospace,
  DASIA 2018

\bibitem[{Lange {et~al.}(2017)Lange, Fiethe, Michel, Michalik, Albert, \&
  Hirzberger}]{lange17}
Lange, T., Fiethe, B., Michel, H., {et~al.} 2017, in 2017 {NASA/ESA} Conference
  on Adaptive Hardware and Systems, {AHS} 2017, Pasadena, CA, USA, July 24-27,
  2017, 186--191

\bibitem[{Lange {et~al.}(2015)Lange, Michel, Fiethe, \& Michalik}]{lange15b}
Lange, T., Michel, H., Fiethe, B., \& Michalik, H. 2015, Xilinx Xcell Journal,
  Issue 90

\bibitem[{{Lange} {et~al.}(2015){Lange}, {Michel}, {Fiethe}, {Michalik}, \&
  {Walter}}]{lange15a}
{Lange}, T., {Michel}, H., {Fiethe}, B., {Michalik}, H., \& {Walter}, D. 2015,
  in \procspie, Vol. 9646, High-Performance Computing in Remote Sensing V,
  96460B

\bibitem[{{Leighton}(1969)}]{leighton69}
{Leighton}, R.~B. 1969, \apj, 156, 1

\bibitem[{{Liang} {et~al.}(2018){Liang}, {Gizon}, {Birch}, {Duvall}, \&
  {Rajaguru}}]{2018arXiv180808874L}
{Liang}, Z.-C., {Gizon}, L., {Birch}, A.~C., {Duvall}, Jr., T.~L., \&
  {Rajaguru}, S.~P. 2018, ArXiv e-prints

\bibitem[{{Liewer} {et~al.}(2017){Liewer}, {Qiu}, \&
  {Lindsey}}]{2017SoPh..292..146L}
{Liewer}, P.~C., {Qiu}, J., \& {Lindsey}, C. 2017, \solphys, 292, 146

\bibitem[{{Lin} \& {Forbes}(2000)}]{lin00}
{Lin}, J. \& {Forbes}, T.~G. 2000, \jgr, 105, 2375

\bibitem[{{Lindsey} \& {Braun}(2000)}]{2000Sci...287.1799L}
{Lindsey}, C. \& {Braun}, D.~C. 2000, Science, 287, 1799

\bibitem[{{Linker} {et~al.}(2017){Linker}, {Caplan}, {Downs}, {Riley}, {Mikic},
  {Lionello}, {Henney}, {Arge}, {Liu}, {Derosa}, {Yeates}, \&
  {Owens}}]{linker17}
{Linker}, J.~A., {Caplan}, R.~M., {Downs}, C., {et~al.} 2017, \apj, 848, 70

\bibitem[{{Linker} {et~al.}(2003){Linker}, {Miki{\'c}}, {Riley}, {Lionello}, \&
  {Odstrcil}}]{linker03}
{Linker}, J.~A., {Miki{\'c}}, Z., {Riley}, P., {Lionello}, R., \& {Odstrcil},
  D. 2003, in American Institute of Physics Conference Series, Vol. 679, Solar
  Wind Ten, ed. M.~{Velli}, R.~{Bruno}, F.~{Malara}, \& B.~{Bucci}, 703--710

\bibitem[{{Lisle} \& {Toomre}(2004)}]{Lisle04}
{Lisle}, J. \& {Toomre}, J. 2004, in ESA Special Publication, Vol. 559, SOHO 14
  Helio- and Asteroseismology: Towards a Golden Future, ed. D.~{Danesy}, 556

\bibitem[{{Lites} {et~al.}(2013){Lites}, {Akin}, {Card}, {Cruz}, {Duncan},
  {Edwards}, {Elmore}, {Hoffmann}, {Katsukawa}, {Katz}, {Kubo}, {Ichimoto},
  {Shimizu}, {Shine}, {Streander}, {Suematsu}, {Tarbell}, {Title}, \&
  {Tsuneta}}]{lites13}
{Lites}, B.~W., {Akin}, D.~L., {Card}, G., {et~al.} 2013, \solphys, 283, 579

\bibitem[{{Lites} {et~al.}(2014){Lites}, {Centeno}, \& {McIntosh}}]{lites14}
{Lites}, B.~W., {Centeno}, R., \& {McIntosh}, S.~W. 2014, \pasj, 66, S4

\bibitem[{{Lites} {et~al.}(2004){Lites}, {Scharmer}, {Berger}, \&
  {Title}}]{lites04}
{Lites}, B.~W., {Scharmer}, G.~B., {Berger}, T.~E., \& {Title}, A.~M. 2004,
  \solphys, 221, 65

\bibitem[{{Lockwood} {et~al.}(1992){Lockwood}, {Skiff}, {Baliunas}, \&
  {Radick}}]{Lockwood92}
{Lockwood}, G.~W., {Skiff}, B.~A., {Baliunas}, S.~L., \& {Radick}, R.~R. 1992,
  \nat, 360, 653

\bibitem[{{Lockwood}(2005)}]{lockwood05}
{Lockwood}, M. 2005, in Saas-Fee Advanced Course 34: The Sun, Solar Analogs and
  the Climate, ed. J.~D. {Haigh}, M.~{Lockwood}, M.~S. {Giampapa},
  I.~{R{\"u}edi}, M.~{G{\"u}del}, \& W.~{Schmutz}, 109--306

\bibitem[{{L{\"o}ptien} {et~al.}(2016{\natexlab{a}}){L{\"o}ptien}, {Birch},
  {Duvall}, {Gizon}, \& {Schou}}]{loeptien16a}
{L{\"o}ptien}, B., {Birch}, A.~C., {Duvall}, T.~L., {Gizon}, L., \& {Schou}, J.
  2016{\natexlab{a}}, \aap, 587, A9

\bibitem[{{L{\"o}ptien} {et~al.}(2016{\natexlab{b}}){L{\"o}ptien}, {Birch},
  {Duvall}, {Gizon}, \& {Schou}}]{loeptien16}
{L{\"o}ptien}, B., {Birch}, A.~C., {Duvall}, T.~L., {Gizon}, L., \& {Schou}, J.
  2016{\natexlab{b}}, \aap, 590, A130

\bibitem[{{L{\"o}ptien} {et~al.}(2015){L{\"o}ptien}, {Birch}, {Gizon}, {Schou},
  {Appourchaux}, {Blanco Rodr{\'{\i}}guez}, {Cally}, {Dominguez-Tagle},
  {Gandorfer}, {Hill}, {Hirzberger}, {Scherrer}, \& {Solanki}}]{loeptien15}
{L{\"o}ptien}, B., {Birch}, A.~C., {Gizon}, L., {et~al.} 2015, \ssr, 196, 251

\bibitem[{{L{\"o}ptien} {et~al.}(2018){L{\"o}ptien}, {Lagg}, {van Noort}, \&
  {Solanki}}]{loeptien18}
{L{\"o}ptien}, B., {Lagg}, A., {van Noort}, M., \& {Solanki}, S.~K. 2018, \aap,
  619, A42

\bibitem[{{Luhmann} {et~al.}(2007){Luhmann}, {Ledvina}, {Krauss-Varban},
  {Odstrcil}, \& {Riley}}]{Luhmann07}
{Luhmann}, J.~G., {Ledvina}, S.~A., {Krauss-Varban}, D., {Odstrcil}, D., \&
  {Riley}, P. 2007, Advances in Space Research, 40, 295

\bibitem[{{Lynch} {et~al.}(2004){Lynch}, {Antiochos}, {MacNeice}, {Zurbuchen},
  \& {Fisk}}]{lynch04}
{Lynch}, B.~J., {Antiochos}, S.~K., {MacNeice}, P.~J., {Zurbuchen}, T.~H., \&
  {Fisk}, L.~A. 2004, \apj, 617, 589

\bibitem[{{Makarov} {et~al.}(2003){Makarov}, {Tlatov}, \&
  {Sivaraman}}]{makarov03}
{Makarov}, V.~I., {Tlatov}, A.~G., \& {Sivaraman}, K.~R. 2003, \solphys, 214,
  41

\bibitem[{{Maksimovic} {et~al.}(2019){Maksimovic}, {Bale}, {Chust}, \&
  et~al.}]{Maksimovic2019a}
{Maksimovic}, M., {Bale}, S.~D., {Chust}, T., \& et~al. 2019, \aap, this voume

\bibitem[{{Marsch}(2006)}]{marsch06}
{Marsch}, E. 2006, Living Reviews in Solar Physics, 3, 1

\bibitem[{{Marsden} {et~al.}(2011){Marsden}, {M{\" u}ller}, {Antonucci},
  {Benz}, {DeForest}, {Harrison}, {Hassler}, {Horbury}, {Howard}, {Lweis},
  {Maksimiovic}, {Marsch}, {Mason}, {Mart{\'\i}nez Pillet}, {Owen}, {Rochus},
  {Rodriguez-Pacheco}, {Solanki}, {Velli}, {Vourlidas}, \&
  {Wimmer-Schweingruber}}]{marsden11a}
{Marsden}, R.~G., {M{\" u}ller}, D., {Antonucci}, E., {et~al.} 2011, {Solar
  Orbiter -- exploring the Sun-heliosphere connection }, Tech. rep., definition
  Study Report (Red Book

\bibitem[{{Mart{\'{\i}}nez-Oliveros} {et~al.}(2007){Mart{\'{\i}}nez-Oliveros},
  {Moradi}, {Besliu-Ionescu}, {Donea}, {Cally}, \& {Lindsey}}]{martinez07}
{Mart{\'{\i}}nez-Oliveros}, J.~C., {Moradi}, H., {Besliu-Ionescu}, D., {et~al.}
  2007, \solphys, 245, 121

\bibitem[{{Mart{\'{\i}}nez Pillet}(2007)}]{martinezpillet07}
{Mart{\'{\i}}nez Pillet}, V. 2007, in ESA Special Publication, Vol. 641, Second
  Solar Orbiter Workshop, 27

\bibitem[{{Mart{\'{\i}}nez Pillet} {et~al.}(2011){Mart{\'{\i}}nez Pillet}, {Del
  Toro Iniesta}, {{\'A}lvarez-Herrero}, {Domingo}, {Bonet}, {Gonz{\'a}lez
  Fern{\'a}ndez}, {L{\'o}pez Jim{\'e}nez}, {Pastor}, {Gasent Blesa}, {Mellado},
  {Piqueras}, {Aparicio}, {Balaguer}, {Ballesteros}, {Belenguer}, {Bellot
  Rubio}, {Berkefeld}, {Collados}, {Deutsch}, {Feller}, {Girela}, {Grauf},
  {Heredero}, {Herranz}, {Jer{\'o}nimo}, {Laguna}, {Meller}, {Men{\'e}ndez},
  {Morales}, {Orozco Su{\'a}rez}, {Ramos}, {Reina}, {Ramos},
  {Rodr{\'{\i}}guez}, {S{\'a}nchez}, {Uribe-Patarroyo}, {Barthol}, {Gandorfer},
  {Knoelker}, {Schmidt}, {Solanki}, \& {Vargas
  Dom{\'{\i}}nguez}}]{martinezpillet11}
{Mart{\'{\i}}nez Pillet}, V., {Del Toro Iniesta}, J.~C., {{\'A}lvarez-Herrero},
  A., {et~al.} 2011, \solphys, 268, 57

\bibitem[{{Mart{\'{\i}}nez Pillet} \& {V{\' a}zquez}(1993)}]{martinezpillet93}
{Mart{\'{\i}}nez Pillet}, V. \& {V{\' a}zquez}, M. 1993, \aap, 270, 494

\bibitem[{{Mathew} {et~al.}(2004){Mathew}, {Solanki}, {Lagg}, {Collados},
  {Borrero}, \& {Berdyugina}}]{mathew04}
{Mathew}, S.~K., {Solanki}, S.~K., {Lagg}, A., {et~al.} 2004, \aap, 422, 693

\bibitem[{{McQuillan} {et~al.}(2014){McQuillan}, {Mazeh}, \&
  {Aigrain}}]{McQuillan14}
{McQuillan}, A., {Mazeh}, T., \& {Aigrain}, S. 2014, \apjs, 211, 24

\bibitem[{{Metcalf} {et~al.}(2006){Metcalf}, {Leka}, {Barnes}, {Lites},
  {Georgoulis}, {Pevtsov}, {Balasubramaniam}, {Gary}, {Jing}, {Li}, {Liu},
  {Wang}, {Abramenko}, {Yurchyshyn}, \& {Moon}}]{metcalf06}
{Metcalf}, T.~R., {Leka}, K.~D., {Barnes}, G., {et~al.} 2006, \solphys, 237,
  267

\bibitem[{{Meunier} {et~al.}(2015){Meunier}, {Lagrange}, {Borgniet}, \&
  {Rieutord}}]{meunier15}
{Meunier}, N., {Lagrange}, A.-M., {Borgniet}, S., \& {Rieutord}, M. 2015, \aap,
  583, A118

\bibitem[{Michel {et~al.}(2013)Michel, Bubenhagen, Gr{\"u}rmann, Lange, Fiethe,
  \& Michalik}]{michel13}
Michel, H., Bubenhagen, F., Gr{\"u}rmann, K., {et~al.} 2013, in Adaptive
  Hardware and Systems (AHS), 2013 NASA/ESA, Torino, Italy

\bibitem[{{M{\"u}ller} {et~al.}(2013){M{\"u}ller}, {Marsden}, {St.~Cyr}, \&
  {Gilbert}}]{Mueller:2013a}
{M{\"u}ller}, D., {Marsden}, R.~G., {St.~Cyr}, O.~C., \& {Gilbert}, H.~R. 2013,
  \solphys, 285, 25

\bibitem[{{M{\"u}ller} {et~al.}(2019){M{\"u}ller}, {Zouganelis}, {St.~Cyr}, \&
  {Gilbert}}]{Mueller2019a}
{M{\"u}ller}, D., {Zouganelis}, I., {St.~Cyr}, O.~C., \& {Gilbert}, H.~R. 2019,
  \aap, this volume

\bibitem[{{Neckel} \& {Labs}(1984)}]{neckel84}
{Neckel}, H. \& {Labs}, D. 1984, \solphys, 90, 205

\bibitem[{{November} \& {Simon}(1988)}]{november88}
{November}, L.~J. \& {Simon}, G.~W. 1988, \apj, 333, 427

\bibitem[{{Ofman}(2010)}]{ofman10}
{Ofman}, L. 2010, Living Reviews in Solar Physics, 7, 4

\bibitem[{{Orozco Su{\'a}rez} \& {Del Toro Iniesta}(2007)}]{orozco}
{Orozco Su{\'a}rez}, D. \& {Del Toro Iniesta}, J.~C. 2007, \aap, 462, 1137

\bibitem[{{Ortiz} {et~al.}(2002){Ortiz}, {Solanki}, {Domingo}, {Fligge}, \&
  {Sanahuja}}]{ortiz02}
{Ortiz}, A., {Solanki}, S.~K., {Domingo}, V., {Fligge}, M., \& {Sanahuja}, B.
  2002, \aap, 388, 1036

\bibitem[{Osterloh {et~al.}(2009)Osterloh, Michalik, \& Fiethe}]{osterloh09}
Osterloh, B., Michalik, H., \& Fiethe, B. 2009, in Architecture of Computing
  Systems - {ARCS} 2009, 22nd International Conference, Delft, The Netherlands,
  March 10-13, 2009. Proceedings, 50--59

\bibitem[{{Owen}(2019)}]{Owen2019a}
{Owen}, C. e.~a. 2019, \aap, this volume

\bibitem[{{Parker}(1988)}]{parker88}
{Parker}, E.~N. 1988, \apj, 330, 474

\bibitem[{{Parnell} {et~al.}(2009){Parnell}, {DeForest}, {Hagenaar},
  {Johnston}, {Lamb}, \& {Welsch}}]{parnell09}
{Parnell}, C.~E., {DeForest}, C.~E., {Hagenaar}, H.~J., {et~al.} 2009, \apj,
  698, 75

\bibitem[{P\'erez-Grande {et~al.}(2016)P\'erez-Grande, Torralbo, Alonso,
  Gomez-Sanjuan, \& Fernandez-Rico}]{perezgrande16}
P\'erez-Grande, I., Torralbo, I., Alonso, G., Gomez-Sanjuan, A., \&
  Fernandez-Rico, G. 2016, in 46$^{th}$ International Conference on
  Environmental Systems, Vienna, Austria

\bibitem[{{Petrie}(2015)}]{petrie15}
{Petrie}, G.~J.~D. 2015, Living Reviews in Solar Physics, 12, 5

\bibitem[{{Petrovay}(2010)}]{petrovay10}
{Petrovay}, K. 2010, Living Reviews in Solar Physics, 7, 6

\bibitem[{{Poletto}(2015)}]{poletto15}
{Poletto}, G. 2015, Living Reviews in Solar Physics, 12, 7

\bibitem[{{Pomoell} \& {Poedts}(2018)}]{pomoell18}
{Pomoell}, J. \& {Poedts}, S. 2018, Journal of Space Weather and Space Climate,
  8, A35

\bibitem[{{Priest}(2014)}]{priest14}
{Priest}, E. 2014, {Magnetohydrodynamics of the Sun}

\bibitem[{{Priest} {et~al.}(2018){Priest}, {Chitta}, \& {Syntelis}}]{priest18}
{Priest}, E.~R., {Chitta}, L.~P., \& {Syntelis}, P. 2018, \apjl, 862, L24

\bibitem[{{Priest} \& {Forbes}(2002)}]{priest02}
{Priest}, E.~R. \& {Forbes}, T.~G. 2002, \aapr, 10, 313

\bibitem[{{Pulkkinen}(2007)}]{pulkkinen07}
{Pulkkinen}, T. 2007, Living Reviews in Solar Physics, 4, 1

\bibitem[{{Rabello-Soares} {et~al.}(1997){Rabello-Soares}, {Roca Cortes},
  {Jimenez}, {Andersen}, \& {Appourchaux}}]{rabello97}
{Rabello-Soares}, M.~C., {Roca Cortes}, T., {Jimenez}, A., {Andersen}, B.~N.,
  \& {Appourchaux}, T. 1997, \aap, 318, 970

\bibitem[{{Radick} {et~al.}(2018){Radick}, {Lockwood}, {Henry}, {Hall}, \&
  {Pevtsov}}]{radick18}
{Radick}, R.~R., {Lockwood}, G.~W., {Henry}, G.~W., {Hall}, J.~C., \&
  {Pevtsov}, A.~A. 2018, \apj, 855, 75

\bibitem[{{Rauer} {et~al.}(2014){Rauer}, {Catala}, {Aerts}, {Appourchaux},
  {Benz}, {Brandeker}, {Christensen-Dalsgaard}, {Deleuil}, {Gizon}, {Goupil},
  {G{\"u}del}, {Janot-Pacheco}, {Mas-Hesse}, {Pagano}, {Piotto}, {Pollacco},
  {Santos}, {Smith}, {Su{\'a}rez}, {Szab{\'o}}, {Udry}, {Adibekyan}, {Alibert},
  {Almenara}, {Amaro-Seoane}, {Eiff}, {Asplund}, {Antonello}, {Barnes},
  {Baudin}, {Belkacem}, {Bergemann}, {Bihain}, {Birch}, {Bonfils}, {Boisse},
  {Bonomo}, {Borsa}, {Brand{\~a}o}, {Brocato}, {Brun}, {Burleigh}, {Burston},
  {Cabrera}, {Cassisi}, {Chaplin}, {Charpinet}, {Chiappini}, {Church},
  {Csizmadia}, {Cunha}, {Damasso}, {Davies}, {Deeg}, {D{\'{\i}}az}, {Dreizler},
  {Dreyer}, {Eggenberger}, {Ehrenreich}, {Eigm{\"u}ller}, {Erikson}, {Farmer},
  {Feltzing}, {de Oliveira Fialho}, {Figueira}, {Forveille}, {Fridlund},
  {Garc{\'{\i}}a}, {Giommi}, {Giuffrida}, {Godolt}, {Gomes da Silva},
  {Granzer}, {Grenfell}, {Grotsch-Noels}, {G{\"u}nther}, {Haswell}, {Hatzes},
  {H{\'e}brard}, {Hekker}, {Helled}, {Heng}, {Jenkins}, {Johansen},
  {Khodachenko}, {Kislyakova}, {Kley}, {Kolb}, {Krivova}, {Kupka}, {Lammer},
  {Lanza}, {Lebreton}, {Magrin}, {Marcos-Arenal}, {Marrese}, {Marques},
  {Martins}, {Mathis}, {Mathur}, {Messina}, {Miglio}, {Montalban}, {Montalto},
  {Monteiro}, {Moradi}, {Moravveji}, {Mordasini}, {Morel}, {Mortier},
  {Nascimbeni}, {Nelson}, {Nielsen}, {Noack}, {Norton}, {Ofir}, {Oshagh},
  {Ouazzani}, {P{\'a}pics}, {Parro}, {Petit}, {Plez}, {Poretti}, {Quirrenbach},
  {Ragazzoni}, {Raimondo}, {Rainer}, {Reese}, {Redmer}, {Reffert},
  {Rojas-Ayala}, {Roxburgh}, {Salmon}, {Santerne}, {Schneider}, {Schou},
  {Schuh}, {Schunker}, {Silva-Valio}, {Silvotti}, {Skillen}, {Snellen}, {Sohl},
  {Sousa}, {Sozzetti}, {Stello}, {Strassmeier}, {{\v S}vanda}, {Szab{\'o}},
  {Tkachenko}, {Valencia}, {Van Grootel}, {Vauclair}, {Ventura}, {Wagner},
  {Walton}, {Weingrill}, {Werner}, {Wheatley}, \& {Zwintz}}]{rauer14}
{Rauer}, H., {Catala}, C., {Aerts}, C., {et~al.} 2014, Experimental Astronomy,
  38, 249

\bibitem[{{Reale}(2014)}]{reale14}
{Reale}, F. 2014, Living Reviews in Solar Physics, 11, 4

\bibitem[{{Rees} \& {Semel}(1979)}]{rees}
{Rees}, D.~E. \& {Semel}, M.~D. 1979, \aap, 74, 1

\bibitem[{{Reinhold} {et~al.}(2013){Reinhold}, {Reiners}, \&
  {Basri}}]{Reinhold13}
{Reinhold}, T., {Reiners}, A., \& {Basri}, G. 2013, \aap, 560, A4

\bibitem[{{Rempel}(2014)}]{rempel14}
{Rempel}, M. 2014, \apj, 789, 132

\bibitem[{{Ricker} {et~al.}(2016){Ricker}, {Vanderspek}, {Winn}, {Seager},
  {Berta-Thompson}, {Levine}, {Villasenor}, {Latham}, {Charbonneau}, {Holman},
  {Johnson}, {Sasselov}, {Szentgyorgyi}, {Torres}, {Bakos}, {Brown},
  {Christensen-Dalsgaard}, {Kjeldsen}, {Clampin}, {Rinehart}, {Deming}, {Doty},
  {Dunham}, {Ida}, {Kawai}, {Sato}, {Jenkins}, {Lissauer}, {Jernigan},
  {Kaltenegger}, {Laughlin}, {Lin}, {McCullough}, {Narita}, {Pepper},
  {Stassun}, \& {Udry}}]{ricker16}
{Ricker}, G.~R., {Vanderspek}, R., {Winn}, J., {et~al.} 2016, in \procspie,
  Vol. 9904, Space Telescopes and Instrumentation 2016: Optical, Infrared, and
  Millimeter Wave, 99042B

\bibitem[{{Rochus} {et~al.}(2019){Rochus}, {Auch{\` e}re}, {Berghmans}, \&
  et~al.}]{Rochus2019a}
{Rochus}, P., {Auch{\` e}re}, F., {Berghmans}, D., \& et~al. 2019, \aap, This
  Volume

\bibitem[{{Rodr{\'\i}guez-Pacheco} {et~al.}(2019){Rodr{\'\i}guez-Pacheco},
  {Wimmer-Schweingruber}, {Mason}, \& et~al.}]{Rodriguez2019a}
{Rodr{\'\i}guez-Pacheco}, J., {Wimmer-Schweingruber}, R.~F., {Mason}, G.~M., \&
  et~al. 2019, \aap, this volume

\bibitem[{{Rouillard} {et~al.}(2019){Rouillard}, {Pinto}, {Vourlidas}, \&
  et~al.}]{Rouillard2019a}
{Rouillard}, A.~P., {Pinto}, R.~F., {Vourlidas}, A., \& et~al. 2019, \aap, this
  volume

\bibitem[{{Rust} {et~al.}(1988){Rust}, {Appourchaux}, \& {Hill}}]{rust88}
{Rust}, D.~M., {Appourchaux}, T., \& {Hill}, F. 1988, in IAU Symposium, Vol.
  123, Advances in Helio- and Asteroseismology, ed. J.~{Christensen-Dalsgaard}
  \& S.~{Frandsen}, 475

\bibitem[{{Rust} {et~al.}(1986){Rust}, {Burton}, \& {Leistner}}]{rust86}
{Rust}, D.~M., {Burton}, C.~H., \& {Leistner}, A.~J. 1986, in \procspie, Vol.
  627, Instrumentation in astronomy VI, ed. D.~L. {Crawford}, 39--49

\bibitem[{{Sanchez} {et~al.}(2019){Sanchez}, {Lodiot}, {De Groof}, \&
  et~al.}]{Sanchez2019}
{Sanchez}, L., {Lodiot}, S., {De Groof}, A., \& et~al. 2019, \aap, this volume

\bibitem[{{Sanchez Almeida} \& {Lites}(1992)}]{SanchezAlmeida}
{Sanchez Almeida}, J. \& {Lites}, B.~W. 1992, \apj, 398, 359

\bibitem[{{Sanchis-Kilders} {et~al.}(2014){Sanchis-Kilders}, {Ferreres},
  {Gasent-Blesa}, {Osorno}, {Gilabert}, {Maset}, {Ejea}, {Esteve}, {Jordan},
  {Garrigos}, \& {Blanes}}]{sanchis14}
{Sanchis-Kilders}, E., {Ferreres}, A., {Gasent-Blesa}, J.~L., {et~al.} 2014, in
  ESA Special Publication, Vol. 719, ESA Special Publication, 22

\bibitem[{{Sanchis Kilders} {et~al.}(2016){Sanchis Kilders}, {Meller}, {Lopez
  Jimenez}, {Hirzberger}, {Laget}, {Gasent Blesa}, {Herranz de la Revila},
  {Osorno Caudet}, {Ferreres Sabater}, {Balaguer Jimenez}, {Jordan Martinez},
  {Esteve Gomez}, {Maset Sancho}, \& {Ejea Marti}}]{sanchis16}
{Sanchis Kilders}, E., {Meller}, R., {Lopez Jimenez}, A., {et~al.} 2016, in ESA
  Special Publication, Vol. 738, ESA Workshop on Aerospace EMS, 2

\bibitem[{{Scharmer} {et~al.}(2003){Scharmer}, {Bjelksjo}, {Korhonen},
  {Lindberg}, \& {Petterson}}]{scharmer03}
{Scharmer}, G.~B., {Bjelksjo}, K., {Korhonen}, T.~K., {Lindberg}, B., \&
  {Petterson}, B. 2003, in \procspie, Vol. 4853, Innovative Telescopes and
  Instrumentation for Solar Astrophysics, ed. S.~L. {Keil} \& S.~V. {Avakyan},
  341--350

\bibitem[{{Schatten}(1993)}]{schatten93}
{Schatten}, K.~H. 1993, \jgr, 98, 18

\bibitem[{{Schatten} {et~al.}(1978){Schatten}, {Scherrer}, {Svalgaard}, \&
  {Wilcox}}]{schatten78}
{Schatten}, K.~H., {Scherrer}, P.~H., {Svalgaard}, L., \& {Wilcox}, J.~M. 1978,
  \grl, 5, 411

\bibitem[{{Scherrer} {et~al.}(1995){Scherrer}, {Bogart}, {Bush}, {Hoeksema},
  {Kosovichev}, {Schou}, {Rosenberg}, {Springer}, {Tarbell}, {Title},
  {Wolfson}, {Zayer}, \& {MDI Engineering Team}}]{scherrer95}
{Scherrer}, P.~H., {Bogart}, R.~S., {Bush}, R.~I., {et~al.} 1995, \solphys,
  162, 129

\bibitem[{{Scherrer} {et~al.}(2012){Scherrer}, {Schou}, {Bush}, {Kosovichev},
  {Bogart}, {Hoeksema}, {Liu}, {Duvall}, {Zhao}, {Title}, {Schrijver},
  {Tarbell}, \& {Tomczyk}}]{scherrer12}
{Scherrer}, P.~H., {Schou}, J., {Bush}, R.~I., {et~al.} 2012, \solphys, 275,
  207

\bibitem[{{Schlichenmaier} \& {Collados}(2002)}]{Schlichenmaier}
{Schlichenmaier}, R. \& {Collados}, M. 2002, \aap, 381, 668

\bibitem[{{Schmidt} \& {Fritz}(2004)}]{schmidt04}
{Schmidt}, W. \& {Fritz}, G. 2004, \aap, 421, 735

\bibitem[{{Schmidt} {et~al.}(2012){Schmidt}, {von der L{\"u}he}, {Volkmer},
  {Denker}, {Solanki}, {Balthasar}, {Bello Gonzalez}, {Berkefeld}, {Collados},
  {Fischer}, {Halbgewachs}, {Heidecke}, {Hofmann}, {Kneer}, {Lagg}, {Nicklas},
  {Popow}, {Puschmann}, {Schmidt}, {Sigwarth}, {Sobotka}, {Soltau}, {Staude},
  {Strassmeier}, \& {Waldmann }}]{schmidt12}
{Schmidt}, W., {von der L{\"u}he}, O., {Volkmer}, R., {et~al.} 2012,
  Astronomische Nachrichten, 333, 796

\bibitem[{{Schou}(2015)}]{Schou15}
{Schou}, J. 2015, \aap, 580, L11

\bibitem[{{Schou} {et~al.}(1998){Schou}, {Antia}, {Basu}, {Bogart}, {Bush},
  {Chitre}, {Christensen-Dalsgaard}, {Di Mauro}, {Dziembowski}, {Eff-Darwich},
  {Gough}, {Haber}, {Hoeksema}, {Howe}, {Korzennik}, {Kosovichev}, {Larsen},
  {Pijpers}, {Scherrer}, {Sekii}, {Tarbell}, {Title}, {Thompson}, \&
  {Toomre}}]{schou98}
{Schou}, J., {Antia}, H.~M., {Basu}, S., {et~al.} 1998, \apj, 505, 390

\bibitem[{{Schou} {et~al.}(2012{\natexlab{a}}){Schou}, {Borrero}, {Norton},
  {Tomczyk}, {Elmore}, \& {Card}}]{schou12a}
{Schou}, J., {Borrero}, J.~M., {Norton}, A.~A., {et~al.} 2012{\natexlab{a}},
  \solphys, 275, 327

\bibitem[{{Schou} {et~al.}(2012{\natexlab{b}}){Schou}, {Scherrer}, {Bush},
  {Wachter}, {Couvidat}, {Rabello-Soares}, {Bogart}, {Hoeksema}, {Liu},
  {Duvall}, {Akin}, {Allard}, {Miles}, {Rairden}, {Shine}, {Tarbell}, {Title},
  {Wolfson}, {Elmore}, {Norton}, \& {Tomczyk}}]{schou12}
{Schou}, J., {Scherrer}, P.~H., {Bush}, R.~I., {et~al.} 2012{\natexlab{b}},
  \solphys, 275, 229

\bibitem[{{Schrijver} \& {De Rosa}(2003)}]{schrijver03}
{Schrijver}, C.~J. \& {De Rosa}, M.~L. 2003, \solphys, 212, 165

\bibitem[{{Schwenn}(2006)}]{schwenn06}
{Schwenn}, R. 2006, Living Reviews in Solar Physics, 3, 2

\bibitem[{{Semel}(1967)}]{semel}
{Semel}, M. 1967, Annales d'Astrophysique, 30, 513

\bibitem[{{Shapiro} {et~al.}(2017){Shapiro}, {Solanki}, {Krivova}, {Cameron},
  {Yeo}, \& {Schmutz}}]{shapiro17}
{Shapiro}, A.~I., {Solanki}, S.~K., {Krivova}, N.~A., {et~al.} 2017, Nature
  Astronomy, 1, 612

\bibitem[{{Shapiro} {et~al.}(2016){Shapiro}, {Solanki}, {Krivova}, {Yeo}, \&
  {Schmutz}}]{shapiro16}
{Shapiro}, A.~I., {Solanki}, S.~K., {Krivova}, N.~A., {Yeo}, K.~L., \&
  {Schmutz}, W.~K. 2016, \aap, 589, A46

\bibitem[{{Sheeley} {et~al.}(1997){Sheeley}, {Wang}, {Hawley}, {Brueckner},
  {Dere}, {Howard}, {Koomen}, {Korendyke}, {Michels}, {Paswaters}, {Socker},
  {St.~Cyr}, {Wang}, {Lamy}, {Llebaria}, {Schwenn}, {Simnett}, {Plunkett}, \&
  {Biesecker}}]{sheeley97}
{Sheeley}, N.~R., {Wang}, Y.-M., {Hawley}, S.~H., {et~al.} 1997, \apj, 484, 472

\bibitem[{{Sheeley}(1991)}]{sheeley91}
{Sheeley}, Jr., N.~R. 1991, \apj, 374, 386

\bibitem[{{Shibata} \& {Magara}(2011)}]{shibata11}
{Shibata}, K. \& {Magara}, T. 2011, Living Reviews in Solar Physics, 8, 6

\bibitem[{{Shiota} {et~al.}(2012){Shiota}, {Tsuneta}, {Shimojo}, {Sako},
  {Orozco Su{\'a}rez}, \& {Ishikawa}}]{shiota12}
{Shiota}, D., {Tsuneta}, S., {Shimojo}, M., {et~al.} 2012, \apj, 753, 157

\bibitem[{{Silva-L\' opez} {et~al.}(2015){Silva-L\' opez}, {Garranzo-Garc\'\i
  a}, {S\' anchez}, {Bonet}, {Nu\~ nez}, \& {Álvarez-Herrero}}]{silvalopez15}
{Silva-L\' opez}, M., {Garranzo-Garc\'\i a}, D., {S\' anchez}, A., {et~al.}
  2015, Optical Engineering, 54, 54

\bibitem[{{Silva-L\'opez} {et~al.}(2017){Silva-L\'opez}, {Bastide}, {Restrepo},
  {Garc\'\i a Parejo}, \& {\' Alvarez-Herrero}}]{silvalopez17}
{Silva-L\'opez}, M., {Bastide}, L., {Restrepo}, R., {Garc\'\i a Parejo}, P., \&
  {\' Alvarez-Herrero}, A. 2017, Sensors and Actuators A, 266, 247

\bibitem[{{Solanki}(2003)}]{solanki03}
{Solanki}, S.~K. 2003, \aapr, 11, 153

\bibitem[{{Solanki} {et~al.}(2010){Solanki}, {Barthol}, {Danilovic}, {Feller},
  {Gandorfer}, {Hirzberger}, {Riethm{\"u}ller}, {Sch{\"u}ssler}, {Bonet},
  {Mart{\'{\i}}nez Pillet}, {del Toro Iniesta}, {Domingo}, {Palacios},
  {Kn{\"o}lker}, {Bello Gonz{\'a}lez}, {Berkefeld}, {Franz}, {Schmidt}, \&
  {Title}}]{solanki10}
{Solanki}, S.~K., {Barthol}, P., {Danilovic}, S., {et~al.} 2010, \apjl, 723,
  L127

\bibitem[{{Solanki} {et~al.}(2006){Solanki}, {Inhester}, \&
  {Sch{\"u}ssler}}]{solanki06}
{Solanki}, S.~K., {Inhester}, B., \& {Sch{\"u}ssler}, M. 2006, Reports on
  Progress in Physics, 69, 563

\bibitem[{{Solanki} {et~al.}(2013){Solanki}, {Krivova}, \& {Haigh}}]{solanki13}
{Solanki}, S.~K., {Krivova}, N.~A., \& {Haigh}, J.~D. 2013, \araa, 51, 311

\bibitem[{{Solanki} {et~al.}(2017){Solanki}, {Riethm{\"u}ller}, {Barthol},
  {Danilovic}, {Deutsch}, {Doerr}, {Feller}, {Gandorfer}, {Germerott}, {Gizon},
  {Grauf}, {Heerlein}, {Hirzberger}, {Kolleck}, {Lagg}, {Meller}, {Tomasch},
  {van Noort}, {Blanco Rodr{\'{\i}}guez}, {Gasent Blesa}, {Balaguer
  Jim{\'e}nez}, {Del Toro Iniesta}, {L{\'o}pez Jim{\'e}nez}, {Orozco Suarez},
  {Berkefeld}, {Halbgewachs}, {Schmidt}, {{\'A}lvarez-Herrero},
  {Sabau-Graziati}, {P{\'e}rez Grande}, {Mart{\'{\i}}nez Pillet}, {Card},
  {Centeno}, {Kn{\"o}lker}, \& {Lecinski}}]{solanki17}
{Solanki}, S.~K., {Riethm{\"u}ller}, T.~L., {Barthol}, P., {et~al.} 2017,
  \apjs, 229, 2

\bibitem[{{Solanki} {et~al.}(1993){Solanki}, {Walther}, \&
  {Livingston}}]{solanki93}
{Solanki}, S.~K., {Walther}, U., \& {Livingston}, W. 1993, \aap, 277, 639

\bibitem[{{SPICE Consortium et al.} {et~al.}(2019){SPICE Consortium et al.},
  {Anderson}, {Appourchaux}, {Auchere}, \& et~al.}]{SpiceConsortium2019}
{SPICE Consortium et al.}, {Anderson}, M., {Appourchaux}, T., {Auchere}, F., \&
  et~al. 2019, \aap, this volume

\bibitem[{{Staub} {et~al.}(2019){Staub}, {Oberdorfer}, {Gandorfer}, {Grauf},
  {Deutsch}, {Hirzberger}, {M{\"u}ller}, \& {Woch}}]{staub19}
{Staub}, J.~M., {Oberdorfer}, D., {Gandorfer}, A., {et~al.} 2019, submitted to
  CEAS Space Journal

\bibitem[{{Stein}(2012)}]{stein12}
{Stein}, R.~F. 2012, Living Reviews in Solar Physics, 9, 4

\bibitem[{{Topka} {et~al.}(1997){Topka}, {Tarbell}, \& {Title}}]{topka97}
{Topka}, K.~P., {Tarbell}, T.~D., \& {Title}, A.~M. 1997, \apj, 484, 479

\bibitem[{{Tsuneta} {et~al.}(2008{\natexlab{a}}){Tsuneta}, {Ichimoto},
  {Katsukawa}, {Lites}, {Matsuzaki}, {Nagata}, {Orozco Su{\'a}rez}, {Shimizu},
  {Shimojo}, {Shine}, {Suematsu}, {Suzuki}, {Tarbell}, \& {Title}}]{tsuneta08a}
{Tsuneta}, S., {Ichimoto}, K., {Katsukawa}, Y., {et~al.} 2008{\natexlab{a}},
  \apj, 688, 1374

\bibitem[{{Tsuneta} {et~al.}(2008{\natexlab{b}}){Tsuneta}, {Ichimoto},
  {Katsukawa}, {Nagata}, {Otsubo}, {Shimizu}, {Suematsu}, {Nakagiri},
  {Noguchi}, {Tarbell}, {Title}, {Shine}, {Rosenberg}, {Hoffmann}, {Jurcevich},
  {Kushner}, {Levay}, {Lites}, {Elmore}, {Matsushita}, {Kawaguchi}, {Saito},
  {Mikami}, {Hill}, \& {Owens}}]{tsuneta08}
{Tsuneta}, S., {Ichimoto}, K., {Katsukawa}, Y., {et~al.} 2008{\natexlab{b}},
  \solphys, 249, 167

\bibitem[{{Tu} {et~al.}(2005){Tu}, {Zhou}, {Marsch}, {Xia}, {Zhao}, {Wang}, \&
  {Wilhelm}}]{tu05}
{Tu}, C.-Y., {Zhou}, C., {Marsch}, E., {et~al.} 2005, Science, 308, 519

\bibitem[{{Uribe-Patarroyo} {et~al.}(2011){Uribe-Patarroyo}, {Alvarez-Herrero},
  {Garc{\'{\i}}a Parejo}, {Vargas}, {Heredero}, {Restrepo}, {Mart{\'{\i}}nez
  Pillet}, {del Toro Iniesta}, {L{\'o}pez}, {Fineschi}, {Capobianco},
  {Georges}, {L{\'o}pez}, {Boer}, \& {Manolis}}]{uribe11}
{Uribe-Patarroyo}, N., {Alvarez-Herrero}, A., {Garc{\'{\i}}a Parejo}, P.,
  {et~al.} 2011, in \procspie, Vol. 8148, Solar Physics and Space Weather
  Instrumentation IV, 814810

\bibitem[{{Velli} {et~al.}(2019){Velli}, {M{\"u}ller}, {Zouganelis}, \& et~al.
  TBD}]{Velli2019a}
{Velli}, M., {M{\"u}ller}, D., {Zouganelis}, I., \& et~al. TBD. 2019, \aap,
  this volume

\bibitem[{{Vieira} {et~al.}(2012){Vieira}, {Norton}, {Dudok de Wit},
  {Kretzschmar}, {Schmidt}, \& {Cheung}}]{vieira12}
{Vieira}, L.~E.~A., {Norton}, A., {Dudok de Wit}, T., {et~al.} 2012, \grl, 39,
  L16104

\bibitem[{{V{\"o}gler} \& {Sch{\"u}ssler}(2007)}]{voegler07}
{V{\"o}gler}, A. \& {Sch{\"u}ssler}, M. 2007, \aap, 465, L43

\bibitem[{{Volkmer} {et~al.}(2012){Volkmer}, {Bosch}, {Feger}, {Gomez},
  {Heidecke}, {Schmidt}, {Scheiffelen}, {Sigwarth}, \& {Soltau}}]{volkmer12}
{Volkmer}, R., {Bosch}, J., {Feger}, B., {et~al.} 2012, in \procspie, Vol.
  8442, Space Telescopes and Instrumentation 2012: Optical, Infrared, and
  Millimeter Wave, 84424P

\bibitem[{{Wang} {et~al.}(1989){Wang}, {Nash}, \& {Sheeley}}]{wang89}
{Wang}, Y.-M., {Nash}, A.~G., \& {Sheeley}, Jr., N.~R. 1989, Science, 245, 712

\bibitem[{{Wang} \& {Sheeley}(1995)}]{wang95}
{Wang}, Y.-M. \& {Sheeley}, Jr., N.~R. 1995, \apj, 452, 457

\bibitem[{{Warner} {et~al.}(2018){Warner}, {Rimmele}, {Mart{\'{\i}}nez Pillet},
  {Casini}, {Berukoff}, {Craig}, {Ferayorni}, {Goodrich}, {Hubbard},
  {Harrington}, {Jeffers}, {Johansson}, {Kneale}, {Kuhn}, {Liang}, {Lin},
  {Marshall}, {Mathioudakis}, {McBride}, {McMullin}, {McVeigh}, {Sekulic},
  {Schmidt}, {Shimko}, {Sueoka}, {Summers}, {Tritschler}, {Williams}, \&
  {W{\"o}ger}}]{warner18}
{Warner}, M., {Rimmele}, T.~R., {Mart{\'{\i}}nez Pillet}, V., {et~al.} 2018, in
  Society of Photo-Optical Instrumentation Engineers (SPIE) Conference Series,
  Vol. 10700, Society of Photo-Optical Instrumentation Engineers (SPIE)
  Conference Series, 107000V

\bibitem[{{Wiegelmann} \& {Sakurai}(2012)}]{wiegelmann12}
{Wiegelmann}, T. \& {Sakurai}, T. 2012, Living Reviews in Solar Physics, 9, 5

\bibitem[{{Wiegelmann} {et~al.}(2014){Wiegelmann}, {Thalmann}, \&
  {Solanki}}]{wiegelmann14}
{Wiegelmann}, T., {Thalmann}, J.~K., \& {Solanki}, S.~K. 2014, \aapr, 22, 78

\bibitem[{{Wittrock} {et~al.}(2003){Wittrock}, {Reiche}, {St{\" o}ckner},
  {Michalik}, \& {Gliem}}]{wittrock03}
{Wittrock}, T., {Reiche}, K.-U., {St{\" o}ckner}, K., {Michalik}, H., \&
  {Gliem}, F. 2003, in 54th International Astronautical Congress of the
  International Astronautical Federation, the International Academy of
  Astronautics, and the International Institute of Space Law, International
  Astronautical Congress (IAF)

\bibitem[{{Witzke} {et~al.}(2018){Witzke}, {Shapiro}, {Solanki}, {Krivova}, \&
  {Schmutz}}]{witzke18}
{Witzke}, V., {Shapiro}, A.~I., {Solanki}, S.~K., {Krivova}, N.~A., \&
  {Schmutz}, W. 2018, ArXiv e-prints

\bibitem[{{Yeo} {et~al.}(2013){Yeo}, {Solanki}, \& {Krivova}}]{yeo13}
{Yeo}, K.~L., {Solanki}, S.~K., \& {Krivova}, N.~A. 2013, \aap, 550, A95

\bibitem[{{Yeo} {et~al.}(2017){Yeo}, {Solanki}, {Norris}, {Beeck}, {Unruh}, \&
  {Krivova}}]{yeo17}
{Yeo}, K.~L., {Solanki}, S.~K., {Norris}, C.~M., {et~al.} 2017, Physical Review
  Letters, 119

\bibitem[{{Zhao} {et~al.}(2012){Zhao}, {Nagashima}, {Bogart}, {Kosovichev}, \&
  {Duvall}}]{Zhao12}
{Zhao}, J., {Nagashima}, K., {Bogart}, R.~S., {Kosovichev}, A.~G., \& {Duvall},
  Jr., T.~L. 2012, \apjl, 749, L5

\bibitem[{{Zouganelis} {et~al.}(2019){Zouganelis}, {De Groof}, {Walsh}, \&
  et~al.}]{Zouganelis2019a}
{Zouganelis}, I., {De Groof}, A., {Walsh}, A., \& et~al. 2019, \aap, this
  volume

\end{thebibliography}

\end{document}